\documentclass[12pt,onecolumn]{report}
\usepackage{amsmath}
\usepackage{amssymb}
\usepackage{epsfig}
\usepackage[english,russian]{babel}
\usepackage{bm}
\usepackage{indentfirst}
\makeatletter
\renewcommand{\@biblabel}[1]{#1.}
\makeatother
\begin{document}

\thispagestyle{empty}
\begin{center}
\vspace{10mm}
\large { Российская академия наук} \\
\large { Уральское отделение} \\
\large { Коми научный центр} \\
\large { Отдел математики} \\
\vspace{15mm}
\hspace {7cm} На правах рукописи \\
\hspace {5cm} УДК 539.144\\
\vspace{15mm}
\large { Попов Константин Геннадьевич} \\
\vspace{15mm}
\large {\tt \bf Теория Ферми конденсатного квантового фазового перехода} \\
\end{center}
\vspace{15mm}
\begin{center}
\large {
Доклад \\
на семинаре отдела теоретической и математической физики \\
Института физики металлов \\
УрО РАН\\
\vspace{10mm}
\vspace{10mm}
Екатеринбург \\
14.10.2008 \\
}
\end{center}
\newpage
\noindent
Abstract\\

\noindent
This report is devoted to the building of the theory and analyzing the phenomenon which take place in such strong correlated Fermi systems as High temperature superconductors, metals with heavy fermions and quasi two dimensional Fermi-systems. The described theory of strong correlated Fermi systems bases on conceptions of Fermi condensate quantum phase transition (FCQPT) and Fermi condensate. It develops the well known Landau paradigm based on quasi particles formalism and order parameter. The description of the FCQPT theory is followed by its verification on the different exactly solvable models, by numerical calculations and by comparing with wide collection of experimental data.\\ 

\noindent
Аннотация.\\

\noindent
Этот доклад посвящен построению теории и анализу явлений, которые имеют место в таких сильно коррелированных Ферми системах, как Высоко температурные сверхпроводники, металлы с тяжелыми фермионами и квазидвухмерные Ферми-системы. Описываемая теория сильно коррелированных Ферми систем базируется на концепциях Ферми конденсатного квантового фазового перехода (ФККФП) и Ферми конденсата. Она развивает хорошо известную парадигму Ландау, основанную на формализме квазичастиц и параметре порядка. Описание теории ФККФП сопровождается ее верификацией на различных точно решаемых моделях, с помощью численных расчетов и путем сравнения с широким набором экспериментальных данных.\\
\newpage
\chapter* {1. Введение}
\label{INTR}
\addcontentsline{toc}{chapter}{ 1. Введение}

Теория ферми-жидкости Ландау (ФЖЛ) имеет давнюю историю и блестящие
результаты в описании многообразных свойств электронной жидкости
обычных металлов и ферми-жидкостей типа $^3$He. Эта теория
построена в предположении, что физику при низких температурах
определяют элементарные возбуждения, которые ведут себя как
квазичастицы, характеризующиеся эффективной массой и находящиеся по
основным своим свойствам в одном классе с квазичастицами слабо
взаимодействующего ферми-газа. Поэтому эффективная масса $M^*$ не
зависит от температуры, давления, магнитного поля и является
параметром теории.

Открытый относительно недавно новый класс сильно коррелированных
ферми-систем, таких как металлы с тяжелыми фермионами (ТФ),
высокотемпературные сверхпроводники (ВТСП) и квази-двумерные
ферми-жидкости, обнаруживает огромное разнообразие физических
свойств \cite{ste,varma,vojta,voj,belkop}. Свойства этих материалов
принципиально отличаются от свойств обычных ферми-систем. Например,
в случае металлов с тяжелыми фермионами сильная корреляция
электронов приводит к перенормировке эффективной массы квазичастиц,
которая может превысить голую массу на несколько порядков или даже
стать неограниченно большой. При этом эффективная масса
демонстрирует сильную зависимость от температуры, давления или
приложенного магнитного поля. Такие металлы имеют аномальное
поведение и необычные степенные законы температурной зависимости
своих термодинамических свойств при низких температурах, а их
поведение принято определять как поведение аномальной
ферми-жидкости Ландау.

Неспособность теории ферми-жидкости Ландау объяснить
экспериментальные наблюдения, связанные с зависимостью $M^*$ от
температуры $T$, магнитного поля $B$, давления и т.д., привело к
заключению, что квазичастицы не выживают в сильно коррелированных
ферми-системах, и тяжелый электрон не сохраняет своей целостности
как возбуждение-квазичастица
\cite{cust,sen,senth1,senth3,col1,col2}.

\section* { 1.1. Квантовые фазовые переходы и аномальное поведение коррелированных ферми-систем }
\addcontentsline{toc}{section}{ 1.1. Квантовые фазовые переходы и аномальное поведение коррелированных ферми-систем }

Предполагается, что необычные свойства и аномальное поведение,
наблюдаемые в ВТСП и металлах с ТФ, определяются разнообразными
магнитными квантовыми фазовыми переходами \cite{vojta}. Поскольку
квантовый фазовый переход происходит при температуре $T=0$, то
контролирующими параметрами являются такие как состав, плотность
числа электронов (или дырок) $x$, давление, напряженность
магнитного поля $B$ и т.д. Квантовый фазовый переход  имеет место в
квантовой критической точке, которая отделяет упорядоченную фазу,
являющуюся результатом квантового фазового перехода, от
неупорядоченной фазы. Обычно предполагается, что магнитные
(например, ферромагнитные и антиферромагнитные) квантовые фазовые
переходы ответственны за аномальное поведение. Критическую точку
такого фазового перехода смещают в абсолютный нуль температур при
помощи указанных выше параметров.

Можно ожидать, что универсальное поведение наблюдается только в том
случае, когда рассматриваемая система находится очень близко к
квантовой критической точке, например, когда длина корреляций
намного больше, чем микроскопический масштаб длины, и критические
квантовые и тепловые флуктуации определяют аномальный вклад в
термодинамические функции металла. Квантовые фазовые переходы
такого вида весьма распространены \cite{varma,vojta,voj}, поэтому
мы будем называть их обычными квантовыми фазовыми переходами. В
этом случае, физику явления определяют тепловые и квантовые
флуктуации критического состояния, а квазичастичные возбуждения
разрушены этими флуктуациями. Считается, что отсутствие
квазичастичных возбуждений является основной причиной аномального
поведения металлов с тяжелыми фермионами и высокотемпературных
сверхпроводников \cite{varma,vojta,voj}. Однако, на этом пути есть
трудности. Критическое поведение, наблюдаемое в экспериментах с
металлами с ТФ, имеет место до довольно высоких температур,
сопоставимых с эффективной температурой Ферми $T_k$. Например,
коэффициент теплового расширения $\alpha(T)$, который в случае
нормальной ферми-жидкости Ландау $\alpha(T)\propto T$, при
измерениях на CeNi$_2$Ge$_2$, демонстрирует зависимость $\sqrt{T}$
при изменении температуры на два порядка, при ее снижение от 6 K до
по крайней мере 50 мК \cite{geg1}. Вряд ли можно объяснить такое
поведение на основе флуктуационной теории критической точки.
Очевидно, что такая ситуация может иметь место только при $T\to0$,
когда критические флуктуации вносят доминирующий вклад в энтропию,
и когда длина корреляций намного больше чем микроскопический
масштаб длины. При некоторой температуре $T\ll T_k$ эта
макроскопически большая длина корреляций должна быть разрушена
обычными тепловыми флуктуациями и соответствующее универсальное
поведение исчезнуть. Следующая трудность связана с объяснением
восстановления поведения ФЖЛ под воздействием магнитного поля $B$,
наблюдаемого в металлах с ТФ и ВТСП \cite{geg,ste,cyr}. В случае
ФЖЛ при $T\to0$, электрическое сопротивление $\rho(T)=\rho_0+AT^2$,
теплоемкость $C(T)=\gamma T$ и магнитная восприимчивость
$\chi=const$. Оказывается, что зависимости от магнитного поля
коэффициента $A(B)$, коэффициента Зоммерфельда $\gamma (B)$ и
восприимчивости $\chi(B)$ таковы, что $A(B)\propto\gamma^2(B)$ и
$A(B)\propto\chi^2(B)$, из чего следует, что соотношение
Кадоваки-Вудса (КВ), $K=A (B)/\gamma^2 (B)$ \cite{kadw} является
$B$-независимым и сохраняется \cite{geg}. Такое универсальное
поведение, вполне естественное при определяющей роли квазичастиц,
едва ли возможно объяснить в рамках подхода, предполагающего
отсутствие квазичастиц, которое имеет место при обычных квантовых
фазовых переходах в окрестности квантовой критической точки.
Например, соотношение КВ не соответствует сценарию волн спиновой
плотности \cite{geg} и результатам исследований квантовой
критичности с помощью ренорм-группового подхода \cite{mill}. Затем,
измерения переноса заряда и тепла показали, что закон Видемана -
Франца (ВФ) выполняется в ряде случаев ВТСП \cite{cyr,cyr1}  и
металлов с ТФ \cite{pag2,ronn1}. Поэтому мы можем заключить, что
квазичастицы существуют в этих металлах, о чем свидетельствуют и
результаты фотоэмиссионной спектроскопии \cite{koral,fujim}.

Невозможность объяснять поведение металлов с тяжелыми фермионами в
рамках теорий, основанных на обычных квантовых фазовых переходах,
ведет к заключению, что другое важное понятие, введенное Ландау,
--- понятие параметра порядка --- также не работает (см., например,
\cite{senth1, senth3, col1, col2}). Таким образом, мы остаемся без
наиболее фундаментальных принципов квантовой физики многих тел, а
широкий круг интересных явлений, связанных с аномальным поведением
сильно коррелированных ферми-систем, остается без объяснений.

\section* { 1.2. Ряд вводных замечаний и ограничений } 
\addcontentsline{toc}{section}{ 1.2. Ряд вводных замечаний и ограничений }

В данном докладе будет показано, что такие разнородные системы, как ВТСП,
металлы с ТФ и квази-двумерные сильно коррелированные
ферми-жидкости, обладают общим универсальным поведением, которое
можно описать в рамках единого подхода, основанного на теории
фермионной конденсации \cite{ks,ksk}. С другой стороны,
сосредоточившись на универсальном поведении и располагая ограниченным
объемом доклада, мы не имеем возможности рассматривать специфику
отдельных сильно коррелированных систем. Например, мы оставляем в
стороне физику таких ферми-систем как нейтронные звезды, атомные
кластеры и ядра, кварковая плазма и ультрахолодные газы в ловушках,
в которых возможно существование фермионного конденсата (ФК)
\cite{khod3,amsh3,khod4,volovik1,volovik2}. Последние интересны
тем, что благодаря их легкой настройке, можно осуществить
параметры, необходимые для наблюдений квантовой критической точки и
ФК.

Экспериментальные исследования свойств квантовых фазовых переходов
и их критических точек являются принципиально важными для понимания
физической природы высокотемпературной сверхпроводимости и металлов
с ТФ. Экспериментальные данные, относящиеся к различным сильно
коррелированным ферми системам, являются взаимодополняющими. В
случае высокотемпературной сверхпроводимости такие эксперименты
практически отсутствуют, поскольку при низких температурах
соответствующие критические точки находятся в области
сверхпроводимости, и физические свойства соответствующего
квантового фазового перехода изменены сверхпроводимостью. Однако
для металлов с тяжелыми фермионами эти исследования вполне
возможны. Недавние экспериментальные данные, относящиеся к
поведению металлов с ТФ, проливают свет на природу критических
точек и фазовых переходов (см., например,
\cite{geg,cyr,cyr1,pag2,ronn1,koral,fujim}). Поэтому существенно
важным является одновременное изучение высокотемпературной
сверхпроводимости и аномального поведения металлов с ТФ.

Чтобы избежать сложностей, связанных с анизотропией, порождаемой
кристаллической решеткой твердых тел, ее спецификой,
нерегулярностями и т.п., мы используем модель однородной тяжелой
электронной (фермионной) жидкости для изучения универсального
поведения электронных (дырочных) систем ВТСП, металлов с ТФ  и
квази-двумерных ферми-систем при низких температурах. Такая модель
вполне работоспособна, так как мы рассматриваем универсальное
поведение, демонстрируемое этими материалами при низких
температурах, которое связано со степенными расходимостями таких
величин как эффективная масса, теплоемкость, тепловое расширение и
т.д. Эти расходимости, или характеризующие их критические индексы,
определяются малой величиной передаваемых импульсов по сравнению с
импульсами порядка величины обратной постоянной решетки, поэтому
вклады этих импульсов не определяют динамику системы, связанную с
малыми импульсами, и могут быть опущены, как это делается,
например, при построении флуктуационной теории критических индексов
\cite{lanl1}. Таким же образом, мы можем игнорировать сложности и
специфику, связанные с конкретным ВТСП или металлом с ТФ.

Универсальные свойства сильно коррелированных ферми-систем мы будем
анализировать в рамках теории  фермионной конденсации
\cite{ks,ksk,shag4}, поскольку само поведение металлов с ТФ
предлагает ассоциировать их необычные свойства с квантовым фазовым
переходом, связанным с неограниченным ростом эффективной массы в
его критической точке. Например, расходимость эффективной массы
наблюдалась в индуцированной магнитным полем квантовой критической
точке \cite{geg,pag,cust,pag2,ronn1}. Таким квантовым фазовым
переходом является ферми-конденсатный квантовый фазовый переход
(ФККФП), ведущий к образованию ФК. Основной особенностью ФККФП
является расходимость эффективной массы $M^*$ в его квантовой
критической точке \cite{ks,ksk,shag4}. Мы будем предполагать, что
тяжелая электронная жидкость находится вблизи ФККФП или уже за
критической точкой этого фазового перехода.

\chapter* {2. Ферми-жидкость с фермионным конденсатом }
\label{FLFC}
\addcontentsline{toc}{chapter}{2. Ферми-жидкость с фермионным конденсатом }

Одной из самых сложных проблем современной физики конденсированного
состояния является вопрос о структуре и свойствах ферми-систем с
большими константами межчастичного взаимодействия. Впервые способ
решения таких проблем был предложен Ландау в теории ферми-жидкости,
позже названной "нормальной",  посредством введения понятия
квазичастиц и амплитуд, которые характеризуют эффективное
взаимодействие между ними \cite{landau,lanl1}. Теория Ландау может
рассматриваться как эффективная теория при низких энергиях, в
которой высокоэнергетичные степени свободы удалены путем введения
амплитуд, определяющих взаимодействия между квазичастицами при
низких температурах вместо сильного межчастичного взаимодействия.
Стабильность основного состояния ферми-жидкости Ландау определяется
условиями стабильности Померанчука: стабильность нарушается, когда
хотя бы одна из амплитуд Ландау становится отрицательной и
достигает критического значения \cite{lanl1,pom}. Отметим, что
новая фаза, в которой условия стабильности восстанавливаются, может
в принципе быть снова описана в рамках той же самой теории Ландау.

\section* { 2.1. Теория Ландау ферми-жидкости } \label{FL}
\addcontentsline{toc}{section}{ 2.1. Теория Ландау ферми-жидкости }

Начнем с краткого напоминания основных положений теории
ферми-жидкости Ландау \cite{landau, lanl1}. Теория ферми-жидкости
Ландау опирается на понятие квазичастиц, которые представляют
элементарные слабо возбужденные состояния ферми-жидкости. Поэтому
они являются специфическими возбуждениями, формирующими
низкотемпературные термодинамические и транспортные свойства
ферми-жидкости. В случае электронной жидкости квазичастицы
характеризуются квантовыми числами электрона и эффективной массой
$M ^*$. Энергия основного состояния рассматриваемой системы
является функционалом чисел заполнения квазичастиц (или функции
распределения квазичастиц) $n({\bf p},T)$, точно так же, как и
свободная энергия $F[n({\bf p},T)]$, энтропия $S[n({\bf p},T)]$ и
другие термодинамические функции. Из условия, что большой
термодинамический потенциал $\Omega[n({\bf p},T)]=F[n({\bf p},T)]-\mu N[n({\bf p},T)]$, где $F[n({\bf p},T)]=E[n({\bf p},T)]-TS[n({\bf p},T)]$     --- свободная энергия,
должен быть минимален, можно найти функцию распределения n({\bf p},T):
\begin{equation}\label{FL1} \frac{\delta(F-\mu N)}{\delta n({\bf p}, T)}=\varepsilon({\bf
p}, T) -\mu (T)-T\ln\frac{1-n({\bf p},T)}{n({\bf p},T)}=0.
\end{equation}
Здесь $\mu$ --- химический потенциал, а
\begin{equation} \varepsilon({\bf p},T) \ = \ \frac {\delta
E[n({\bf p},T)]} {\delta n({\bf p},T)}\,\label{FL2} \end{equation}
есть энергия квазичастицы. Эта энергия является функционалом
$n({\bf p},T)$ точно так же, как энергия $E[n({\bf p},T)]$.
Изменение энергии квазичастицы при малой вариации функции
распределения $\delta n({\bf p},T)$ должно иметь вид:
\begin{equation}\label{FL2IN1}
\delta \varepsilon[n({\bf p},T)] =
\int f({\bf p},{\bf p'})\delta n({\bf p'},T)
\frac{d{\bf p'}}{(2\pi)^3}.
\end{equation}
Здесь $ f({\bf p},{\bf p'}) $ функция или амплитуда взаимодействия
квазичастиц (в ферми-газе $ f({\bf p},{\bf p'})=0 $). По
своему определению эта функция представляет собой вторую вариационную производную от полной энергии системы $E[n({\bf p},T)]$:
\begin{equation} f({\bf p},{\bf p'}) = \ \frac {\delta^{2}
E[n({\bf p},T)]} {\delta n({\bf p},T)\delta n({\bf p'},T)}, \,\label{FL2IN2} \end{equation}
и поэтому симметрична по переменным $({\bf p},{\bf p'})$ и
соответствующим им парам спиновых индексов (мы будем по возможности опускать спиновые индексы для улучшения восприятия формул, хотя для однородных систем и малых магнитных полей влияние спиновой структуры на вычисляемые величины --- незначительно). С учетом взаимодействия энергия квазичастиц вблизи поверхности ферми-сферы дается суммой:
\begin{equation}\label{FL2IN3}
\varepsilon[n({\bf p},T)] = \varepsilon_{F} + v_{F}(p-p_{F}) +
\int f({\bf p},{\bf p'})\delta n({\bf p'},T)
\frac{d{\bf p'}}{(2\pi)^3}.
\end{equation}
В частности, для термодинамически равновесных распределений второй член в формуле (\ref{FL2IN3}) определяет зависимость энергии квазичастицы от температуры. Отклонение $\delta n({\bf p'},T)$ заметно отлично от нуля только в узком слое значений $  p' $ вблизи поверхности ферми-сферы, и в таком же слое находятся импульсы реальных квазичастиц. Поэтому функцию
$ f({\bf p},{\bf p'})=0 $ в формулах (\ref{FL2IN1}), (\ref{FL2IN3})
фактически можно заменить ее значением на самой этой поверхности, т.е.
положить $ {\bf p}={\bf p'}={\bf p_{F}}$, так что $ f({\bf p},{\bf p'})=0 $
будет зависеть только от направлений векторов  $ {\bf p},{\bf p'} $.

Энтропия $S[n({\bf p},T)]$ задается известным выражением
\cite{landau,lanl1}
\begin{eqnarray}
\nonumber
S[n({\bf p},T)]&=& -2\int[n({\bf p},T) \ln (n({\bf p},T))+(1-n({\bf p},T))\\
&\times&\ln (1-n({\bf p},T))]\frac{d{\bf p}}{(2\pi) ^3},\label{FL3}
\end{eqnarray}
которое следует из чисто комбинаторного рассмотрения. Уравнению
(\ref{FL1}) обычно придают стандартный вид распределения
Ферми-Дирака
\begin{equation} n({\bf p},T)=
\left\{1+\exp\left[\frac{(\varepsilon({\bf p},T)-\mu)}
{T}\right]\right\}^{-1}.\label{FL4} \end{equation} При $T\to 0 $,
из уравнений (\ref{FL1}) и (\ref{FL4}) получают стандартное решение
$n(p,T\to0)\to\theta(p_F-p)$, здесь $\theta (p_F-p)$ является
ступенчатой функцией, $\varepsilon(p\simeq
p_F)-\mu=p_F(p-p_F)/M^*_L$, где $M ^*_L$ --- эффективная масса
квазичастицы Ландау \cite{landau, lanl1}
\begin{equation}
\frac1{M^*_L}=\frac1p\, \frac{d\varepsilon(p,T=0)}{dp}|_{p=p_F}\
\label{FL5},\end{equation}
а $ v_{F} $ --- скорость квазичастиц на поверхности Ферми:
\begin{equation}
v_{F} = \frac{d\varepsilon(p,T=0)}{dp}|_{p=p_F}\
\label{FL5IN4}.\end{equation} Подразумевается, что $M ^*_L$ является
положительной и конечной на поверхности Ферми. В результате
$T$-зависящие поправки к $M ^*_L$, к энергии квазичастиц
$\varepsilon({\bf p})$ и к другим величинам начинаются с члена
пропорционального $T^2$.

В основе вывода соотношения, связывающего эффективную массу квазичастиц с массой фермионов системы, лежит принцип относительности Галилея. Функция взаимодействия квазичастиц удовлетворяет интегральному соотношению, следующему из этого принципа. А именно, прямым следствием этого принципа является совпадение импульса единицы объема квазичастиц ферми-жидкости
$ \int {\bf p}n({\bf p})\frac{\textstyle{d{\bf p}}}{\textstyle{(2\pi)^3}} $ с плотностью потока массы $ М $ реальных частиц $ \int M n({\bf p}) \frac{\textstyle{d\varepsilon(p)}}{\textstyle{dp}}
\frac{\textstyle{d{\bf p}}}{\textstyle{(2\pi)^3}} $. Таким образом, получим следующее равенство:
\begin{equation}\label{FL2IN5}
\int {\bf p}n({\bf p})\frac{d{\bf p}}{(2\pi)^3}=
\int M n({\bf p}) \frac{d\varepsilon(p)}{dp}
\frac{d{\bf p}}{(2\pi)^3}.
\end{equation}
После варьирования обеих частей равенства (\ref{FL2IN5}) и взятия по частям второго интеграла в правой части результирующего выражения получим:
\begin{eqnarray}\label{FL2IN6}
\int {\bf p} \delta n({\bf p})\frac {d{\bf p}}{(2\pi)^3} =
M \int \delta n({\bf p}) \frac{\partial \varepsilon(p)}{\partial p}
\frac {d{\bf p}}{(2\pi)^3}+
M \int \int \delta n({\bf p}) n({\bf p'})
\frac{\partial f({\bf p},{\bf p'})}{\partial p'}
\frac{d{\bf p}}{(2\pi)^3} \frac{d{\bf p'}}{(2\pi)^3} &=& \nonumber \\ =
M \int \delta n({\bf p}) \frac{\partial \varepsilon(p)}{\partial p}
\frac{d{\bf p}}{(2\pi)^3}-
M \int \int \delta n({\bf p})f({\bf p},{\bf p'})
\frac{\partial n({\bf p'})}{\partial p'}
\frac{d{\bf p}}{(2\pi)^3} \frac{d{\bf p'}}{(2\pi)^3}.
\end{eqnarray}
Ввиду произвольности $ \delta n({\bf p}) $ получаем искомое соотношение:
\begin{equation}\label{FL2IN7}
\frac {{\bf p}}{M} = \frac{\partial \varepsilon(p)}{\partial p}-
\int f({\bf p},{\bf p'})\frac{\partial n({\bf p'})}{\partial p'}
\frac{d{\bf p'}}{(2\pi)^3}.
\end{equation}
В окрестности ферми-поверхности можно воспользоваться выражением (\ref{FL5}) и после несложных преобразований получить хорошо известное уравнение Ландау для эффективной массы:
\begin{equation}\label{FL6}
\frac{1}{M^*_L} = \frac{1}{M}+\int \frac{({\bf
p}_F{\bf p'})}{p_F^3} F({\bf p_F},{\bf p'})
\frac{\partial n({\bf p'},T)}{\partial {p'}} \frac{d{\bf p'}}{(2\pi)^3}.
\end{equation}
Используя уравнение (\ref{FL6}) при $T=0$ и учитывая, что $n({\bf
p},T=0)$ становится ступенчатой функцией $\theta (p_F-p) $, мы
получаем известный результат \cite{pfw}
$$ \frac{M^*_L}{M}=\frac{1}{1-N_0f^1(p_F,p_F)/3}.$$
Здесь $N_0$ --- плотность состояний свободного ферми-газа и
$f^1(p_F,p_F)$ --- $p$-волновая компонента амплитуды взаимодействия
Ландау. Поскольку в теории ферми-жидкости Ландау $x=p_F^3/3\pi^2$,
амплитуда Ландау
может быть записана как $f^1(p_F,p_F)=f^1(x)$. Предположим, что в
некоторой критической точке $x_{FC}$ знаменатель $(1-N_0f^1(p_F,
p_F)/3)$ стремится к нулю, то есть,
$(1-N_0f^1(x)/3)\propto(x-x_{FC})+a(x-x_{FC})^2 + ...\to 0$. В
результате получаем, что $M^*_L(x)$ ведет себя как
\cite{shag1,khod1}
\begin{equation}
\frac{M^*_L(x)}{M}\simeq A+\frac{B}{x-x_{FC}}\propto\frac{1}{r}.
\label{FL7}\end{equation} Здесь $A$ и $B$ --- константы, а
$r=(x-x_{FC})/x_{FC}$ --- "расстояние" от квантовой критической
точки $x_{FC}$, в которой $M^*_L(x\to x_{FC})\to\infty$.
Наблюдаемое поведение находится в хорошем согласии с результатами
экспериментов \cite{cas1,skdk} и вычислениями
\cite{krot,sarm1,sarm2}. В случае
электронных систем уравнение (\ref{FL7}) справедливо при
$x>x_{FC}$, когда $r>0$ \cite{ksk,ksz}. Такое поведение эффективной
массы может наблюдаться в металлах с ТФ с достаточно плоской и
узкой зоной проводимости, соответствующей большой эффективной массе
$M^*_L(x\simeq x_{FC})$, с сильной электронной корреляцией и
эффективной температурой Ферми $T_k\sim p_F^2/M ^*_L (x) $ порядка
нескольких десятков градусов Кельвина или еще ниже (см., например,
\cite{ste}).

\section* { 2.2. Ферми-конденсатный квантовый фазовый переход }
\addcontentsline{toc}{section}{ 2.2. Ферми-конденсатный квантовый фазовый переход }

Не так давно было показано, что условия стабильности Померанчука не
охватывают все возможные типы неустойчивости, и, по крайней мере,
один из них пропущен \cite{ks}. Он соответствует ситуации, когда
эффективная масса --- наиболее важная характеристика квазичастиц
Ландау --- может стать бесконечно большой. В результате,
кинетическая энергия квазичастиц  становится пренебрежимо малой у
ферми-поверхности, и функция $n({\bf p})$ определяется
потенциальной энергией. Это приводит к новому классу сильно
коррелированных ферми-систем:  ферми-жидкость с фермионным конденсатом
(ФК) \cite{ks,vol,ksk}, который отделен от нормальной ферми-жидкости ферми-конденсатным квантовым фазовым переходом (ФККФП) \cite{ms,shb}.

Из (\ref{FL7}) следует, что при $T=0$ и $r=(x-x_{FC})\to0$
эффективная масса расходится, $M^*_L(r)\to\infty$. За критической
точкой $x_{FC}$ расстояние $r$ становится отрицательным, и,
соответственно, эффективная масса становится отрицательной. Для
того, чтобы предотвратить нестабильное и принципиально
бессмысленное состояние с отрицательным значением эффективной
массы, в системе должен осуществиться квантовый фазовый переход в
критической точке $x=x_{FC}$, который, как мы увидим ниже, есть
ФККФП \cite{ms,shag3}. Так как кинетическая энергия квазичастиц,
находящихся вблизи поверхности Ферми, пропорциональна обратной
величине эффективной массы, то при $x\to x_{FC}$ потенциальная
энергия квазичастиц, расположенных у ферми-поверхности, определяет
энергию основного состояния. Поэтому фазовый переход, понижая
энергию системы, должен вести к перестройке функции распределения
квазичастиц, и за точкой фазового перехода при $x\leq x_{FC}$
распределение квазичастиц определяется обычным уравнением для
поиска минимума функционала энергии \cite{ks}
\begin{equation} \frac{\delta E[n({\bf p})]}{\delta
n({\bf p},T=0)}=\varepsilon({\bf p})=\mu; \, p_i\leq p\leq p_f.
\label{FL8}\end{equation} Уравнение (\ref{FL8}) задает функцию
распределения квазичастиц $n_0({\bf p}) $, которая минимизирует
энергию основного состояния $E$. Будучи определенной из уравнения
(\ref{FL8}), $n_0({\bf p})$ не совпадает со ступенчатой функцией в
области $(p_f-p_i)$ так, что $0<n_0({\bf p})<1$, вне этой области
$n_0 ({\bf p})$ совпадает со ступенчатой функцией. Из уравнения
(\ref{FL8}) также следует, что одночастичный спектр имеет абсолютно
"'плоскую"' форму в этой области.
\begin{figure} [ht]
\begin{center}
\includegraphics [width=0.47\textwidth]{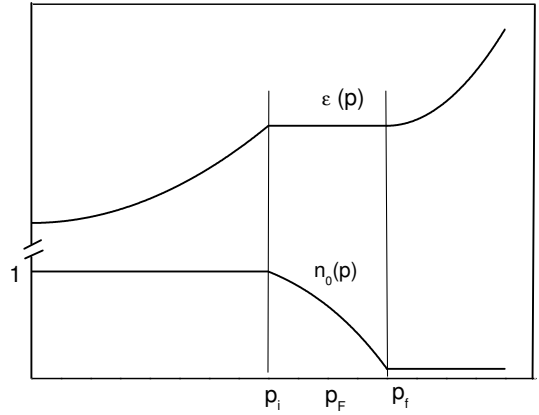}
\end{center}
\caption {Функция распределения квазичастиц $n_0(p)$ и
одночастичный спектр $\varepsilon(p)$. Поскольку $n_0(p)$ является
решением уравнения (\ref{FL8}), то $n_0(p<p_i)=1$,
$0<n_0(p_i<p<p_f)<1$, $n_0(p>p_f)=0$ и $\varepsilon
(p_i<p<p_f)=\mu$. Импульс Ферми $p_F$ удовлетворяет условию
$p_i<p_F<p_f$.} \label{Fig1}
\end{figure}
\begin{figure} [ht]
\begin{center}
\includegraphics [width=0.47\textwidth]{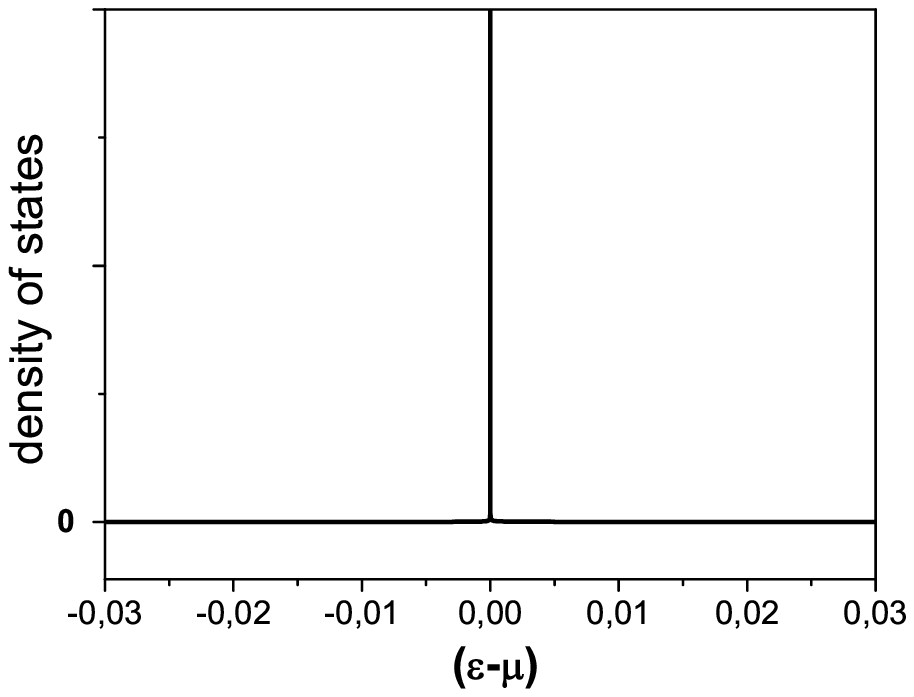}
\end{center}
\caption {Функция распределения плотности состояний в области,
занятой ФК} \label{Fig1111}
\end{figure}
Возможное решение $n_0({\bf p})$ уравнения (\ref{FL8}) и
соответствующий одночастичный спектр $\varepsilon({\bf p})$
показаны на Рис. (\ref{Fig1}). Квазичастицы с импульсами в области
$(p_f-p_i)$ имеют одинаковые одночастичные энергии, равные
химическому потенциалу $\mu$, и образуют ФК, а сами распределения
$n_0({\bf p})$ описывают новое состояние ферми-жидкости с ФК
\cite{ksk,ks,vol}.
Плотность состояний в области занятой ФК, Рис. (\ref{Fig1111}), описывается
дельта-функцией, что свидетельствует о высокой степени вырождения этого состояния.

Отметим, что решения $n_0({\bf p})$ уравнения (\ref{FL8}) являются
новыми решениями известных уравнений теории ферми-жидкости Ландау.
Действительно, при $T=0$, стандартное решение, представленное
ступенчатой функцией, $n({\bf p},T\to0)\to\theta(p_F-p)$ --- не
единственно возможное. Аномальные решения $\varepsilon({\bf
p})=\mu$ уравнения (\ref{FL1}) могут существовать, если логарифм в
правой части уравнения (\ref{FL1}) конечен. Это возможно при
условии, что $0<n_0({\bf p})<1$ в некоторой области $(p_i\leq p
\leq p_f)$. Поэтому в рассматриваемой области логарифм остается
конечным при $T\to0$, а произведение $T\ln[(1-n_0({\bf
p}))/n_0({\bf p})]_{|T\to0}\to0$, и мы снова получаем уравнение
(\ref{FL8}). Таким образом, при $T\to0$, функция распределения
квазичастиц $n_0({\bf p})$, будучи решением уравнения (\ref{FL8}),
не стремится к ступенчатой функции $\theta(p_F-p)$, соответственно,
как следует из уравнения (\ref{FL3}), энтропия $S_{NFL}(T)$ этого
состояния стремится к конечной величине $S_0$ при $T\to0$:
\begin{equation}
S_{NFL}(T\to0)\to S_0.\label{snfl} \end{equation} Вопрос о том, как
работает теорема Нернста в системах с ФК треует особого рассмотрения.

Предположим, что при уменьшении плотности (или с ростом силы
взаимодействия) мы достигли точки $x\simeq x_{FC}$, когда возникает
ФК. Это означает, что $p_i\to p_f\to p_F$, и отклонение $\delta
n({\bf p})=n_0({\bf p})-\theta(p_F-p)$ является малым. Раскладывая
функционал $E[n({\bf p})]$ в ряд Тейлора по $\delta n({\bf p})$ и
сохраняя старшие члены, из уравнения (\ref{FL8}) можно получить
следующее соотношение в интервале $p_i\leq p\leq p_f$
\begin{equation} \mu=\varepsilon({\bf p})=\varepsilon_0({\bf
p})+\int F({\bf p},{\bf p}_1)\delta n({\bf p_1})\frac{d{\bf
p}_1}{(2\pi)^2},\label{FL9}\end{equation} где $F({\bf p},{\bf
p}_1)=\delta^2E/\delta n({\bf p})\delta n({\bf p}_1)$ --- амплитуда
Ландау. Обе величины, амплитуда и одночастичная энергия
$\varepsilon_0({\bf p})$, вычислены при $n({\bf p})=\theta(p_F-p)$.
Уравнение (\ref{FL9}) имеет нетривиальные решения при плотностях
$x\leq x_{FC}$, если соответствующая амплитуда Ландау, зависящая от
плотности, положительна и достаточно велика, чтобы потенциальная
энергия была больше, чем кинетическая. Например, такое состояние
реализуется в электронной жидкости при малых плотностях. Тогда
преобразование ступенчатой функции Ферми $n({\bf p})=\theta(p_F-p)$
в гладкую, определяемую из уравнения (\ref{FL9}), становится
возможным \cite{ks,ksk,ksz}.

Система с ФК может считаться сильно коррелированной ферми-жидкостью
при плотностях $x<x_{FC}$. Из уравнения (\ref{FL9}) видно, что
квазичастицы фермионного конденсата формируют коллективное
состояние, поскольку их состояние определено макроскопическим
числом квазичастиц с импульсами $p_f<p<p_i$. Форма одночастичного
спектра, связанного с фермионным конденсатом, не зависит от
взаимодействия Ландау, которое, вообще говоря, определяется
свойствами системы как целого, включая коллективные состояния,
нерегулярность структуры, примеси, состав и т.д. Единственная
характеристика, определяемая взаимодействием Ландау,
--- ширина области $(p_f-p_i)$, занятая ФК; разумеется,
взаимодействие должно быть достаточной силы, чтобы ФККФП вообще был
возможен. Значит, можно заключить, что спектры, связанные с ФК,
имеют универсальную форму. Можно показать, что они зависят от температуры и сверхпроводящей щели, но
эта зависимость имеет также универсальный характер. Существование
таких спектров может рассматриваться, как характерная особенность
"'квантового протектората"', при котором свойства материала, включая
и термодинамические свойства, определяются некоторым
фундаментальным принципом \cite{rlp,pa}. В нашем случае, состояние
вещества с ФК также является "`квантовым протекторатом"', поскольку
новый тип квазичастиц этого состояния задает особые универсальные
термодинамические и транспортные свойства ферми-жидкости с ФК.

\section* { 2.3. Структура функции Грина ферми-конденсата }\label{SFG}
\addcontentsline{toc}{section}{ 2.3. Структура функции Грина ферми-конденсата }

Рассмотрим  общую  топологическую  структуру  функций   Грина  для таких систем как: нормальная ферми-жидкость, маргинальная ферми-жидкость латтинжеровская ферми-жидкость и ферми-конденсатное состояние ферми-жидкости. Покажем, что латтинжеровская и маргинальная ферми-жидкости
относятся к тому же топологическому классу, что и обычная ферми-жидкость, т.е. их функция Грина обладает в импульсном пространстве вихревой особенностью, и геометрическое место точек, в которых имеется особенность, образует поверхность Ферми. В состоянии фермионного конденсата вихревая особенность размыта в полосу конечной ширины (вихревой лист), что по аналогии с вихрями в сверхтекучей жидкости соответствует расщеплению вихря с одним квантом циркуляции на два полувихря, соединенных вихревым листом.

Нормальная ферми-жидкость Ландау характеризуется функцией Грина с хорошо определенным полюсом вблизи ферми-поверхности:
\begin{equation} G(\omega,{\bf p}) \simeq \frac{Z}{\omega-v_{F}(p-p_{F})+i\gamma(p)sign(p-p_{F})}\label{FLVOL1},\end{equation}
здесь Z --- вычет. Вычет Z маргинальной ферми-жидкости обращается в нуль, как логарифмическая функция частоты \cite{var}: $Z \propto 1/ln(\omega_{c}/\omega)$. Это связано с учетом взаимодействия электронов с попереными фотонами \cite{reizvol}.
В ферми-жидкости Латтинжера $Z$ спадает степенным образом \cite{haldvol}. Последнее имеет место в одномерных ферми-системах, но по-видимому возможно и в системах с большей размерностью \cite{wenvol}.

Нас интересует одночастичная функция Грина $ G(\Omega,{\bf p})$ на мнимой полуоси частот $ \omega = i\Omega $. Согласно общим аналитическим свойствам $G(\Omega,{\bf p})$ может иметь особенности только при $ \Omega = 0 $ \cite{abrvol}. В обычной ферми-жидкости эти особенности расположены на поверхности Ферми и характеризуются следующим инвариантом
\cite{vol}
\begin{equation} N=tr\oint_C\frac{dl}{2\pi i}G(i\omega,{\bf p})
\partial_lG^{-1}(i\omega,{\bf p})\label{FLVOL},\end{equation} где
$tr$ обозначает след по спиновым или зонным индексам функции Грина, а интеграл берется по произвольному контуру $C$, охватывающему особенность
функции Грина.

\begin{figure} [ht]
\begin{center}
\includegraphics [width=0.47\textwidth] {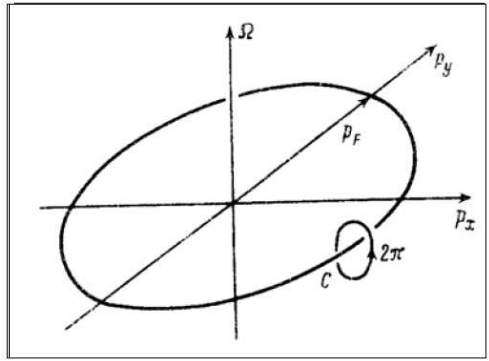}
\end{center}
\caption {Ферми-поверхность в двумерной (трехмерной) ферми-жидкости представляет собой топологически устойчивую особую вихревую линию (поверхность) в трехмерном (четырехмерном) пространстве $\Omega,p_{x},p_{y}$ $(\Omega,p_{x},p_{y},p_{z})$, при обходе вокруг которой фаза функции Грина меняется на $ 2\pi $. Такая структура сохраняется в маргинальной и латтинжеровской ферми-жидкостях. } \label{Fig2222}
\end{figure}

На (\ref{Fig2222}) для простоты изображен случай двумерной ферми-жидкости, где контур С охватывает линейную особенность --- ферми-линию. Если особенность имеет аналитический характер, то $N$ совпадает с кратностью полюса в особенности. Так в обычной ферми-жидкости функция Грина имеет в особенности полюс первого порядка, $ G(\Omega,{\bf p}) \sim Z/(z-v_{F}p_{F}) $, где $ z = v_{F}p-i\Omega $, поэтому $N = 1$ для каждой из двух проекций спина, так что суммарный заряд $N = 2$. Нам важно, что инвариант $N$ остается целочисленным даже в том случае, когда особенность не является полюсной. Этот индекс описывает набег фазы $\Phi$ функции Грина при обходе по контуру, равный $2\pi N$, поэтому он не может меняться непрерывно и тем самым сохраняется при малых шевелениях функции Грина. В этом отношении имеется аналогия с топологически устойчивыми особенностями у фазы конденсата сверхтекучей жидкости --- можно сказать, что ферми-линия на (\ref{Fig2222}) для одной из компонент спина соответствует в этой аналогии линии квантованного вихря с квантом циркуляции $N = 1$. (В случае трехмерной ферми-жидкости квантовый вихрь в четырехмерном пространстве $ (\Omega,{\bf p})$ образует двумерную поверхность -поверхность Ферми; для одномерной ферми-жидкости вихрь является точкой в двумерном $(\Omega, p)$ пространстве).

Инвариант N не меняется при переходе от обычной ферми-жидкости к неполюсным (маргинальной или латтинжеровской) ферми-жидкостям. Действительно, функция Грина для одномерной латтинжеровской бесспиновой ферми-жидкости имеет вблизи каждой из двух ферми-точек $ \pm p_{F}$ следующий вид \cite{wenvol}:
$G(\omega, k = p\pm p_{F}) \sim (v^{2}k^{2}-\omega^{2})^{g}/(\omega \pm vk)$. Продолжая $G$ аналитически на мнимую полуось, где
$G(\Omega,p) \sim (z-v_{F}p_{F})^{g-1}(z^{\ast}-v_{F}p_{F})^{g}$, и вычисляя интеграл (\ref{FLVOL}), получаем как и в обычном случае ту же вихревую особенность при $z = v_{F}p_{F}$ с $N = 1$ независимо от величины $g$. То же получается и для маргинальной ферми-жидкости \cite{var, reizvol} с $G \sim (i\Omega ln(i\Omega)-v_{F}(p- p_{F}))^{-1}$. Таким образом переход к маргинальным и латтинжеровским состояниям не меняет топологическую структуру функции Грина ферми-жидкости, сохраняя понятие ферми-поверхности, как поверхности вихревых сингулярностей функции Грина. Ситуация не меняется при учете спиновой структуры. В этом случае интеграл (\ref{FLVOL}) для обычной ферми-жидкости дает суммарный заряд $N = 2$ для двух компонент спина, и самое большое, что может произойти при переходе к неполюсным ферми-жидкостям, это расщепление вырожденной по спину ферми-поверхности с $N = 2$ на две хорошо определенных ферми-поверхности с $N = 1$ каждая, которые соответствуют холонам и спинонам \cite{andvol}. С топологической точки зрения это эквивалентно расщеплению ферми-поверхности паулиевским магнитным полем, которое раздвигает энергии фермионов с разными проекциями спина, что также приводит к двум невырожденным ферми-поверхностям с $N = 1$.

\begin{figure} [ht]
\begin{center}
\includegraphics [width=0.47\textwidth] {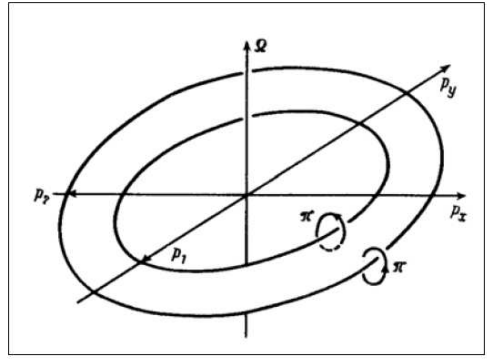}
\end{center}
\caption {В фермионном конденсате ферми-поверхность расплывается в ферми-полосу, границы которой $p = p_{1}$ и $p = p_{2}$ представляют собой полуквантовые вихри. При обходе вокруг такого полувихря фаза функции Грина меняется на $\pi$. } \label{Fig2223}
\end{figure}

Совсем другая ситуация возникает в системе с фермионным конденсатом. В этом случае вихревая линия с $N = 1$, которая не может исчезнуть в силу топологической устойчивости, растягивается в вихревой лист (ферми-полосу, см. (\ref{Fig2223}) для размерности $D = 2$), на котором фаза $\Phi$ функции Грина испытывает скачок. В модели хаотических фаз \cite{ks}, в которой на ферми-полосе $p_{1} < p < p_{2}$ энергия $ (\varepsilon ({ \bf p})- \mu) = 0$ и, следовательно, $G = 1/i\Omega$, скачок фазы $\Phi$ при прохождении через
$\Omega = 0$ постоянен и равен $\pi$. Это означает, что границы вихревого листа представляют собой вихри с полуцелым квантом циркуляции $N = 1/2$, поскольку набег фазы вокруг них составляет $\pi$. Это свойство является грубым, т.е. должно сохраниться и в точном решении уравнений для функции Грина, которое вообще говоря не должно совпадать с результатом модели. Можно предположить, что при достаточно малом расщеплении $p_{2}-p_{1}$, когда функция Грина ведет себя одинаково вблизи каждого из полувихрей, оно имеет простой степенной вид:
\begin{equation} G(\omega,{\bf p}) = \frac{Z}{(z-v_{F}p_{1})^{1/2}(z-v_{F}p_{2})^{1/2}}\label{FLVOL3},\end{equation}
что соответствует корневому разрезу на отрезке $ p_{1}<p<p_{2} $. Фаза $\Phi$ функции Грина по обе стороны разреза отличается на $\pi$.

Итак, в отличии от маргинальной и латтинжеровской ферми-жидкостей, система с ФК по своей топологической структуре является существенно новым классом ферми-жидкостей. Переход от ферми-поверхности в обычной ферми-жидкости к ферми-полосе в фермионном конденсате связан с изменением топологической характеристики функции Грина и представляет собой разновидность топологических фазовых переходов Лифшица, происходящих при нулевой температуре \cite{vol,volovik1,volovik2}. Это явление может иметь место не только в нормальной ферми-жидкости, но и в сверхпроводниках, где достаточно сильное спаривательное взаимодействие может привести к открытию ферми-поверхности даже в сверхпроводящем состоянии \cite{volovikvol}; последняя представляет собой поверхность особенностей у расширенной функции Грина сверхпроводника (функции Горькова).

\section* { 2.4. Электронная жидкость с фермионным конденсатом в магнитных полях }\label{HFL}
\addcontentsline{toc}{section}{ 2.4. Электронная жидкость с фермионным конденсатом в магнитных полях }

\begin{figure} [ht]
\begin{center}
\includegraphics [width=0.47\textwidth] {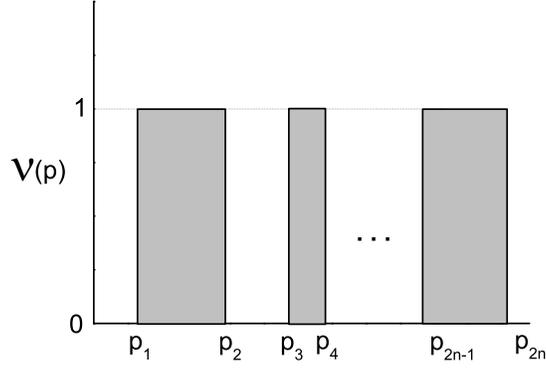}
\end{center}
\caption{Функция $\nu({\bf p})$ для многосвязного распределения,
заменяющего функцию $n_0({\bf p})$ в области $(p_f-p_i)$, занятой
фермионным конденсатом. Импульсы  удовлетворяют соотношению
$p_i<p_F<p_f$, здесь $p_F$ --- импульс нормальной ферми-жидкости
Ландау. Внешняя ферми-поверхность при $p\simeq p_{2n}\simeq p_f$
имеет вид ферми-ступеньки, поэтому при $T<T^*(B)$ система ведет
себя как ферми-жидкость Ландау.}\label{Fig4}
\end{figure}

В этом разделе рассмотрим поведение тяжелой электронной жидкости с
ФК в магнитном поле. Предположим, что константа связи отлична от
нуля, $\lambda_0\neq0$, но бесконечно мала. При $T=0$ сверхпроводящий параметр порядка $\kappa({\bf
p})$ конечен в области занятой ФК, а максимальная величина
сверхпроводящей щели $\Delta_1\propto \lambda_0$ --- бесконечно
мала. Поэтому любое малое магнитное поле $B\neq0$ является
критическим и разрушает $\kappa({\bf p})$ и сам ФК. Для определения
типа перестройки состояния с ФК достаточно простых энергетических
аргументов. С одной стороны, выигрыш энергии $\Delta E_B$ из-за
разрушения состояния с ФК равен $\Delta E_B\propto B^2$ и стремится
к нулю при $B\to 0 $. С другой стороны, занимая конечный интервал
$(p_f-p_i)$ в пространстве импульсов, новая функция $n_0({\bf p})$ должна приводить к
конечному выигрышу в энергии основного состояния по сравнению с
нормальной ферми-жидкостью \cite{ks}. Таким образом, в слабых
магнитных полях новое основное состояние без ФК должно иметь почти
ту же энергию, что и состояние с фермионным конденсатом. Такое
состояние образуется многосвязными сферами Ферми, напоминающими
луковицу, где гладкая функция распределения квазичастиц $n_0 ({\bf
p}) $ в области $ (p_f-p_i) $ заменяется распределением $\nu({\bf
p})$ \cite{asp,pogsh}
\begin{equation}
\nu ({\bf p})=\sum\limits_{k=1}^n\theta(p-p_{2k-1})
\theta(p_{2k}-p).\label{HF1}
\end{equation}
Здесь параметры $p_i\leq p_1<p_2<\ldots<p_{2n}\leq p_f$ подобраны
так, чтобы удовлетворить условиям нормировки и сохранения числа
частиц:
$$
\int_{p_{2k-1}}^{p_{2k+3}}\nu({\bf p})\frac{d{\bf p}}{(2\pi)^3}=
\int_{p_{2k-1}}^{p_{2k+3}}n_0({\bf p})\frac{d{\bf p}}{(2\pi)^3}.
$$
Соответствующее многосвязное распределение показано на Рис.
\ref{Fig4}.
Для определенности приведем наиболее интересный случай трехмерной
системы, рассмотрение двумерной системы может быть выполнено
аналогично. Отметим, что возможность существования многосвязных
сфер Ферми была предложена в  работах \cite{llvp,zb}. Предположим,
что толщина $\delta p$ каждого внутреннего блока ферми-сферы
приблизительно равна $\delta p\simeq p_{2k+1}-p_{2k}$, и $\delta p$
определяется величиной магнитного поля $B$. Используя известное
правило для оценки погрешности при вычислении интегралов, получим,
что минимальная потеря в энергии основного состояния из-за
формирования блоков приблизительно равна $(\delta p)^4$. Этот
результат становится ясным, если принять во внимание, что
непрерывные ферми-конденсатные функции $n_0 ({\bf p})$ обеспечивают
минимальную величину функционала энергии $E [n({\bf p})]$, в то
время как приближение $\nu({\bf p})$ ступеньками шириной $\delta p$
приводит к минимальной ошибке, порядок которой $(\delta p)^4$.
Учитывая, что выигрыш из-за магнитного поля пропорционален $B^2$, и
приравнивая оба вклада, получаем
\begin{equation} \delta p\propto\sqrt{B}.\label{HF2}
\end{equation} Таким образом, при $T\to 0 $, когда $B\to0$, толщина
$\delta p$ стремится к нулю ($\delta p\to0$), и поведение
ферми-жидкости с ФК заменяется поведением ферми-жидкости Ландау с
импульсом Ферми $p_f$. Из уравнения (\ref{FL8}) следует, что
$p_f>p_F$, а плотность $x$ электронов остается постоянной. При этом
ферми-импульс многосвязной ферми-сферы $p_{2n}\simeq p_f>p_F$, см.
Рис. \ref{Fig4}. Как мы увидим далее, это наблюдение играет важную
роль при рассмотрении коэффициента Холла $R_H(B)$ как функции $B$
при низких температурах в металлах с ТФ.

\chapter* {3. Другие теоретические доказательства существования фермионного конденсата }
\label{DSFC}
\addcontentsline{toc}{chapter}{ 3. Другие теоретические доказательства существования фермионного конденсата }

В случае ФККФП, как и при рассмотрении других фазовых переходов, мы
имеем дело с сильным межчастичным взаимодействием, где абсолютно
надежный ответ нельзя дать на основе чистого теоретического
рассмотрения, базирующегося на первых принципах. Поэтому
единственный способ проверки реальности ФК состоит в изучении этого
состояния в рамках точно решаемых моделей и в рассмотрении
экспериментальных фактов, которые можно интерпретировать как прямое
подтверждение существования ФК. В данной главе рассматриваются теоретические свидетельства, доказывающие существование ФК. Обзор результатов экспериментов, объяснение которых чрезвычайно затруднительно, вне теории ФККФП, будет проведен в последующих главах.

\section* { 3.1. Точно решаемые модели фермионного конденсата. }\label{TRM}
\addcontentsline{toc}{section}{ 3.1. Точно решаемые модели фермионного конденсата. }

Точно решаемые модели однозначно
свидетельствуют, что ферми-системы с ФК являются объективной реальностью (см., например,
\cite{khv,dzyal,lid,irk,ksk}). Принимая во внимание топологическое
рассмотрение проведенное в разделе \ref{SFG}, можно заключить, что новый класс ферми-жидкостей с ФК
не является пустым, действительно существует и представлен обширным
семейством новых состояний ферми-систем.

\subsection* { 3.1.1. Точно решаемые модели ФК в рамках функционального подхода.} \label{TRMKS}
\addcontentsline{toc}{subsection}{ 3.1.1. Точно решаемые модели ФК в рамках функционального подхода. }

Поведение ферми-системы, совершающей ФККФП, в рамках подхода,  использующего модельный функционал Ландау, впервые были выполнены В. Ходелем и В. Шагиняном, а также Ф. Нозьером  \cite{ksk,ksn}:
\begin{eqnarray}
\nonumber E[n(p)]&=&\int\frac{{\bf p}^2}{2M}\frac{d{\bf
p}}{(2\pi)^3}+\frac{1}
{2}\int U({\bf p}_1-{\bf p}_2)\\
&\times&n({\bf p}_1)n({\bf p}_2) \frac{d{\bf p}_1d{\bf
p}_2}{(2\pi)^6}.\label{TRMKS1}
\end{eqnarray}
Записанный в виде (\ref{TRMKS1}), функционал $E[n(p)]$ имеет форму простейшего функционала энергии Ландау с некоторым эффективным потенциалом двухчастичного взаимодействия $V({\bf p}_1-{\bf p}_2)$.
Отметим, что данный вид функционала является лишь первым приближением для энергии системы, удобным для анализа систем в рамках точно решаемых моделей или численных расчетов.
Существенно, что использование этого функционала возможно только для изучения эффектов, связанных с каналом взаимодействия частиц и дырок в окрестности точки фазового перехода, и не может быть использован для получения характеристик канала взаимодействия частица-частица, например, для вычисления спаривательных сил \cite {ksk}.

Подставляя (\ref{NMC1}) в (\ref{FL2}), получаем выражение для энергии квазичастиц:
\begin{equation}
\varepsilon[n(p)] = \frac{{\bf p}^2}{2M}+
\frac{1}
{2}\int U({\bf p}_1-{\bf p}_2)n({\bf p}_1)
\frac{d{\bf p}_1}{(2\pi)^3}.\label{TRMKS2}
\end{equation}
Подбирая различные формы потенциала $V({\bf p}_1-{\bf p}_2)$, можно получить все известные модели, которые могут быть решены аналитически в рамках рассматриваемого подхода.

Если воспользоваться соотношением (\ref{FL4}), связывающим
$n({\bf p},T)$ и $\varepsilon({\bf p},T)$,
\begin{equation}
\nonumber n({\bf p},T)=
\left\{1+\exp\left[\frac{(\varepsilon({\bf p},T)-\mu)}
{T}\right]\right\}^{-1}, \label{TRMKS8}
\end{equation}
и учесть, что уравнение (\ref{TRMKS2}) нужно решать с учетом условия нормировки, которое фактически служит уравнением для определения химического потенциала $\mu$:
\begin{equation}
\rho = 2\int n({\bf p}) \frac{d{\bf p}}{(2\pi)^3},\label{TRMKS3}
\end{equation}
где $\rho$ плотность системы, то уравнения (\ref{TRMKS2}), (\ref{TRMKS8}), (\ref{TRMKS3}) образуют замкнутую систему, позволяющую вычислить аналитически   $n({\bf p},T)$, $\varepsilon({\bf p},T)$ и $\mu$.

\vspace{5mm}
В качестве примера, рассмотрим несколько потенциалов взаимодействия квазичастиц.

1. Простейший потенциал, впервые рассмотренный в \cite{noz, ksk}, который соответствует ограниченному дальнодействующему взаимодействию в координатном пространстве, имеет вид:
\begin{equation} U({\bf p}-{\bf p}_1) =
(2\pi)^{3}U\delta({\bf p}-{\bf p}_1), \label{TRMKS5}
\end{equation}
здесь $U$ --- постоянная взаимодействия. Этому взаимодействию соответствует химический потенциал $\mu_{F} = \varepsilon(p_{F},n_{F}(p))$ основного состояния:
\begin{equation} \mu_{F} = \frac{p_{F}^{2}}{2M}+\frac{U}{2}, \label{TRMKS9}
\end{equation}
и энергия квазичастиц:
\begin{equation} \varepsilon(p,n(p)) = \frac{p^{2}}{2M}+Un(p), \label{TRMKS10}
\end{equation}
Условие стабильности для основного состояния выполняется при $U>0$. В этой модели существуют два критических значения константы взаимодействия. Первое,  $U_{2c}=0$, когда ФК возникает, и  второе,  $U_{1c}=(5/2)^{2/3}\varepsilon_{F}^{0}$, когда $p_{i}$ обращается в ноль (при $U>U_{1c}$ все частицы принадлежат конденсату). В рассматриваемой модели распределение квазичастиц по импульсам можно записать в виде:
\begin{equation} \left \{ \begin{array}{l}
n_{p}=1 \hspace{3mm} \mbox{при} \hspace{3mm}  p<p_{i} \\
n_{p} =  \frac{\textstyle{p_{f}^{2}-p^{2}}}{\textstyle{p_{f}^{2}-p_{i}^{2}}},
\hspace{3mm}  \mbox{при} \hspace{3mm}  p_{i}<p<p_{f} \\
n_{p}=0 \hspace{3mm} \mbox{при} \hspace{3mm}  p>p_{f}
 \end{array} \right. \label{TRMKS11},
\end{equation}
Стандартные графики функций $\varepsilon(p,n(p))$ и $n_{p}$ имеют вид как  на Рис.\ref{Fig1}.

\begin{figure} [ht]
\begin{center}
\includegraphics [width=0.47\textwidth] {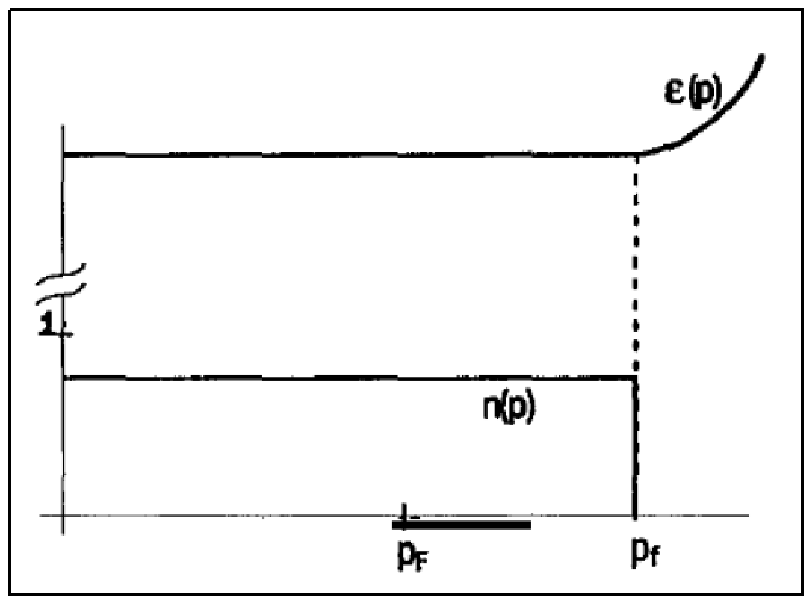}
\end{center}
\caption {Функция распределения  $n_{k}$ и эергия квазичастиц основного состояния $\varepsilon(p)$ для модели с потенциалом взаимодействия $U({\bf p}-{\bf p}_1) =
\frac{\textstyle{g}} {\textstyle{\mid{\bf p}-{\bf p}_1\mid}}$ } \label{Fig2237}
\end{figure}

2. Вторая модель связана с потенциалом эффективного взаимодействия:
\begin{equation} U({\bf p}-{\bf p}_1) =
\frac{g} {\mid{\bf p}-{\bf p}_1\mid},
\label{TRMKS12}
\end{equation}
и соответствует движению квазичастиц в координационном пространстве под влиянием внешнего квазиупругого поля $V_{ex}(r)=(1/(2M))r^{2}$ и взаимодействующих друг с другом посредством Кулоновского потенциала:
\begin{equation} U({\bf r}-{\bf r}_1) \approx
\frac{1} {\mid {\bf r}-{\bf r}_1\mid},
\label{TRMKS13}
\end{equation}
Кроме ограничения, наложенного на число частиц системы условием нормировки (\ref{TRMKS3}), существует еще дополнительное условие, связанное с принципом Паули $n(p)<(2\pi)^{-3}$. Введем в данной задаче безразмерный параметр $\xi=1-M/M^{*}$. Пока величина параметра $\xi$ не превосходит критического значения $\xi_{c} = gM/6\pi^{2}$, влияние внешнего поля $V_{ex}(r)$ доминирует и ферми-ступенька является функцией распределения квазичастиц.
Для $\mu_{F}$ и $\varepsilon(p)$ имеем следующие выражения:
\begin{equation} \mu_{F} = \frac { p_{F}^{2}(1+2\xi)}{2M},
\label{TRMKS15}
\end{equation}
\begin{equation}  \varepsilon(p)= \left \{ \begin{array}{l}
\frac {\textstyle{p^{2}(1-\xi)}}{\textstyle{2M}} + \frac {\textstyle{3p_{F}^{2}\xi}}{\textstyle{2M}}  \hspace{3mm} \mbox{при} \hspace{3mm}  p<p_{F} \\
\frac {\textstyle{p^{2}}}{\textstyle{2M}} + \frac {\textstyle{p_{F}^{3}\xi}}{\textstyle{pM}} \hspace{3mm} \mbox{при} \hspace{3mm}  p>p_{F}
 \end{array} \right. \label{TRMKS16}.
\end{equation}
Видно, что при $\xi_{c}=1$ эффективная масса становится бесконечной, а при $\xi>1$ кулоновское отталкивание превышает квазиупругие связи системы, и распределение Ферми должно быть заменено:
\begin{equation} n(p) = \theta(p-p_{f})/\xi,
\label{TRMKS14}
\end{equation}
где $p_{f}=p_{F}\xi^{1/3}$. Для этого распределения новый химический потенциал   $\mu = 3p_{f}^{2}/2M$, а новая энергия квазичастиц:
\begin{equation}  \varepsilon(p)= \left \{ \begin{array}{l}
\mu_{f}  \hspace{3mm} \mbox{при} \hspace{3mm}  p<p_{f} \\
\mu_{f} + A (p-p_{f})^{2} \hspace{3mm} \mbox{при} \hspace{3mm}  p>p_{f}
 \end{array} \right. \label{TRMKS17},
\end{equation}
где $A$ --- константа. На Рис. \ref{Fig2237} приведены графики функций распределения  $n_{k}$ и эергии квазичастиц основного состояния $\varepsilon(p)$.

\begin{figure} [ht]
\begin{center}
\includegraphics [width=0.47\textwidth] {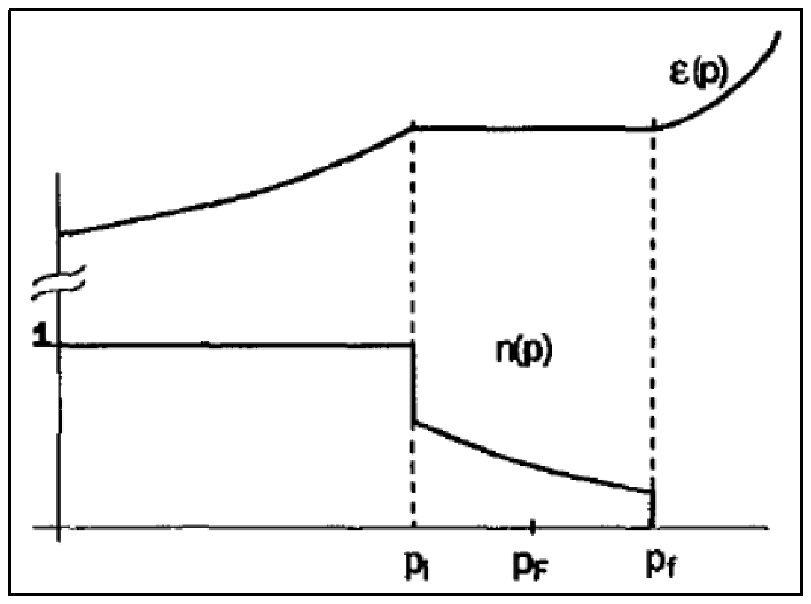}
\end{center}
\caption {Функции распределения квазичастиц по импульсам $n(p)$ и энергия квазичастиц $\varepsilon(p)$ для модели с потенциалом взаимодействия между частицами $U({\bf p}-{\bf
p}_1)=g\frac{\textstyle{\exp(-\beta\mid{\bf p_{1}}-{\bf
p_{2}}\mid}} {\textstyle{\mid{\bf p_{1}}-{\bf
p_{2}}\mid}}$} \label{Fig2238}
\end{figure}

3. Аналитические вычисления также возможны для модели с межчастичным взаимодействием:
\begin{equation}
U({\bf p}-{\bf
p}_1)=g\frac{\exp(-\beta\mid({\bf p_{1}}-{\bf
p_{2}}\mid} {\mid({\bf p_{1}}-{\bf
p_{2}}\mid}.
\label{TRMKS18} \end{equation}
Здесь $g$ --- постоянная спаривания, $\beta$ --- константа взаимодействия. Эта модель включает в себя обе модели, рассмотренные выше в этом разделе. Модель "`2"', с "`Кулоновскими силами"', получается при $\beta=0$ в (\ref{TRMKS18}). Модель "`1"', с силами, характеризующимися ограниченным дальнодействием, получается, если положить $g/2\beta^{2}\pi^{2}=const$ и перейти к пределу в следующем выражении:
\begin{equation} \lim_{\beta \rightarrow \infty}
\beta^{2}\frac{\exp(-\beta\mid({\bf p_{1}}-{\bf
p_{2}}\mid} {\mid({\bf p_{1}}-{\bf
p_{2}}\mid}=4\pi\delta(\mid({\bf p_{1}}-{\bf
p_{2}}\mid).
\label{TRMKS19} \end{equation}
Подставляя (\ref{TRMKS18}) в (\ref{TRMKS2}) и решая его относительно $n(p)$ и $\varepsilon(p)$ получим следующие выражения для этих величин:
\begin{equation}  \varepsilon(p) = \left  \{ \begin{array}{l}
p^{2}/2M-g[1-(p_{F}\beta + 1)\sinh{(p\beta)}\exp{(-p_{F}\beta)}/p\beta]/2\pi^{2}\beta^{2}, \hspace{3mm} \mbox{при} \hspace{3mm}  p < p_{F} \\
p^{2}/2M + g[p_{F}\beta \cosh{(p_{F}\beta)}-\sinh{(p_{F}\beta)}]\exp{(-p\beta)}/2\pi^{2}p\beta^{3}, \hspace{3mm} \mbox{при} \hspace{3mm}  p < p_{F}  \end{array} \right.  \label{TRMKS20},
\end{equation}
\begin{equation}
n(p) = n_{0}-n_{2}p^{2}/p_{f}^{2},  \label{TRMKS21},
\end{equation}
где $\xi = 1-M/M^{*}$, $\mu = p_{f}^{2}(\beta p_{f}+3)/2M(\beta p_{f}+1)$, $n_{0} = (1 + \beta^{2}M\mu/3)/\xi$,
$n_{2} = \beta^{2}p_{f}^{2}/6\xi$. На Рис. \ref{Fig2238} изображены графики функции распределения квазичастиц по импульсам $n(p)$ и энергия квазичастиц $\varepsilon(p)$.

\subsection* { 3.1.2. Точно решаемая модель ФК Хацугаи-Комото.} \label{LTRM}
\addcontentsline{toc}{subsection}{ 3.1.2. Точно решаемая модель ФК Хацугаи-Комото. }

В работе \cite{lid} Д. Лидским, Дж. Шираиши, Дж. Хацугаи и М. Комото использована точно решаемая двухмерная обобщенная модель Хацугаи-Комото \cite{hatkohm} для исследования основного состояния ферми-жидкости при $T=0$. Эта модель анадогична модели Шеррингтона-Киркпатрика \cite{sherkirk} для спинового стекла и эквивалентна модели Хаббарда, в той части, когда обе допускают перескакивания электронов на любые расстояния \cite{nogand}.

Рассматривается гамильтониан системы $H$:
\begin{equation}
H = \sum_{k} \varepsilon({\bf k})n_{k} +  \frac{1}{2 V} \sum_{k,k_{'}}f_{k,k^{'}}n_{k}n_{k^{'}}\label{LTRM1},
\end{equation}
где $n_{k} = c_{k}^{+}c_{k}$ --- оператор распределения частиц по импульсам, $c_{k}^{+}c_{k}$ --- операторы рождения и уничтожения, $V$ --- объем системы, $f_{k,k^{'}}$ --- оператор взаимодействия частиц.

В термодинамическом пределе $V \to \infty $, функция взаимодействия в (\ref{LTRM1}) может быть записана в виде:
\begin{equation}
f(k,k^{'}) = (2\pi)^{2}U(k)\delta(k-k^{'})\frac{g(\phi)}{k}\label{LTRM2},
\end{equation}
где $k=\mid {\bf k} \mid $, $k^{'}=\mid {\bf k^{'}} \mid $, $\phi$ --- угол между ${\bf k}$ и ${\bf k^{'}}$. Функции $U(k)$ и $g(\phi)$ выбираются в рассматриваемой модели. Функция $g(\phi)$ $2\pi$-периодична и четна. Будем для определенности считать, что  $g(\phi)$ достигает максимума при $ \phi = \pi$ ( ${\bf k} =-{\bf k^{'}}$ ) и монотонно убывает при $\phi \to 0$. Для описанного класса моделей можно аналитически вычислить энергию основного состояния и  функцию распределения электронов по импульсам $n_{k}$, имеющую непрерывную зависимость от $ k $, что свидетельствует о не-ферми-жидкостном поведении описываемых систем или о ферми-конденсатном состоянии этих систем.
Функция распределения $n_{k}$ вычисляется из соотношения:
\begin{equation}
n_{k} = \frac{\phi_{k}}{2\pi}\label{LTRM3},
\end{equation}
где $\phi_{k}$ является решением уравнения:
\begin{equation}
\varepsilon(k) + U(k)\int_{0}^{\phi_{k}} g(\phi)d\phi = \mu \label{LTRM4}.
\end{equation}
Предполагая, что $\varepsilon({\bf k})$ не зависит от $\phi$ и учитывая вклад химического потенсиала $\mu$, ожидаемое значение энергии основного состояния $\xi(k) = \langle H-\mu N \rangle_{G}$, определяется выражением:
\begin{equation}
\xi(k) = \frac{V}{(2\pi)^{2}} \int_{0}^{k_{c}} kdk [(\varepsilon(k)-\mu)\phi_{k} + U(k)\int_{0}^{\phi_{k}} d\phi \int_{0}^{\phi_{k^{'}}} g(\phi^{'})d\phi^{'}] \label{LTRM5},
\end{equation}
где $k_{c}$ --- импульс, соответствующий обрезанию потенциала взаимодействия при малых $r$.
Хотя энергия основного состояния однозначно определяется $\phi_{k}$, она сильно вырождена из-за большого числа вариантов угловых дуг $\phi_{k}$ для каждого $k$.

Рассмотрим несколько примеров использования моделных функций $g(\phi)$ для вычисления распределений $n_{k}$, считая $U_{k}$ постоянной величиной $U$.

\begin{figure} [ht]
\begin{center}
\includegraphics [width=0.47\textwidth] {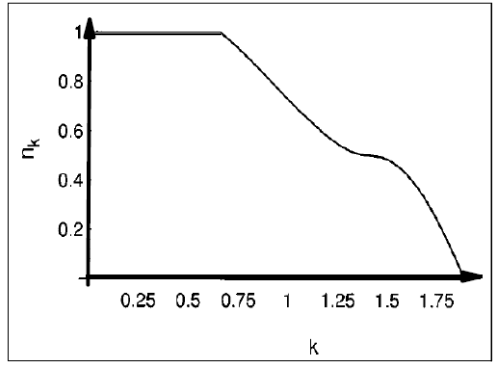}
\end{center}
\caption {Функция распределения основного состояния $n_{k}$ при $g(\phi)= (\pi^{2}-\phi^{2})^{-1/2}$ и $U=1/2$ } \label{Fig2231}
\end{figure}

1.
\begin{equation}  g(\phi) = \left  \{ \begin{array}{l}
\frac {\textstyle{1}}{\textstyle{(\pi^{2}-\phi^{2})^{1/2}}} \hspace{3mm} \mbox{при} \hspace{3mm}  0 < \phi < \pi \\
\frac {\textstyle{1}}{\textstyle{(\pi^{2}-(2\pi-\phi)^{2})^{1/2}}} \hspace{3mm} \mbox{при} \hspace{3mm}  \pi < \phi < 2\pi \end{array} \right.  \label{LTRM6},
\end{equation}
Взаимодействие имеет интегрируемую сингулярность при $\phi = \pi$, следствием которой является то, что касательная к $n_{k}$ при $n_{k}=1/2$ --- горизонтальна. Функция $n_{k}$ имеет излом в точках $k_{1}$ и $k_{0}$, см. Рис. \ref{Fig2231}:
\begin{equation}  n_{k} = \left  \{ \begin{array}{l}
1-\frac {\textstyle{1}}{\textstyle{2}}sin\frac {\textstyle{\mu-\varepsilon(k)}}{\textstyle{U}} \hspace{3mm} \mbox{при} \hspace{3mm}  k_{1} < k < k_{1/2} \\
\frac {\textstyle{1}}{\textstyle{2}}sin\frac {\textstyle{\mu-\varepsilon(k)}}{\textstyle{U}} \hspace{3mm} \mbox{при} \hspace{3mm}  k_{1/2} < k < k_{0} \end{array} \right.  \label{LTRM7},
\end{equation}

\begin{figure} [ht]
\begin{center}
\includegraphics [width=0.47\textwidth] {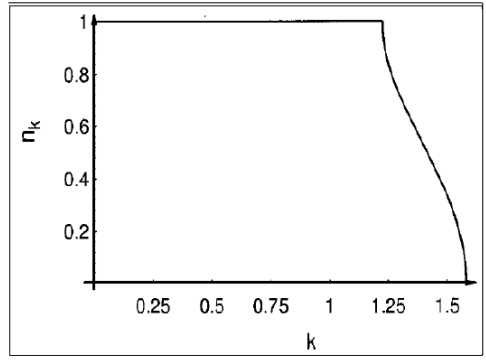}
\end{center}
\caption {Функция распределения основного состояния $n_{k}$ при $g(\phi)= sin(\phi/2)$ и $U=1/8$ } \label{Fig2232}
\end{figure}

2.
\begin{equation}  g(\phi) = sin ( \frac{\phi}{2}) \label{LTRM8},
\end{equation}
Функция $n_{k}$ опять имеет излом в точках $k_{1}$ и $k_{0}$. При $n_{k}=1/2$ вторая производная меняет знак, $n_{k}$ имеет точку перегиба. см. Рис. \ref{Fig2232}:
\begin{equation}  n_{k} = \left  \{ \begin{array}{l}
1-\frac {\textstyle{1}}{\textstyle{\pi}}arccos(\frac {\textstyle{\mu-\varepsilon(k)}}{\textstyle{2U}}-1) \hspace{3mm} \mbox{при} \hspace{3mm}  k_{1} < k < k_{1/2} \\
\frac {\textstyle{1}}{\textstyle{\pi}}arccos(1-\frac {\textstyle{\mu-\varepsilon(k)}}{\textstyle{2U}}) \hspace{3mm} \mbox{при} \hspace{3mm}  k_{1/2} < k < k_{0} \end{array} \right.  \label{LTRM9},
\end{equation}

\begin{figure} [ht]
\begin{center}
\includegraphics [width=0.47\textwidth] {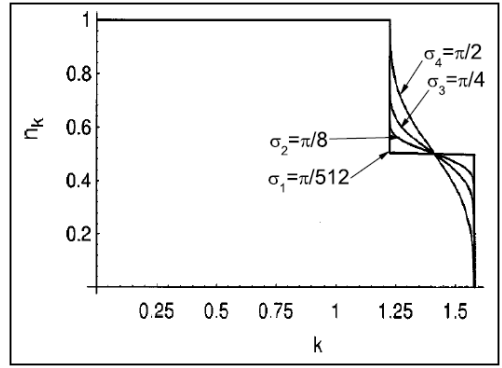}
\end{center}
\caption {Функция распределения основного состояния $n_{k}$ при $g(\phi)= (1/\sigma\pi^{1/2})exp[(\pi-\phi)^{2}/\sigma^{2}]$ и $U=1/2$, при $\sigma_{1}=\pi/512$, $\sigma_{2}=\pi/8$, $\sigma_{3}=\pi/4$, $\sigma_{4}=\pi/2$ } \label{Fig2233}
\end{figure}

3.
В этом примере $\varepsilon(k) = k^{2}/(2m)$. При $\sigma \to 0$ функция распределения $n_{k}$ имеет две псевдо ферми-поверхности. Это можно наблюдать на Рис. \ref{Fig2233} для наименьшего значения $\sigma=\pi/512$.
\begin{equation}  g(\phi) = \frac{1}{\sigma \pi^{1/2}}exp[-\frac{(\pi-\phi)^{2}}{\sigma^{2}}] \label{LTRM10},
\end{equation}
Функция $\phi_{k}$ может быть определена из уравнения (\ref{LTRM10}), а затем $n_{k}$ из (\ref{LTRM3}):
\begin{equation}  n_{k} = \left  \{ \begin{array}{l}
\sqrt {(2m[\mu-\frac {\textstyle{U}}{\textstyle{2}} erf(\frac {\textstyle{\pi-\phi_{k}}}{\textstyle{\sigma}},\frac {\textstyle{\pi}}{\textstyle{\sigma}})])} \hspace{3mm} \mbox{при} \hspace{3mm} \phi_{k} < \pi \\
\sqrt {(2m[\mu-\frac {\textstyle{U}}{\textstyle{2}} erf(0,\frac {\textstyle{\pi}}{\textstyle{\sigma}})-\frac {\textstyle{U}}{\textstyle{2}} erf(0,\frac {\textstyle{\phi_{k}-\pi}}{\textstyle{\sigma}})])} \hspace{3mm} \mbox{при} \hspace{3mm} \phi_{k} > \pi \end{array} \right.  \label{LTRM11},
\end{equation}

\subsection* { 3.1.3. Точно решаемая модель ФК при магнитном взаимодействии между фермионами.} \label{KTRM}
\addcontentsline{toc}{subsection}{ 3.1.3. Точно решаемая модель ФК при магнитном взаимодействии между фермионами. }

\begin{figure} [ht]
\begin{center}
\includegraphics [width=0.47\textwidth] {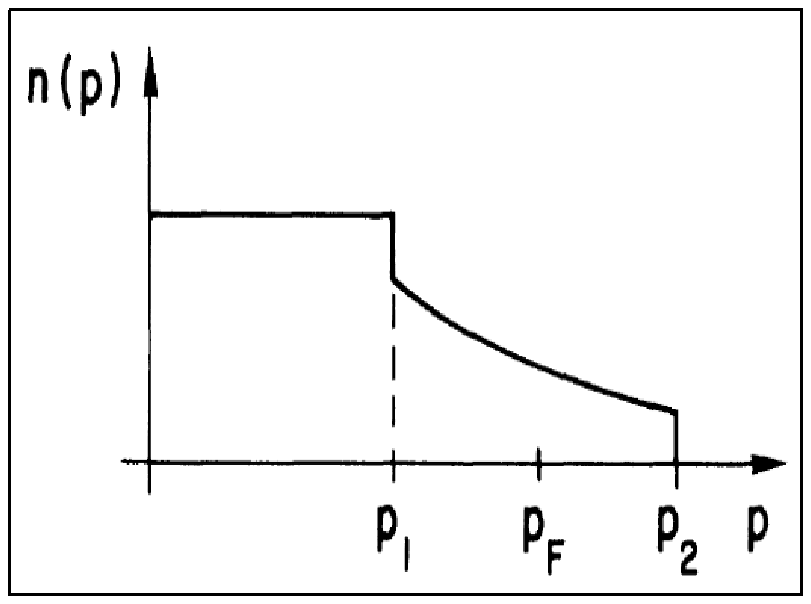}
\end{center}
\caption {Функция распределения $n_{k}$, полученная в рамках функционального подхода для двухмерного взаимодействия  $f({\bf p},{\bf p^{'}}) = g\frac{\textstyle{({\bf p},{\bf p^{'}})}-\frac{\textstyle{({\bf p}{\bf q})({\bf p^{'}}{\bf q})}}{\textstyle{q^{2}}}}
{\textstyle{\mid{\bf p}\mid\mid{\bf p^{'}}\mid}} $  } \label{Fig2234}
\end{figure}

Рассмотрим точно решаемую модель для расчета функции распределения квазичастиц двухмерной ферми-системы в основноном состоянии в рамках функционального подхода, предложенную Д.В. Хвещенко, Р. Хлюбина и Т.М. Райсом (\cite{khv}).
Рассмотрим функционал энергии:
\begin{equation}
E[n_{p}]=\int \varepsilon(p)n_{p}d^{2}p+
\frac{1}{2\Omega}\int f(p,p^{'})n_{p}n_{p^{'}}d^{2}pd^{2}p^{'},\label{KTRM1}
\end{equation}
где $\varepsilon(p)$ --- одночастичный спектр невзаимодействующих фермионов, $n_{p}$ --- их функция распределения, $f(p,p^{'})$ --- амплитуда взаимодействия Ландау и $\Omega$ --- объем системы. Электронное распределение находится из условия экстремума:
\begin{equation}
\frac{\delta E[n_{p}]}{\delta n_{p}}=0 ,\label{KTRM2}
\end{equation}
при дополнительном ограничении $0<n_{p}<1$.
Основная идея используемой модели состоит в том, чтобы описать взаимодействие между фермионами с помощью магнитных (Амперовских) сил, которые действуют между токами электрических зарядов. Не останавливаясь на обосновании деталей модели, выразим амплитуду взаимодействия $f(p,p^{'})$ через "`экранированную"' вершинную функцию, описывающую токовое взаимодействие электронов:
\begin{equation}
f({\bf p},{\bf p^{'}}) =-\Gamma({\bf p},{\bf p^{'}};{\bf q})=-g\frac{\frac{\textstyle{({\bf p}{\bf q})({\bf p^{'}}{\bf q})}}{\textstyle{q^{2}}}-({\bf p},{\bf p^{'}})}
{\mid{\bf p}\mid\mid{\bf p^{'}}\mid} ,\label{KTRM3}
\end{equation}
здесь ${\bf q} = {\bf p}-{\bf p^{'}}$. Подставляя (\ref{KTRM3}) в (\ref{KTRM1}) и решая уравнение (\ref{KTRM2}), для $g>g_{c}= 6/\pi $ находим $n_{p}$:
\begin{equation} n_{p}= \left  \{ \begin{array}{l}
1 \hspace{3mm} \mbox{при} \hspace{3mm}  p<p_{1}= \frac{\textstyle{p_{0}}}{\textstyle{\sqrt{(2\pi g/3-1)}}} \\
\frac{\textstyle{1}}{\textstyle{2\pi g}} \frac{\textstyle{p_{0}^{2}+3p^{2}}}{\textstyle{p^{2}}},
\hspace{3mm}  \mbox{при} \hspace{3mm}  p_{1}= \frac{\textstyle{p_{0}}}{\sqrt{\textstyle{(2\pi g/3-1)}}}<p<p_{2}=\frac{\textstyle{p_{0}}}{\sqrt{\textstyle{3}}} \\
0 \hspace{3mm} \mbox{при} \hspace{3mm}  p>p_{2}=\frac{\textstyle{p_{0}}}{\textstyle{\sqrt{3}}}
 \end{array} \right. \label{KTRM4},
\end{equation}
здесь $p_{0}=p_{F}\sqrt{\frac{\textstyle{2\pi g}}
{\textstyle{4+\ln{\frac{\textstyle{2\pi g-3}}{\textstyle{g}}}}}}$. 
На Рис. \ref{Fig2234} представлен график полученной функции распределения $n_{p}$.

\subsection* { 3.1.4. Точно решаемая модель ФК в рамках модели Хаббарда. } \label{IKH}
\addcontentsline{toc}{subsection}{ 3.1.4. Точно решаемая модель ФК в рамках модели Хаббарда. }

Изучение сильно коррелированных электронных систем с помощью модели Хаббарда имеет долгую историю и значительные успехи \cite{izyumov}. Поэтому получение характеристик ФК состояния в рамках модели Хаббарда представляет значительный интерес. Эта задача была в принципе решена в работе В.Ю. Ирхина, А.А. Катанина и М.И. Кацнельсона \cite{irk} для двухмерной системы. Основная идея выполненных расчетов состоит в  учете решающей роли сингулярностей Ван Хова (СВХ) и захвата ими Ферми поверхности. Рассмотрение велось в рамках ренормогруппового (РГ) подхода.

Стартуем с $t-t^{'}$ модели Хаббарда на двухмерной решетке с Гамильтонианом:
\begin{equation}
H =  \sum_{k}\varepsilon_{k} c_{k \sigma}^{+} c_{k \sigma} +
U \sum_{i}  n_{i \uparrow} n_{i \downarrow} ,\label{IKH1}
\end{equation}
где $c_{k \sigma}^{+}, c_{k \sigma} $ --- Фермиевские операторы рождения и уничтожения электрона со спином $\sigma$ ($\sigma=\uparrow,\downarrow$) на $k$-том узле решетки, $U$ --- величина кулоновского отталкивания на узле, $n_{i\sigma}= c_{i\sigma}^{+} c_{i\sigma}$ --- оператор плотности электронов на $i$-м узле со спином $\sigma$. Будем использовать модельный  одноэлектронный спектр $\varepsilon_{k}$ вида:
\begin{equation}
\varepsilon_{k} = -2t(cos{k_{x}}+ cos{k_{y}}) -
4t^{'}(cos{k_{x}} cos{k_{y}} + 1)-\mu  .\label{IKH2}
\end{equation}
Ограничимся рассмотрением, допированной дырками, системы, когда $t,t^{'}$ --- операторы перескакивания удовлетворяют следующим условиям: $t>0, t_{'}<0$
и $0<\mid t_{'} \mid /t < 1/2$. Для произвольного значения $t^{'}/t$ спектр (\ref{IKH2}) содержит сингулярности Ван Хова, связанные с точками $A=(\pi,0)$ и $B(0,\pi)$. Химический потенциал $\mu$, в данном случае, будет служить мерой энергктического расстояния между сингулярностью Ван Хова и энергией Ферми. При $\mu=0$ (СВХ) лежит на Ферми поверхности. Не останавливаясь на деталях вычислений, описанных в оригинальной работе \cite{irk}, приведем ее основной результат, а именно уплощение одноэлектронного спектра в области Ферми поверхности. На Рис. \ref{Fig2234} приведены графики одноэлектронных энергий рассмотренной модельной системы для трех значений ренормализованного химического потенциала $\bar{\mu}$ ($\bar{\mu} = 0, -0.2t, -0.4t$). Можно отметить, что по мере приближения сингулярности Ван Хова к Ферми поверхности $\bar{\mu} \to 0$ область плато увеличивается в $k$-пространстве. Наблюдаемая картина вполне соответствует ферми-конденсатному состоянию системы сильно коррелированных электронов.
\begin{figure} [ht]
\begin{center}
\includegraphics [width=0.47\textwidth] {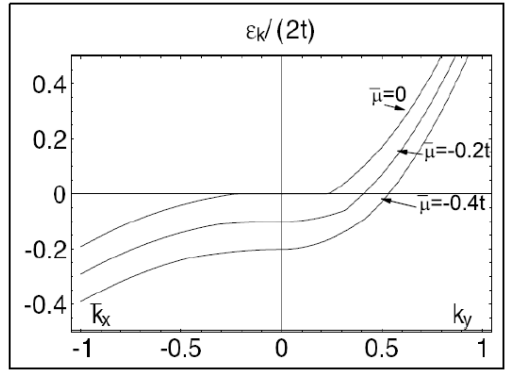}
\end{center}
\caption {Дисперсия квазичастиц для $t^{'}/t=-0.3$ и $U=4t$, полученная в рамках модели Хаббарда с использованием техники (РГ) преобразований. Величина  ренормализованного химического потенциала $\bar{\mu}$ принимала значения  ($\bar{\mu} = 0, -0.2t, -0.4t$)  } \label{Fig2235}
\end{figure}

\subsection* { 3.1.5. Расчет ферми-системы с ФК в рамках простейшего функционала Ландау. } \label{NHF}
\addcontentsline{toc}{subsection}{ 3.1.5. Расчет ферми-системы с ФК в рамках простейшего функционала Ландау. }

Вскоре после появления основополагающей работы В. Ходеля и В. Шагиняна \cite{ks}, описывающей ферми-коненсатное состояние ферми-жидкости в рамках функционального подхода, Ф. Нозьером был проведен анализ этого явления с использованием простейшего функционала Ландау \cite{noz}.

Рассмотрим энергию основного состояния ферми-жидкости как функционал распределения квазичастиц $n_{k}$:
\begin{equation} E[n_{k}]-\mu N = \sum_{k}(\xi_{k}-\mu)n_{k} +  \frac{1}{2}\sum_{kk^{'}}V_{kk^{'}}n_{k}n_{k^{'}}\label{NHF1},
\end{equation}
где $ \xi_{k} $ --- кинетическая энергия, $ \mu $ --- химический потенциал. $ V_{kk^{'}}(\mid{\bf k}-{\bf k^{'}}\mid) $ --- оператор энергии взаимодействия квазичастиц, характеризуется видом зависимости от $ \sigma = (\mid{\bf k}-{\bf k^{'}}\mid) $ и силой взаимодействия. Силу взаимодействия можно оценить с помощью величины $ U $
\begin{equation} U = \sum_{k^{'}}V_{kk^{'}(\sigma),\sigma = (\mid{\bf k}-{\bf k^{'}}\mid)}\label{NHF2},
\end{equation}
которая имеет смысл энергии взаимодействия при нулевом расстоянии между частицами. Обычно при $U>0$ частицы отталкиваются, а при $U<0$ --- притягиваются. Отметим, что при $V_{kk^{'}}=V=const$ рассматриваемая модель сводится к модели Хаббарда. Энергия квазичастиц $\varepsilon_{k}$ легко выражается из (\ref{NHF1})
\begin{equation} \varepsilon_{k} = \frac{\delta E[n_{k}]}{\delta n_{k}} = (\xi_{k}-\mu) +  \sum_{k^{'}}V_{kk^{'}}n_{k^{'}} \label{NHF3},
\end{equation}

Не стремясь представить здесь все результаты, которые можно получить в рамках рассматриваемого подхода, ограничимся только качественным анализом, демонстрирующим существование ферми-конденсатного состояния.
Анализ уравнений (\ref{NHF1}), (\ref{NHF3}) в зависимости от величины $\sigma$ начнем с рассмотрения предельных случаев. Если $\sigma \to \infty$ (контактное взаимодействие), то энергия взаимодействия $ V_{kk^{'}}n_{k^{'}} $ равна константе $VN$, которая может быть включена в состав химического потенциала. Это ситуация соответствующая поведению ферми-газа. В противоположном случае, когда $\sigma \to 0$, но $\sigma \neq 0$ (дальнодействующее взаимодействие), уравнения (\ref{NHF1}), (\ref{NHF3}) сводятся к виду:
\begin{equation} \left \{ \begin{array}{c}
E-\mu N = \sum_{k} \{ (\xi_{k}-\mu)n_{k} +  \frac{\textstyle{1}}{\textstyle{2}}U n_{k}^{2} \} \\
\varepsilon_{k} = \xi_{k}-\mu +  Un_{k} \end{array} \right. \label{NHF4},
\end{equation}
\begin{figure} [ht]
\begin{center}
\includegraphics [width=0.47\textwidth] {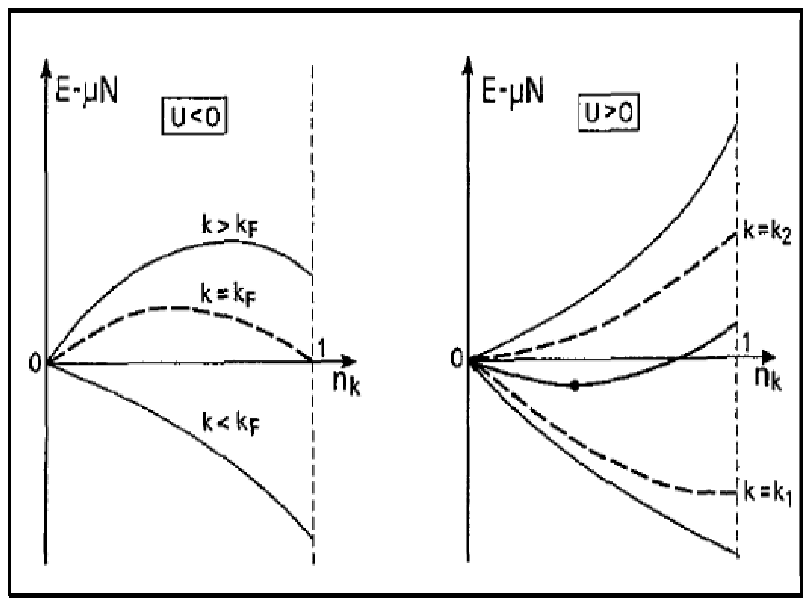}
\end{center}
\caption {Энергия основного состояния системы как функция $n_{k}$ для взаимодействия с $\sigma \to 0$. } \label{Fig2229}
\end{figure}
Энергия основного состояния системы  существенно зависит от знака величины $U$. Когда $U<0$ (левая панель, Рис. \ref{Fig2229} ), энергия минимальна при двух значениях $n_{k}$: либо при $n_{k}=0$, либо при $n_{k}=1$. В этом случае имеет место фазовый переход первого рода при некотором значении $k=k_{F}$. Ему соответствует ступенчатое распределение Ферми. В противоположном случае, $U>0$ (правая панель, Рис. \ref{Fig2229} ), минимум энергии достигается при некотором промежуточном распределении  $n_{k}$, в ограниченной области изменения $k$, $k_{1}<k<k_{2}$. Ферми-ступенька размазывается при $T=0$. Результат имеет следующий вид:
\begin{equation} n_{k}= \left \{ \begin{array}{l}
1 \hspace{3mm} \mbox{при} \hspace{3mm}  k<k_{1} \\
\frac{\textstyle{\xi_{k}-\mu}}{\textstyle{U}},\hspace{3mm} \varepsilon_{k} = 0 \hspace{3mm} \mbox{при} \hspace{3mm}  k_{1}<k<k_{2} \\
0\hspace{3mm} \mbox{при} \hspace{3mm}  k>k_{2}
 \end{array} \right. \label{NHF5},
\end{equation}
где $k_{1}$ и $k_{2}$ соответствуют значениям $\xi_{k}$, для которых $n_{k}$ равно 1 и 0.
\begin{figure} [ht]
\begin{center}
\includegraphics [width=0.47\textwidth] {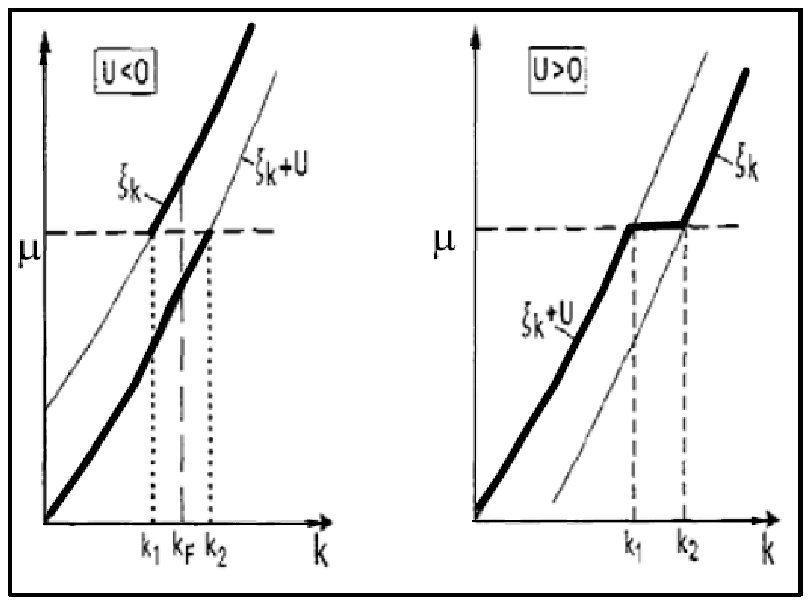}
\end{center}
\caption {Энергия основного состояния системы как функция $n_{k}$ для взаимодействия с $\sigma \to 0$. } \label{Fig2230}
\end{figure}
Графически зависимость $n_{k}$ в выражении (\ref{NHF5}) выглядит примерно так, как изображено на Рис. \ref{Fig1}. Немного другая интерпретация уравнений (\ref{NHF4}), позволяющая понять поведение энергии квазичастиц, может быть получена из Рис. \ref{Fig2230}. На нем изображены энергии квазичастиц $\varepsilon_{k}+\mu$ в виде двух ветвей, одна из которых соответствует  $n_{k}=0$ а другая $n_{k}=1$. Жирные линии соответствуют состояниям, которые реализуются при $T=0$. Когда $U<0$ между $k_{1}$ и $k_{2}$ существуют два решения, одно из которых реализуется, так как минимизирует энергию системы. Переход от одного решения ко второму происходит скачком в некоторой точке $k_{F}$. Когда $U>0$ между $k_{1}$ и $k_{2}$ нет решений соответствующих ни $n_{k}=0$, ни $n_{k}=1$, поэтому мы вынуждены искать решения в этой области, соответствующие некоторым промежуточным  $n_{k}$. Решение имеет вид плато при $\varepsilon_{k}=0$.

Проведенное рассмотрение хотя и весьма наглядно, но требует некоторой аккуратности в понимании области его справедливости. Поскольку предел  $\sigma = 0$ физически нереализуем, то следует учитывать, что реально представляет интерес ситуация, когда $\sigma << p_{F}$, то есть когда речь идет о взаимодействиях, приводящих к малым изменениям импульсов. С другой стороны $\sigma $ должно быть много больше расщепления уровней в импульсном пространстве $\propto 1/N$. Другими словами, суммирование в (\ref{NHF2}) должно проводиться по достаточно большому набору состояний $k^{'}$ для того, чтобы точность усреднения $n_{k}$ была достаточно высокой. При этом $\sigma $ должна быть достаточно малой, чтобы в ее пределах $n_{k}$ менялась слабо.

С учетом вышесказанного, существенным является то, что в рамках данного подхода к описанию фирми-систем ферми-конденсатное состояние возникает как естественное следствие из уравнений, описывающих это состояние.

\section* { 3.2. Численные расчеты. } \label{CR}
\addcontentsline{toc}{section}{ 3.2. Численные расчеты. }

Численные расчеты параметров систем, испытывающих ФККФП, из первых принципов, а также на основе модельных представлений возможны и демонстрируют качественно неплохие результаты, однако их применение связано с необходимостью использовать значительные вычислительные ресурсы.

\subsection* { 3.2.1. Расчет характеристик ферми-конденсатного состояния из первых принципов. } \label{ZRPP}
\addcontentsline{toc}{subsection}{ 3.2.1. Расчет характеристик ферми-конденсатного состояния из первых принципов. }

В основу расчетов одночастичных энергий электронов 2D системы, описанных в работе М.И. Зверева \cite{zverev}, положена микроскопическая первопринципная теория ферми-систем, развитая В.А. Ходелем и В.Р. Шагиняном \cite{ksk}.
Не излагая здесь этой теории, сформулируем ее основные результаты, положенные в основу вычислительных алгоритмов.

\begin{figure} [ht]
\begin{center}
\includegraphics [width=0.47\textwidth] {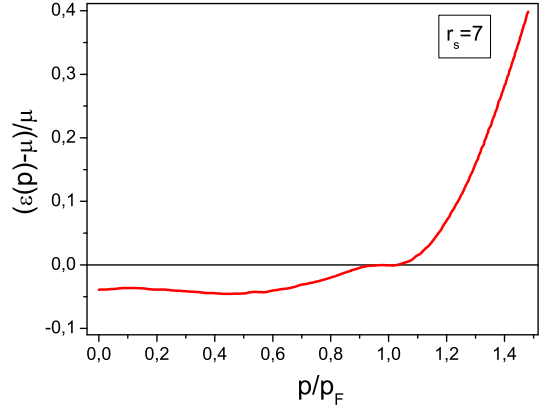}
\end{center}
\caption {Одночастичная энергия квазичастиц $\varepsilon(p)$ в $2D$ ферми-системе для радиуса Зейтца $r_{s}=7$. } \label{Fig2236}
\end{figure}

Рассматривается 2D электронный газ в модели "`желе"' с потенциалом парного взаимодействия $V(q)=2\pi g/q$, где $q$ --- импульс, $g$ --- константа спаривания ($g_{max}=e^{2}$), $e$ --- заряд электрона. Степень разреженности этой системы определяется значением параметра Зейтца $r_{s}=\sqrt{2}Me^{2}/p_{F}$.
По определению одночастичный спектр электронов описывается сооношением:
\begin{equation}
\varepsilon(p) = \frac {\delta E_{0}}{\delta n(p)} \label{ZRPP1},
\end{equation}
из которого в рамках рассматриваемой теории получено равенство:
\begin{equation}
\varepsilon(p) = \frac {p^{2}}{2M} + \varepsilon_{1}(p) + \varepsilon_{2}(p) \label{ZRPP2}.
\end{equation}
В выражении (\ref{ZRPP2}) $\varepsilon_{1}(p)$ и $\varepsilon_{2}(p)$ описываются соотношениями:
\begin{equation}
\varepsilon_{1}(p) =-\frac {1}{2} \int \frac{d^{2}q}{2\pi} \frac {e^{2}}{\mid{\bf p}- {\bf q}\mid}
- \frac {1}{2} \int_{0}^{e^{2}}dg \int \frac{d^{2}q}{2\pi q} \int_{0}^{\infty}\frac {d\omega}{\pi} Im[ \varphi^{2}({\bf q},\omega) \frac{\delta \chi_{0}({\bf q},\omega)}{\delta n(p)}] \label{ZRPP3},
\end{equation}
\begin{equation}
\varepsilon_{2}(p) =
- \frac {1}{2} \int_{0}^{e^{2}}dg \int \frac{d^{2}q}{2\pi q} \int_{0}^{\infty}\frac {d\omega}{\pi} Im[ \chi^{2}({\bf q},\omega) \frac{\delta R({\bf q})}{\delta n(p)}] \label{ZRPP31},
\end{equation}
здесь $\varphi(q,\omega) = [1-R(q)\chi_{0}(q,\omega)]^{-1}$, $ R(q)$ --- функция эффективного взаимодействия, $\chi_{0}(q,\omega)$ --- функция отклика невзаимодействующей системы, $\chi (q,\omega)$ --- функция отклика системы со взаимодействием. Не останавливаясь на технических сложностях, связанных с вычислением $ R(q)$, $\chi_{0}(q,\omega)$, $\chi(q,\omega)$ и их вариационных производных, приведем результаты расчетов. Отметим, что как и в любых численных вычислениях, выполненных $"ab \hspace{2mm} inicio"$, они требуют больших вычислительных ресурсов. Например, решение данной задачи потребовало использования распределенных вычислений на 600 процессорах.
На Рис. \ref{Fig2236} приведен график одночастичной энергии квазичастиц $\varepsilon(p)$ в $2D$ ферми-системе для радиуса Зейтца $r_{s}=7$.

\subsection* { 3.2.2. Модельные расчеты параметров ферми-системы с ФК. } \label{SPCMR}
\addcontentsline{toc}{subsection}{ 3.2.2. Модельные расчеты параметров ферми-системы с ФК. }

Поведение ферми-системы, совершающей ФККФП, с помощью  численных вычислений, использующих модельный функционал Ландау \cite{ksk,ksn}
\begin{eqnarray}
\nonumber E[n(p)]&=&\int\frac{{\bf p}^2}{2M}\frac{d{\bf
p}}{(2\pi)^3}+\frac{1}
{2}\int V({\bf p}_1-{\bf p}_2)\\
&\times&n({\bf p}_1)n({\bf p}_2) \frac{d{\bf p}_1d{\bf
p}_2}{(2\pi)^6},\label{NMC1}
\end{eqnarray}
исследовано К. Поповым и В. Шагиняном \cite{sp,sap}.
В выражении (\ref{NMC1}) $V({\bf p}_1-{\bf p}_2)$ --- эффективный потенциал двухчастичного взаимодействия.
Отметим, что данный вид функционала является лишь первым приближением для энергии системы, удобным для анализа систем в рамках точно решаемых моделей или численных расчетов.
Существенно, что использование этого функционала возможно только для изучения эффектов, связанных с каналом взаимодействия частиц и дырок в окрестности точки фазового перехода, и не может быть использован для получения характеристик канала взаимодействия частица-частица, например, для вычисления спаривательных сил \cite {ksk}.

Подставляя (\ref{NMC1}) в (\ref{FL2}), получаем выражение для энергии квазичастиц:
\begin{equation}
\varepsilon[n(p)] = \frac{{\bf p}^2}{2M}+
\frac{1}
{2}\int V({\bf p}_1-{\bf p}_2)n({\bf p}_1)
\frac{d{\bf p}_1}{(2\pi)^3}.\label{NMC2}
\end{equation}
Подбирая различные формы потенциала $V({\bf p}_1-{\bf p}_2)$, можно  реализовать целый ряд вычислительных алгоритмов для численного решения проблемы.

Если воспользоваться соотношением (\ref{FL4}), связывающим
$n({\bf p},T)$ и $\varepsilon({\bf p},T)$,
\begin{equation}
\nonumber n({\bf p},T)=
\left\{1+\exp\left[\frac{(\varepsilon({\bf p},T)-\mu)}
{T}\right]\right\}^{-1},
\end{equation}
 и учесть, что уравнение (\ref{NMC2}) нужно решать с учетом условия нормировки, которое фактически служит уравнением для определения химического потенциала $\mu$:
\begin{equation}
\rho = 2\int n({\bf p}) \frac{d{\bf p}}{(2\pi)^3},\label{NMC3}
\end{equation}
где $\rho$ плотность системы, то уравнения (\ref{NMC2}), (\ref{FL4}), (\ref{NMC3}) образуют замкнутую систему, позволяющую для вычисления   $n({\bf p},T)$, $\varepsilon({\bf p},T)$ и $\mu$ реализовать итерационный процесс.

В качестве примера, рассмотрим потенциал взаимодействия квазичастиц, напоминающий потенциал Юкавы:
\begin{equation}
V({\bf p}-{\bf
p}_1)=g_0\frac{\exp(-\beta_0\sqrt{({\bf p}-{\bf
p}_1)^2+\gamma_0^2}} {\sqrt{({\bf p}-{\bf p}_1)^2+\gamma_0^2}}.
\label{NMC4} \end{equation}
Здесь $g_{0}$ --- постоянная спаривания, $\beta_{0}$ --- константа взаимодействия, $\gamma_{0}$ --- дисперсионная константа. Это довольно общий вид потенциала, позволяющий за счет изменения параметров $g_{0}$,  $\beta_{0}$ и $\gamma_{0}$ моделировать различные принципиально важные частные взаимодействия. Например, при
$\gamma_{0}=0$, $g_{0}/(2\beta_{0}^{2}\pi^{2})=const$ и $\beta_{0} \to \infty $ получаем:
\begin{equation} V({\bf p}-{\bf p}_1) =
(2\pi)^{3}V\delta({\bf p}-{\bf p}_1), \label{NMC5}
\end{equation}
здесь $V$ --- постоянная, который соответствует ограниченному дальнодействующему взаимодействию в координатном пространстве. В этом случае модель допускает точное решение \cite{noz, ksk} и была рассмотрена  ранее в разделе \ref{TRMKS}.
При
$ \gamma_{0}=0 $ и $ \beta_{0} = 0 $ получаем потенциал:
\begin{equation} V({\bf p}-{\bf p}_1) =
\frac{g_0} {\mid{\bf p}-{\bf p}_1\mid},
\label{NMC6}
\end{equation}
также рассмотренный в разделе \ref{TRMKS}, допускающий точное аналитическое решение \cite{ksk} и соответствующий движению квазичастиц в координационном пространстве под влиянием внешнего квазиупругого поля $V_{e}(r)=(1/(2M))r^{2}$ и взаимодействующих друг с другом посредством Кулоновского потенциала:
\begin{equation} V({\bf r}-{\bf r}_1) \approx
\frac{1} {\mid {\bf r}-{\bf r}_1\mid},
\label{NMC7}
\end{equation}

Преимуществом потенциала (\ref{NMC4}) является то, что при подстановке его в уравнение (\ref{NMC2}), интеграл в правой части последнего можно проинтегрировать по угловым переменным. Выполняя интегрирование по угловым переменным в (\ref{NMC2}) и (\ref{NMC3}), а также переходя к безразмерным переменным:
\begin{equation} z = p/p_{F}, T = T\frac{2M}{p_{F}^{2}},
\varepsilon(z,T) = \varepsilon(p,T)\frac{2M}{p_{F}^{2}},
\mu = \mu\frac{2M}{p_{F}^{2}}, \tilde{\rho}=\rho\frac{3\pi^{2}}{p_{F}^{3}},
\label{NMC8}
\end{equation}
и вводя безразмерные параметры:
\begin{equation} \beta = \beta_{0}p_{F}, \gamma = \gamma_{0}p_{F},
g = \frac{g_{0}M}{2\pi^{2}}
\label{NMC9}
\end{equation}
здесь $p_{F}$ --- импульс Ферми, M --- голая масса фермиона, получаем из (\ref{NMC2}), (\ref{FL4}), (\ref{NMC3}) рекурентные соотношения, положенные в основу итерационного процесса для вычисления   $n(z,T)$, $\varepsilon(z,T)$ и $\mu$:
\begin{equation}
n_{k}(z,T) =
\left\{1+\exp\left[\frac{(\varepsilon_{k}(z,T)-\mu_{k})}
{T}\right]\right\}^{-1},\label{NMC10}
\end{equation}
\begin{equation}
1 = \tilde{\rho}_{k} = 3\int n_{k}(z,T) z^{2} dz,\label{NMC11}
\end{equation}
\begin{equation}
\varepsilon_{k+1}(z,T) = z^2+\frac{g}{\beta
z}\int \left[
\exp(-\beta\sqrt{(z-z_1)^2+\gamma^2})-\exp(-\beta\sqrt{(z+z_1)^2+\gamma^2})\right]
n_{k}(z_1,T)z_1dz_1.\label{NMC12}
\end{equation}
Предложенный алгоритм как правило хорошо сходится, позволяя достигать разницы вычисления условия нормировки (\ref{NMC11}) на двух соседних шагах  итерации в интервале температур $[1,10^{-7}]$ не хуже $10^{-6}$. Полученная функция $n(z,T)$ использовалась для вычисления энтропии $S(T)$
\begin{equation} S(T)=-\frac{p_F^3}{\pi^2}\int[n(z,T)
\ln(n(z,T))+(1-n(z,T))\ln(1-n(z,T))] z^2dz, \label{NMC13}
\end{equation}
эффективной массы $M^{*}(T)$
\begin{equation} M^{*}(T)=3\frac{S(T)}{p_{F}T}, \label{NMC14}
\end{equation}
и других термодинамических и кинетических характеристи системы.

\begin{figure} [ht]
\begin{center}
\includegraphics [width=0.47\textwidth] {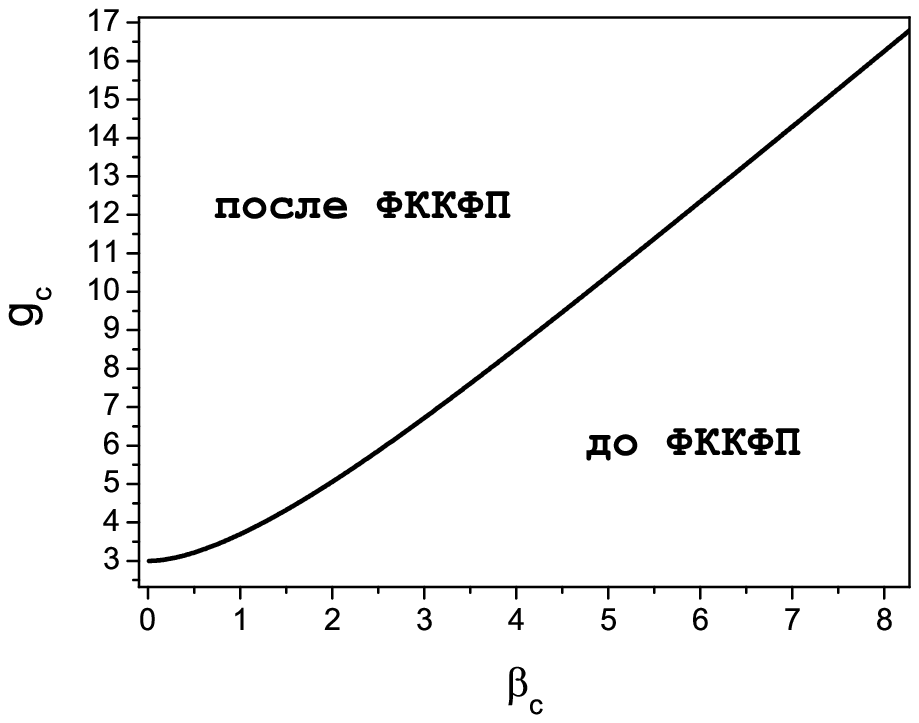}
\end{center}
\caption {Фазовая диаграмма ${\bf g_{c}}-{\bf \beta_{c}} $, разделяющая неупорядоченное состояние системы до ФККФП и упорядоченное состояние системы после  ФККФП} \label{Fig2224}
\end{figure}

ФККФП происходит, когда
параметры потенциала взаимодействия (\ref{NMC9}) достигают критических значений
\begin{equation} \beta=\beta_{c}, \gamma=\gamma_{c},  g=g_c. \label{NMC15}
\end{equation}
С другой стороны, ФККФП имеет место, когда эффективная масса
$M^{*}\to\infty$. Это условие позволяет для любой пары параметров (\ref{NMC15}) подобрать третий так, что при $T=0$  система попадает в критическую точку ФККФП. Для случая $\gamma=0$ можно аналитически получить выражение, связывающее $b_c$ и $g_c$ \cite{ksk,ksn}:
\begin{equation} \frac{g_c}{b_c^3}(1+b_c)\exp(-b_c)[b_c\cosh(b_c)-\sinh(b_c)]=1. \label{NMC16}
\end{equation}
На рисунке (\ref{Fig2224}) приведен график функции (\ref{NMC16}), иллюстрирующий характер зависимости $g_c$ от $b_c$

\begin{figure} [ht]
\begin{center}
\includegraphics [width=0.47\textwidth] {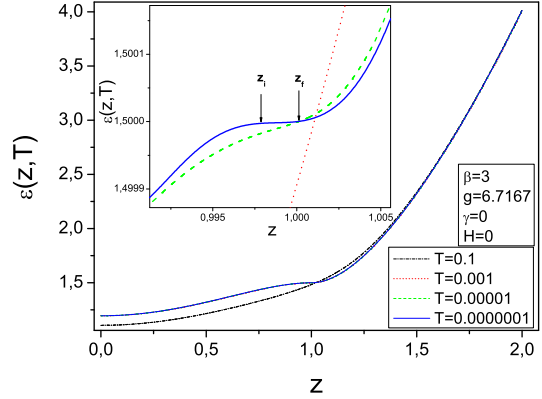}
\end{center}
\caption {Одночастичная энергия квазичастиц $\varepsilon (z,T)$ для значений безразмерной температуры в интервале от 0.1 до $10^{-7}$. } \label{Fig2225}
\end{figure}

\begin{figure} [ht]
\begin{center}
\includegraphics [width=0.47\textwidth] {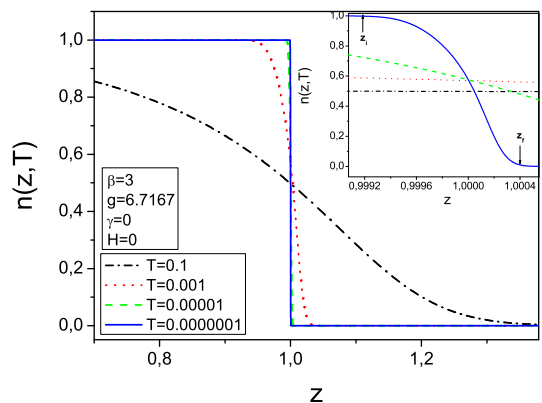}
\end{center}
\caption {Функции распределения квазичастиц $n (z,T)$ для значений безразмерной температуры в интервале от 0.1 до $10^{-7}$. На вставке область быстрого изменения $n (z,T)$ в увеличенном виде.}
\label{Fig2226}
\end{figure}

\begin{figure} [ht]
\begin{center}
\includegraphics [width=0.47\textwidth] {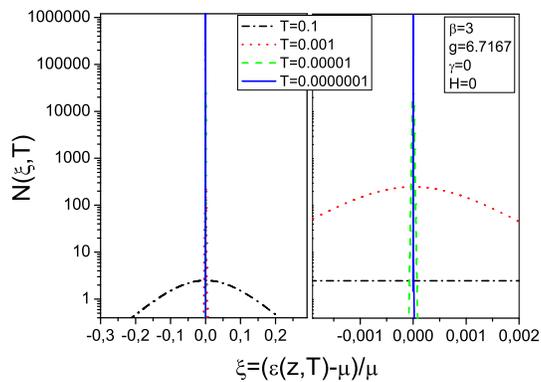}
\end{center}
\caption {Функции плотности  состояний $ N(\xi ,T) $ для значений безразмерной температуры в интервале от 0.1 до $10^{-7}$ (левая панель). На правой панели область быстрого изменения $ N(\xi ,T)$ в увеличенном виде. }  \label{Fig2227}
\end{figure}

В качестве примера, демонстрирующего достижение системой критической точки
ферми-конденсатного квантового фазового перехода, рассмотрим результаты вычислений для $ \gamma_{с}=0 $, $ \beta_{с} = 3 $ и $ g_{с} = 6.7167... $ при $T\to 0$. На рисунке (\ref{Fig2225}) изображено семейство одночастичных энергий квазичастиц ферми-системы $\varepsilon (z,T)$ для стремящейся к нулю последовательности значений температуры. Наблюдается форирование плоского участка спектра в окрестности Ферми-поверхности $z=0$.
На рисунке (\ref{Fig2226}) изображено семейство функций распределения (чисел заполнения) квазичастиц ферми-системы $n (z,T)$ для, стремящейся к нулю, последовательности значений температуры, соответствующее энергиям на рисунке (\ref{Fig2225}). Можно заметить, что форма функций распределения не стремится к ступеньке при $ T \to 0$ и имеет асимметричную относительно $z_{F} \approx 1$ форму. Область занимаемая ферми-конденсатной фазой в рамках рассматриваемого примера приблизительно равна $(z_{f}-z_{i})=0.0012$.
На рисунке (\ref{Fig2227}) изображено семейство функций плотности  состояний $ N( \xi , T) $ в области ферми-поверхности для, стремящейся к нулю, последовательности значений температуры. Видно, что по мере приближения системы к критической точке ФККФП форма функции $ N(\xi ,T) $ стремится к   $ \delta $-функции.
\pagebreak

\chapter* {4. Обзор экспериментальных фактов, доказывающих существование фермионного конденсата. }
\label{OEF}
\addcontentsline{toc}{chapter}{ 4. Обзор экспериментальных фактов, доказывающих существование фермионного конденсата. }

В этом разделе будут приведены различные экспериментальные свидетельства, прямо или косвенно доказывающие правомерность существования концепции фермионного конденсата. Результат каждого эксперимента будет интерпретироваться с позиций развиваемой теории ФККФП. Параллельно с физическими экспериментами для иллюстрации различных положений теории ФККФП будут использоваться результаты модельных компьютерных экспериментов.

\section* { 4.1. Появление ФККФП в ферми-системах при стремлении плотности фермионов к критическому значению. }\label{PF1}
\addcontentsline{toc}{section}{ 4.1. Появление ФККФП в ферми-системах при стремлении плотности фермионов к критическому значению. }

В этом разделе мы рассмотрим поведение эффективной
массы $M^*$ как функции плотности системы $x$, когда $x\to x_{FC}$.

Экспериментальные факты по двумерному $^3$He высокой плотности
\cite{mor,cas1,cas} показывают, что эффективная масса расходится,
когда достигается плотность, при которой 2D жидкость $^3He$
начинает затвердевать \cite{cas}. Далее, наблюдалось резкое
увеличение эффективной массы в металлической 2D электронной системе
при уменьшении плотности $x$, когда $x$ стремится к критической
плотности перехода металл-изолятор \cite{skdk}. Отметим, что
ферромагнитная неустойчивость отсутствует в рассматриваемых
ферми-системах, и соответствующая амплитуда Ландау $F^a_0>-1$
\cite{skdk, cas}, что согласуется с моделью почти локализованных
фермионов \cite{pfw}.

Рассмотрим расходимость эффективной массы в 2D и трехмерной (3D)
высоко коррелированной ферми-жидкости при $T=0$, когда плотность
$x$ приближается к ФККФП со стороны нормальной ферми-жидкости
Ландау, т.е., со стороны неупорядоченной фазы. Сначала вычислим
зависимость $M^*$ как функцию разности $(x-x_{FC})$ в случае 2D
ферми-жидкости. Для этого воспользуемся уравнением для $M ^*$,
полученным в \cite{ksz}, где была предсказана расходимость
эффективной массы $M ^*$, связанная с возникновением волны
плотности в различных ферми-жидкостях. При $x\to x_{FC}$
эффективная масса $M^*$ может быть приближенно представлена как
\begin{eqnarray}
\nonumber \frac{1}{M^{*}}&\simeq&\frac{1}{M}+\frac{1}{4\pi^{2}}
\int\limits_{-1}^{1}\int\limits_0^{g_0}\frac{ydydg}{\sqrt{1-y^{2}}} \\
&\times& \frac{v(q(y))}{\left[1-R(q(y),g) \chi_0(q(y))\right]^{2}}.
\label{DL1}\end{eqnarray} Здесь мы используем обозначение
$p_F\sqrt{2(1-y)}=q(y)$, где $q(y)$ --- переданный импульс, $v(q)$
--- парное взаимодействие, и интеграл взят по константе связи $g$
от нуля до реальной величины $g_0$. В уравнении (\ref{DL1})
$\chi_0(q,\omega)$
 --- линейная функция отклика невзаимодействующей
ферми-жидкости  и $R(q,\omega)$ --- эффективное взаимодействие, обе
функции взяты при нулевой частоте, $\omega=0$. $R$ и $\chi_0$
определяют линейную функцию отклика рассматриваемой системы
\begin{equation}
\chi(q,\omega,g)=\frac{\chi_0(q,\omega)}
{1-R(q,\omega,g)\chi_0(q,\omega)}.\label{DL2}
\end{equation}
В окрестности неустойчивости, связанной с возникновением волны
плотности и происходящей при плотности $x_{cdw}$, сингулярная часть
функции отклика $\chi$ имеет известную форму (см. например
\cite{varma})
\begin{equation}
\chi^{-1}(q,\omega,g)\simeq
a(x_{cdw}-x)+b(q-q_c)^2+c(g_0-g),\label{DL3}
\end{equation}
где $a$, $b$, $c$ --- константы, а $q_c\simeq 2p_F$ --- импульс
волны плотности. После подстановки уравнения (\ref{DL3}) в
уравнение (\ref{DL1}) и интегрирования, уравнение для эффективной
массы $M ^*$ может быть представлено в следующей форме \cite{shag1,
khod1}
\begin{equation}\frac{1}{M^*(x)}=\frac{1}{M}-\frac{C}
{\sqrt{x-x_{cdw}}},\label{DL4} \end{equation} где $C$ --- некоторая
положительная постоянная. Из уравнения (\ref{DL4}) следует, что
$M^*(x)$ расходится как функция разности $(x-x_{FC})$, и  при $x\to
x_{FC}$ эффективная масса $M^*(x)\to\infty$ \cite{shag1, khod1},
\begin{equation}
\frac{M^*(x)}{M}\simeq A+\frac{B}{x-x_{FC}},\label{DL5}
\end{equation}
где $A$ и $B$ --- константы. Из вывода уравнений (\ref{DL4}) и
(\ref{DL5}) замечаем, что их вид не зависит от взаимодействия
$v(q)$. Последнее, однако, влияет на величины $A$, $B$ и $x_{FC}$.
Этот результат находится в согласии с уравнением (\ref{FL7}),
которое определяет тот же самый, независимый от взаимодействия,
универсальный тип расходимости. Поэтому оба этих уравнения
применимы к 2D $^3$He, электронной жидкости и другим
ферми-жидкостям. Из уравнений (\ref{DL4}) и (\ref{DL5}) также
видно, что ФККФП предшествует формированию волн плотности (или
зарядовой плотности) в ферми-системах. Отметим, что в обоих случаях
разность $(x-x_{FC})$ должна быть положительной, потому что
плотность $x$ приближается к $x_{FC}$, когда система находится на
неупорядоченной стороне ФККФП с эффективной массой $M^*(x)>0$. В
случае $^3$He ФККФП имеет место при росте плотности, когда
потенциальная энергия начинает вносить определяющий вклад в энергию
основного состояния. Таким образом, при рассмотрении 2D $^3$He
жидкости заменяем $(x-x_{FC})$ на $(x_{FC}-x)$ в правой части
уравнения (\ref{DL5}). Как показывает эксперимент, эффективная
масса действительно расходится при высоких плотностях в случае 2D
$^3$He и при низких плотностях в случае 2D электронных систем
\cite{skdk,cas}.

\begin{figure} [ht]
\begin{center}
\includegraphics [width=0.47\textwidth] {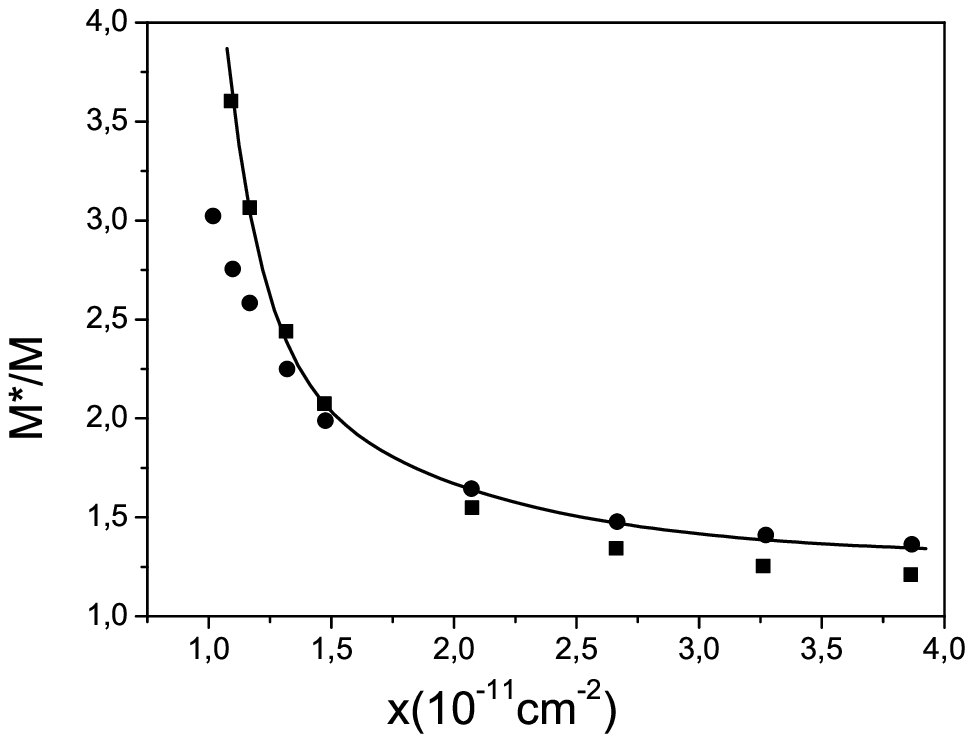}
\end{center}
\caption {Отношение $M ^*/M$ в кремниевом MOSFET как функция
плотности электронов $x$. Квадраты обозначают экспериментальные
данные осцилляций Шубникова-де-Хааза. Данные, полученные при
наложении параллельного магнитного поля, показаны кругами
\cite{skdk,krav}. Сплошная линия представляет функцию (\ref{DL4}).}
\label{Fig5}
\end{figure}

\begin{figure} [! ht]
\begin{center}
\includegraphics [width=0.47\textwidth] {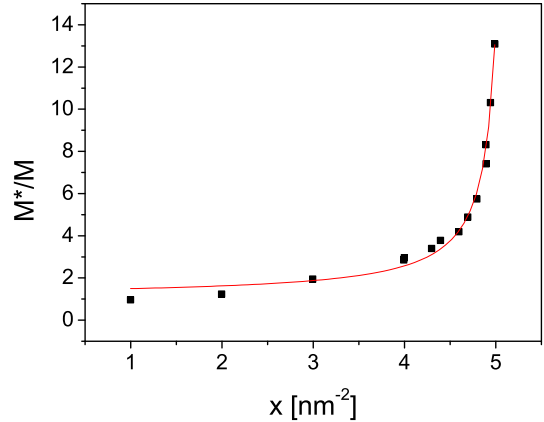}
\end{center}
\caption {Отношение $M ^*/M$ в двумерном $^3$He в зависимости от
плотности жидкости $x$, полученное из измерений теплоемкости и
намагниченности, экспериментальные значения показаны квадратами
\cite{cas1,cas}. Сплошная линия представляет функцию
$M^*(x)/M=A+\frac{B}{(x_{FC}-x)}$, где A=1.09, B=1.68 $\rm nm^{-2}$
и $x_{FC}=5.11$ $\rm nm^{-2}$.} \label {Fig6}
\end{figure}

Поведение эффективной массы как функции электронной плотности $x$ в
кремниевом MOSFET аппроксимировано при помощи (\ref{DL4}) и
показано на Рис. \ref{Fig5}. Соответствующие параметры
аппроксимации: $x_{cdw}=0.7\times10^{-11}$cm$^{-2}$,
$C=2.14\times10^{-6}$cm$^{-1}$ и
$x_{FC}=0.9\times10^{-11}$cm$^{-2}$ \cite{khod1}. Из Рис.
\ref{Fig5} видно, что уравнение (\ref{DL4}) хорошо описывает
эксперимент.
Расходимость эффективной массы $M^*(x)$, обнаруженная в измерениях
на 2D $^3$He \cite{cas1,cas}, показана на Рис. \ref{Fig6}. Как
видно из Рис. \ref{Fig5} и \ref{Fig6}, описание, даваемое формулами
(\ref{DL4}), (\ref{DL5}) и (\ref{FL7}) находится в хорошем согласии
с экспериментом.

В случае 3D систем, при $x\to x_{FC}$, эффективная масса дается
выражением \cite{ksz}
\begin{equation}
\frac {1}{M^{*}}\simeq\frac{1}{M}+\frac{p_F}{4\pi^{2}}\int\limits_
{-1}^{1}\int\limits_0^{g_0}\frac{v(q(y))ydydg}{\left[1-R(q
(y),g)\chi_0(q(y))\right]^{2}}.\label{DL6}
\end{equation}
Сравнение уравнений (\ref{DL1}) и (\ref{DL6}) показывает, что нет
никакого качественного различия между ними несмотря на то, что эти
уравнения описывают 2D и 3D случаи. Для 3D случая  таким же образом
мы можем получить уравнения (\ref{DL4}) и (\ref{DL5}), разумеется,
численные значения коэффициентов изменятся, поскольку они зависят и
от размерности пространства. Единственное различие между 2D и 3D
электронными системами состоит в том, что в последних ФККФП
происходит при плотностях значительно ниже тех, которые
соответствуют 2D системам. Для массивного 3D $^3$He ФККФП не
происходит, поскольку он поглощается фазовым переходом первого рода
жидкость --- твердая фаза \cite{cas1,cas}.

\section* { 4.2. Энтропия ферми-системы в окрестности ФККФП. \label{EVOT }}
\addcontentsline{toc}{section}{ 4.2. Энтропия ферми-системы в окрестности ФККФП. }

В окрестности ФККФП, в области высококоррелированной ферми-жидкости, 
когда система находится на неупорядоченной стороне от критической точки, 
зависимость энтропии от температуры определяется соотношением \cite{sap}:
\begin{equation}
S(T)=c \sqrt{T},\label{EVOT1}
\end{equation}
где $c$ --- константа. 
Такое поведение соответствует свойствам ферми-жидкости Ландау.

В сильнокоррелированной ферми-системе, при $T\ll T_f$ функция $n_0({\bf p})$, заданная уравнением
(\ref{FL8}), определяет энтропию $S_{NFL}(T)$ тяжелой электронной
жидкости, находящейся на упорядоченной стороне ФККФП. Как следует
из формулы (\ref{snfl}), энтропия содержит вклад, не зависящий от
температуры, $S_0\sim(p_f-p_i)/p_F\sim|r|$, где
$r=(x-x_{FC})/x_{FC}$. Другой специфический вклад связан со
спектром $\varepsilon({\bf p})$, который обеспечивает связь между
бездисперсионной областью $(p_f-p_i)$, занятой фермионным
конденсатом, и нормальными квазичастицами, расположенными в $p<p_i$
и в $p>p_f$. Этот спектр имеет форму $\varepsilon({\bf
p})\propto(p-p_f)^2\sim(p_i-p)^2$. Такая форма спектра
подтверждается в точно решаемых моделях для систем с фермионным
конденсатом и приводит к вкладу в теплоемкость $C\sim\sqrt{T/T_f}$
\cite{ks}. Таким образом, при $0<T\ll T_f$ энтропия может быть
аппроксимирована функцией
\begin{equation}
S_{NFL}(T)\simeq
S_0+a\sqrt{\frac{T}{T_f}}+b\frac{T}{T_f},\label{SL1}
\end{equation}
где  $a$ и $b$ --- константы. Третье слагаемое в правой части
уравнения (\ref{SL1}) возникает вследствие вклада независимой от
температуры части спектра $\varepsilon({\bf p})$ и дает
относительно малую добавку в энтропию. Как мы увидим, независящее
от температуры слагаемое $S_0$ определяет универсальные
термодинамические и транспортные свойства тяжелой электронной
жидкости с ФК, которую мы называем сильно коррелированной
ферми-жидкостью. Свойства этой жидкости кардинально отличаются от
свойств высоко коррелированных систем. В результате мы можем
рассматривать ФККФП как фазовый переход, который разделяет высоко
коррелированную и сильно коррелированную ферми-жидкости. Поскольку
высоко коррелированная жидкость при $T\to0$ имеет поведение
ферми-жидкости Ландау, то ФККФП отделяет ферми-жидкость Ландау от
сильно коррелированной ферми-жидкости.

\begin{figure} [ht]
\begin{center}
\includegraphics [width=0.47\textwidth] {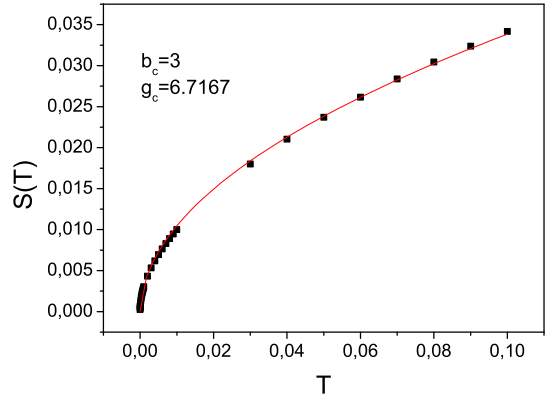}
\end{center}
\caption {Энтропия $S(T)$ высоко коррелированной ферми-жидкости,
находящейся в критической точке ФККФП. Сплошная линия представляет
$S(T)\propto \sqrt{T}$, квадраты --- результаты вычислений.} \label
{Fig10}
\end{figure}

Проиллюстрируем поведение $S(T)$ вычислениями, используя модельный
функционал Ландау \cite{ksk,ksn} с потенциалом (\ref{NMC4}), описанными ранее. 
На Рис. \ref{Fig10}
показан график зависимости энтропии от температуры в случае высоко коррелированной ферми-жидкости.
Расчеты  $S(T)$, проведенные со значениями параметров $g=g_c=6.7167$ и
$\beta=b_c=3$ показывают, что 
$S(T)\propto \sqrt{T}$ и $S(T) \to 0$ при $T \to 0$. 

\begin{figure} [ht]
\begin{center}
\includegraphics [width=0.47\textwidth] {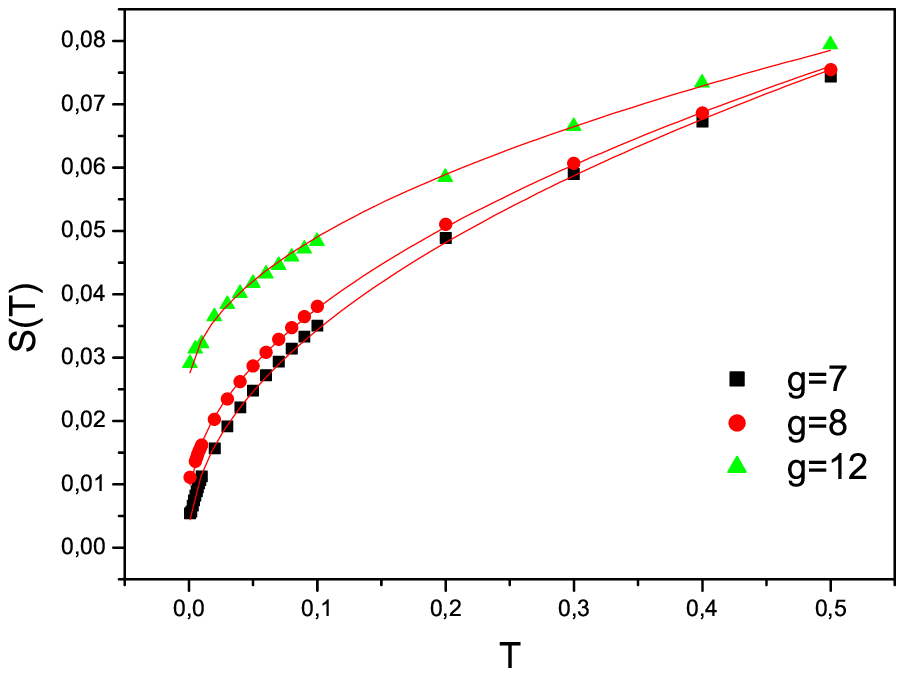}
\end {center}
\caption {Энтропия $S(T)$ как функция температуры. Сплошные линии
представляют аппроксимацию вычисленных значений $S(T)$, основанную
на уравнении (\ref{SL1}).} \label {Fig13}
\end{figure}

Для сильно коррелированной ферми-жидкости
температурная зависимость $S(T)$ представлена в виде треугольников,
кругов и квадратов на Рис. \ref{Fig13}. Вычисления были выполнены
для $g=7,8,12$ и $\beta=b_c=3$. Напомним, что критическая величина
$g=g_c=6.7167$.
Из Рис. \ref{Fig13} видно, что $S_0$ увеличивается по мере того как
система удаляется от ФККФП. Очевидно, что слагаемое $S_0$, не
зависящее от температуры, не вносит вклад в теплоемкость. Однако
корневое слагаемое в (\ref{SL1}) вносит вклад, и теплоемкость имеет
аномальное поведение $C(T)\propto\sqrt{T}$, как это подтверждается
нашими вычислениями.
\pagebreak

\section* { 4.3. Скачек коэффициента Холла. }\label{SKH}
\addcontentsline{toc}{section}{ 4.3. Скачек коэффициента Холла. }

\begin{figure} [ht]
\begin{center}
\includegraphics [width=0.47\textwidth] {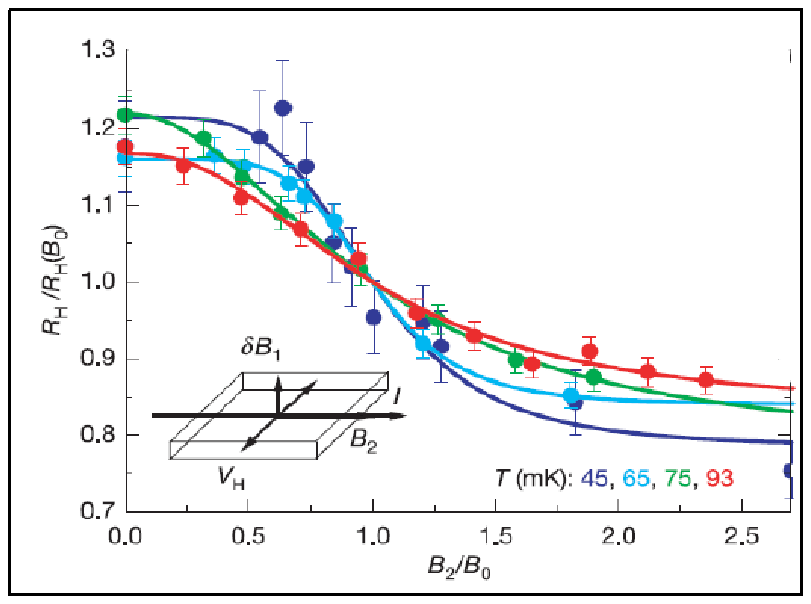}
\end {center}
\caption {Коэффициент Холла для $\rm YbRh_2Si_2$, нормализованный на значение $R_{H}(B_{0})$ при критическом значении $B=B_{c0}=B_{0}$, соответствующее точке перехода от одного режима изменения $R_{H}$ к другому, как функция нормализованной индукции магнитного поля $B/B_{0}$. Измерения проводились при следующих значениях температуры: $45, 65, 75$ и $93 mK$. Непрерывные линии соответствуют наилучшим апроксимациям экспериментальных точек.} \label {Fig1301}
\end{figure}

Прямым следствием поведения энтропии при совершении системой ФККФП 
является скачок коэффициента Холла, обнаруженный в
измерениях на $\rm YbRh_2Si_2$ \cite{pash} Рис. \ref{Fig1301}.  
Коэффициент Холла $R_H(B)$ как
функция $B$ меняется скачком при $T\to0$,  когда  наложенное
магнитное поле $B$ достигает критического значения $B=B_{c0}$ и
затем превосходит это критическое значение так, что
$B=B_{c0}+\delta B$, где $\delta B$ бесконечно малое магнитное поле
\cite{pash}. Как было показано в разделе \ref{HFL}, при $T=0$
наложение критического магнитного поля $B_{c0}$, подавляющего
антиферромагнитную фазу (с импульсом Ферми $p_{AF}\simeq p_F$),
восстанавливает ферми-жидкость Ландау с импульсом Ферми $p_f>p_F$.
При $B<B_{c0}$ энергия основного состояния антиферромагнитной фазы
ниже энергии основного состояния парамагнитной электронной
ферми-жидкости Ландау, но при $B>B_{c0}$ мы имеем дело с
противоположным случаем, когда парамагнитная ферми-жидкость Ландау
имеет меньшую энергию. При $B=B_{c0}$ и $T=0$ обе фазы имеют
одинаковую энергию основного состояния, будучи вырождены. Таким
образом, при $T=0$ и $B=B_{c0}$ бесконечно малое увеличение
магнитного поля $\delta B$ приводит к конечному скачку в импульсе
Ферми, поскольку функция распределения становится многосвязной, см.
Рис. \ref{Fig4}, однако число подвижных электронов не изменяется. В
результате коэффициент Холла испытывает резкий скачок, поскольку в
антиферромагнитной фазе $R_H (B) \propto1/p_F^3$, а в парамагнитной
фазе  $R_H (B) \propto1/p_f^3$ . Здесь предполагаем, как и в
моделях с простым односвязным ферми-объемом, что $R_H(B)$ может
рассматриваться как мера импульса Ферми \cite{pash}, и  получаем
\begin{equation}
\frac {R_H(B=B_{c0}-\delta)}{R_H(B=B_{c0}+\delta)}\simeq1+3\frac
{p_f-p_F}{p_F}\simeq1+d\frac{S_0}{x_{FC}},\label{SL7}
\end{equation}
где $S_0/x_{FC}$ --- энтропия на один тяжелый электрон, а $d$
--- константа, $d\sim 5 $. Из уравнения (\ref{SL7}) следует, что
скачкообразное изменение коэффициента Холла определяется аномальным
поведением энтропии, связанным с членом $S_0$. Поэтому скачок
стремится к нулю при $r\to0$, и исчезает, когда рассматриваемая
система находится на неупорядоченной стороне от ФККФП, где энтропия
не имеет не зависящего от температуры вклада \cite{spa}.

\section* { 4.4. Зависимость эффективной масси  $M^*$ от магнитного поля. }\label{ZEMOP}
\addcontentsline{toc}{section}{ 4.4. Зависимость эффективной масси  $M^*$ от магнитного поля. }

Когда ферми-система приближается к ФККФП со стороны неупорядоченной
фазы, она остается ферми-жидкостью Ландау с эффективной массой
$M^*$, сильно зависящей от расстояния $r = (x-x_{FC})/x_{FC}$,
температуры $T$ и магнитного поля $B$. Это состояние системы с
$M^*$, существенно зависящей от $T$, $r$ и $B$, напоминает сильно
коррелированную жидкость, описанную в разделе \ref{FLFC}. Однако в
отличие от сильно коррелированной жидкости для нее не существует
энергетического масштаба $E_0=4T$ \cite{sap}, и рассматриваемая система становится ферми-жидкостью
Ландау при достаточно низких температурах с эффективной массой
$M^*\propto1/r$ (см. уравнения (\ref{FL7}) и (\ref{DL5})). Такую
жидкость можно назвать высоко коррелированной жидкостью, которая,
как мы увидим, имеет необычные свойства, не совпадающие со
свойствами сильно коррелированных ферми-систем \cite{shag2,shag1}.

Используем уравнение Ландау для изучения поведения эффективной
массы $M^*(T,B)$ как функции температуры и магнитного поля. В
случае однородной жидкости и при конечных температурах и магнитных
полях это уравнение имеет вид \cite{lanl1}
\begin{eqnarray}
\nonumber \frac{1}{M^*(T,
B)}&=&\frac{1}{M}+\sum_{\sigma_1}\int\frac{{\bf p}_F{\bf
p}}{p_F^3}F_
{\sigma,\sigma_1}({\bf p_F},{\bf p}) \\
&\times&\frac{\partial n_{\sigma_1} ({\bf
p},T,B)}{\partial{p}}\frac{d{\bf p}}{(2\pi)^3}. \label{HC1}
\end{eqnarray}
Здесь $F_{\sigma,\sigma_1}({\bf p_F},{\bf p})$ --- амплитуда
Ландау, зависящая от импульсов $p_F$, $p$ и спина $\sigma$. Так как
металлы с ТФ являются преимущественно трехмерными, мы считаем
тяжелую электронную жидкость также 3D жидкостью. Для упрощения,
опускаем спиновую зависимость эффективной массы, поскольку в случае
слабых магнитных полей, $M^*(T, B)$ заметно не зависит от спинов.
Функция распределения квазичастиц имеет вид
\begin{equation} n_{\sigma}({\bf p},T)=\left\{ 1+\exp
\left[\frac{(\varepsilon({\bf p},T)-\mu_{\sigma})}T\right]\right\}
^{-1},\label{HC2}
\end{equation}
где $\varepsilon({\bf p},T)$  определяется уравнением (\ref{FL2}).
В нашем случае одночастичный спектр слабо зависит от спина, но
химический потенциал может иметь зависимость из-за расщепления
Зеемана. Мы будем специально показывать зависимость от спина
физических величин, когда эта зависимость будет важна для
понимания. Представим $n_{\sigma}({\bf p},T,B) \equiv\delta
n_{\sigma}({\bf p},T,B)+n_{\sigma}({\bf p},T=0,B=0)$, и уравнение
(\ref{HC1}) принимает вид
\begin{eqnarray}
\nonumber
&&\frac{M}{M^*(T,B)}=\frac{M}{M^*(x)}+\frac{M}{p_F^2}\sum_{\sigma_1}
\int\frac{{\bf p}_F{\bf p_1}}{p_F}\\
&\times&F_{\sigma,\sigma_1}({\bf p_F},{\bf
p}_1)\frac{\partial\delta n_{\sigma_1}({\bf p}_1,T,B)}
{\partial{p}_1}\frac{d{\bf p}_1}{(2\pi) ^3}. \label{HC3}
\end{eqnarray}
Мы предполагаем, что высоко коррелированная электронная жидкость
находится вблизи ФККФП, и расстояние $r$ мало, так что
$M/M^*(x)\ll1$, как это следует из уравнения (\ref{FL7}). В случае
нормальных металлов, когда электронная жидкость ведет себя как
ферми-жидкость Ландау с эффективной массой порядка нескольких
"голых" электронных масс $M/M^*(x)\sim1$, до температур $T\sim
1000$ K второе слагаемое в правой части уравнения (\ref{HC3}) имеет
порядок $T^2/\mu^2$ и намного меньше, чем первое слагаемое. Можно
проверить, что то же самое справедливо, когда накладывается
магнитное поле $B\lesssim$ 100 T. Таким образом, при
$M/M^*(x)\sim1$ рассматриваемая система имеет поведение
ферми-жидкости Ландау с эффективной массой, фактически независимой
от температуры или магнитного поля, а сопротивление $\rho(T)\propto
T^2$. Это значит, что поправки к ней, определяемые вторым слагаемым
в правой части уравнения (\ref{HC3}), пропорциональны $(T/\mu)^2$
или $(\mu_B B/\mu)^2$, где $\mu_{B}$
--- магнетон Бора.

\begin{figure} [ht]
\begin{center}
\includegraphics [width=0.47\textwidth] {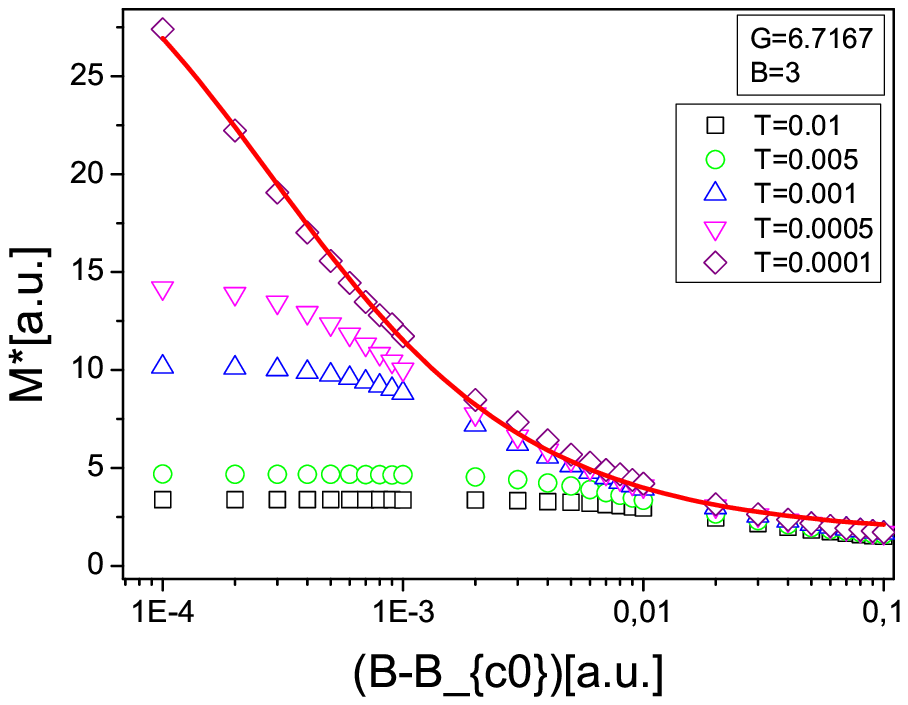}
\end {center}
\caption {Графики функций $M^{*}(B,T)$ для модельной системы с критическими параметрами $\beta$ и $g$. При $T \to 0$ система приближается к ферми-конденсатному состоянию. Сплошная красная линия --- результат апроксимации варианта расчета для Т=0.0001 по формуле (\ref{HC5}).} \label {Fig1302}
\end{figure}

Вблизи критической точки $x_{FC}$, когда $M/M^*(x\to x_{FC})\to0$,
поведение эффективной массы кардинально изменяется, потому что
первое слагаемое в правой части уравнения (\ref{HC3}) исчезает,
второе слагаемое становится главным, и эффективная масса
определяется однородным уравнением (\ref{HC3}), которое задает
эффективную массу как функцию $B$ и $T$. Обратимся к качественному
анализу решений уравнения (\ref{HC3}) при $x\simeq x_{FC}$ и $T=0$.
Наложение магнитного поля приводит к расщеплению Зеемана
поверхности Ферми, и расстояние $\delta p$ между поверхностями
Ферми со "спином вверх" и "спином вниз" становится равным $\delta
p=p_F^{\uparrow}-p_F^{\downarrow}\sim\mu_{B}BM ^*(B)/p_F$. Заметим,
что второе слагаемое в уравнении (\ref{HC3}) пропорционально
$(\delta p)^2\propto(\mu_{B}BM^*(B)/p_F)^2$ и уравнение (\ref{HC3})
принимает вид \cite{shag4,shag,ckhz}
\begin{equation}
\frac{M}{M^*(B)}=\frac{M}{M^*(x)}+c\frac{(\mu_{B}BM^*(B))^2}
{p_F^4},\label{HC4}
\end{equation}
где $c$ --- константа. Отметим, что эффективная масса $M^*(B)$
зависит также от $x$, и эта зависимость исчезает при $x=x_{FC}$. В
точке $x=x_{FC}$ слагаемое $M/M^*(x)$ обнуляется, уравнение
(\ref{HC4}) становится однородным и может быть решено аналитически
\cite{shag1,shag,ckhz}
\begin{equation}
M^*(B)\propto\frac{1}{(B-B_{c0})^{2/3}}.\label{HC5}
\end{equation}
Здесь $B_{c0}$ --- критическое магнитное поле, которое следует
рассматривать как параметр.
Уравнение (\ref{HC5}) задает универсальное степенное поведение
эффективной массы, которое не зависит от межчастичного
взаимодействия. 
В качестве примера такого поведения эффективной массы в зависимости от напряженности поля на Рис. \ref{Fig1302} приведено семейство графиков функций $M^{*}(B,T)$ для модельной системы с критическими параметрами $\beta$ и $g$. При $T \to 0$ система приближается к ферми-конденсатному состоянию. Сплошная красная линия --- результат аппроксимации паттерна для Т=0.0001 по формуле (\ref{HC5})
При плотностях $x>x_{FC}$ эффективная масса
$M^*(x)$ конечна, и мы имеем дело с обычными квазичастицами Ландау
при условии, что магнитное поле мало настолько, что
$M^*(x)/M^*(B)\ll1$, где $M^*(B)$ задается уравнением (\ref{HC5}).
Второе слагаемое в правой части уравнения (\ref{HC4})
пропорционально $(BM^*(x))^2$ и представляет малую поправку. В
противоположном случае, когда $M^*(x)/M^*(B)\gg1$, электронная
жидкость ведет себя так, как будто она находится в квантовой
критической точке. Поскольку в режиме ферми-жидкости Ландау
основные термодинамические и транспортные свойства системы
определяются эффективной массой, то из уравнения (\ref{HC5})
следует, что мы получаем уникальную возможность управлять ее
магнетосопротивлением, сопротивлением, теплоемкостью,
намагничиванием, тепловым объемным расширением и т.д. Здесь
необходимо отметить, что большая эффективная масса приводит к
высокой плотности состояний, вызывающих появление и конкуренцию
друг с другом большого числа состояний и фазовых переходов. Мы
предполагаем, что они могут быть подавлены внешним магнитным полем,
и рассматриваем термодинамические свойства системы без учета этой
конкуренции.

\section* { 4.4. Зависимость эффективной масси  $M^*$ от температуры. }\label{ZEMOT}
\addcontentsline{toc}{section}{ 4.4. Зависимость эффективной масси  $M^*$ от температуры. }

Для изучения качественного поведения $M^*(T)$ при увеличении
температуры, упростим уравнение (\ref{HC3}), опуская переменную $B$
и имитируя влияние внешнего магнитного поля конечной эффективной
массой в знаменателе первого слагаемого в правой части уравнения
(\ref{HC3}). Эта эффективная масса становится функцией расстояния
$r$, $M^*(r)$, которое определяется и величиной магнитного поля
$B$. Если магнитное поле исчезает, то $r=(x-x_{FC})/x_{FC}$.
Проинтегрируем второе слагаемое по угловым переменным, затем
возьмем интеграл по переменной $p$ по частям и заменим переменную
$p$ на $z$, $z=(\varepsilon(p)-\mu)/T$. В случае плоской и узкой
зоны, можем воспользоваться приближением $(\varepsilon(p)-\mu)
\simeq p_F (p-p_F)/M^*(T)$. В результате всех преобразований
получаем
\begin{eqnarray}
\nonumber \frac{M}{M^*(T)}&=&\frac{M}{M^*(r)}+
\alpha\int^{\infty}_{0}\frac{F(p_F,p_F(1 +\alpha z))dz}{1+e^z} \\
&-\alpha&\int^{1/\alpha}_{0}F(p_F, p_F(1-\alpha
z))\frac{dz}{1+e^z}.\label{HC8}
\end{eqnarray}
Здесь были использованы следующие обозначения: $F\sim M
d(F^1p^2)/dp$, переменная $\alpha=TM^*(T)/p_F^2=TM^*(T)/
(T_kM^*(r))$, $T_k=p_F^2/M^*(r)$ и импульс Ферми определен из
условия $\varepsilon(p_F)=\mu$.

Сначала предположим, что $\alpha\ll 1 $. Тогда, опуская слагаемые
порядка $\exp (-1/\alpha) $, можем положить верхний предел второго
интеграла в правой части уравнения (\ref{HC8}) равным бесконечности
и видим, что сумма второго и третьего слагаемых есть четная функция
$\alpha$. Полученные интегралы --- типичные выражения с функцией
Ферми-Дирака в качестве подынтегрального выражения --- могут быть
вычислены с использованием стандартной процедуры (см., например,
\cite{lanl2}). Так как нам необходима только оценка интегралов,
представим уравнение (\ref{HC8}) как
\begin{eqnarray}
\frac{M}{{M^* (T)}}&=&\frac{M}{{M^*(r)}}+\alpha f(0)\ln\left\{
{1+\exp(-1/\alpha)}\right\}\nonumber \\
&+&\lambda _1\alpha^2+\lambda_2 \alpha^4 + ...,\label{HC9}
\end{eqnarray}
где $\lambda_1$ и $\lambda_2$ --- константы порядка единици. Здесь логарифмическое слагаемое есть результат суммирования основного неаналитического (при $T\to 0$) вклада, пропорционального $\exp (-1/\alpha)$. 
Уравнение
(\ref{HC9}) может рассматриваться как типичное уравнение теории
ферми-жидкости Ландау. Единственным исключением при этом является
эффективная масса $M^*(r)$, сильно зависящая от расстояния $r$ и
расходящаяся при $r\to0$. Тем не менее, из уравнения (\ref{HC9})
следует, что при $T\to0$ поправки к $M^*(r)$ начинаются с членов
порядка $T^2$ при условии выполнения неравенства,
\begin{equation} \frac{M}{M^*(r)}\gg\left(\frac{TM^*(T)}{T_kM^*(r)}
\right) ^2\simeq\frac{T^2}{T_k^2},\label{HC10} \end{equation} и
система демонстрирует  поведение ферми-жидкости Ландау. Из
уравнения (\ref{HC9}) видно, что поведение ферми-жидкости Ландау
исчезает, когда $r\to0$ и $M^*(r)\to\infty$. Тогда свободный член в
правой части уравнения (\ref{HC9}) оказывается пренебрежимо малым
$M/M^*(r)\to0$, и уравнение (\ref{HC8}), становясь однородным,
определяет универсальное поведение эффективной массы $M^*(T)$.

При некоторой температуре $T_1\ll T_k$ величина суммы в правой
части уравнения (\ref{HC9}) определяется вторым слагаемым, и  уравнение
(\ref{HC10}) более не справедливо. Удерживая только второе
слагаемое в уравнении (\ref{HC9}), получаем уравнение для
определения $M^*(T)$ в переходной области \cite{ckhz,shag5}
\begin{equation}
M^*(T)\propto\frac{1}{T^{2/3}}.\label{HC11}
\end{equation}
Температурная зависимость вида $T^{-2/3}$ в выражении (\ref{HC11}))
заслуживает комментария. Уравнение (\ref{HC11}) справедливо, если
второе слагаемое в уравнении (\ref{HC9}) намного больше первого:
\begin{equation}
\frac{T^2}{T_k^2}\gg \frac{M}{M^*(r)},\label{HC12}
\end{equation}
и это слагаемое больше, чем третье,
\begin{equation} \frac{T}{T_k}\ll
\frac{M^*(r)}{M^*(T)}\simeq1.\label{HC13}
\end{equation}
Очевидно, что оба уравнения (\ref{HC12}) и (\ref{HC13})
одновременно удовлетворяются, если $M/M^*(r)\ll1$ и $T$ конечна.
Диапазон температур, в котором уравнение (\ref{HC11}) справедливо,
сжимается в ноль, как только $r\to0$, потому что $T_k\to0$. Таким
образом, если система очень близка к квантовой критической точке,
$x\to x_{FC}$, то  поведение эффективной массы, даваемое уравнением
(\ref{HC11}), в широком диапазоне температур возможно, только если
величина эффективной массы $M^*(r)$ значительно уменьшена
наложением магнитных полей. Можно сказать, что расстояние $r$
становится больше под действием $B$. Когда магнитное поле $B$
конечно, $T^{-2/3}$ зависимость может наблюдаться при относительно
высокой температуре $T>T_1(B)$. Для оценки температуры перехода
$T_1(B)$, при которой эффективная масса начинает зависеть от
температуры, заметим, что эффективная масса является непрерывной
функцией температуры и величины магнитного поля: $M^*(B)\sim
M^*(T_1)$. Принимая во внимание уравнения (\ref{HC5}) и
(\ref{HC11}), получаем $T_1(B)\propto B$.

С ростом температуры система переходит в другой режим. Коэффициент
$\alpha$ становится порядка единицы, $\alpha\sim 1 $, верхний
предел второго интеграла в уравнении (\ref{HC8}) не может быть
распространен до бесконечности, и начинают играть роль нечетные
слагаемые. В результате уравнение (\ref{HC9}) более несправедливо,
и сумма первого и второго интегралов в правой части уравнения
(\ref{HC8}) пропорциональна $M^*(T)T$. Пренебрегая первым слагаемым
$M/M^*(r)$ и приближая сумму интегралов выражением $M^*(T)T$,
получаем из (\ref{HC8})
\begin{equation}
M^*(T)\propto\frac{1}{\sqrt{T}}.\label{HC14}
\end{equation}

\begin{figure} [ht]
\begin{center}
\includegraphics [width=0.47\textwidth] {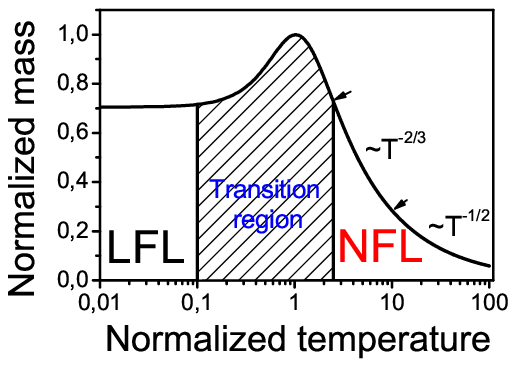}
\end {center}
\caption {Зависимость нормированной на максимальное значение эффективной массы $M^{*}$ от $T$ нормированной на значение температуры, при которой  $M^{*}$ достигает максимума.} \label {Fig1303}
\end{figure}

Таким образом, можно заключить, что при повышении температуры и
когда $x\simeq x_{FC}$, система демонстрирует три типа режимов:
поведение ферми-жидкости Ландау при $\alpha\ll 1$, когда уравнение
(\ref{HC10}) справедливо, и поведение эффективной массы задано
(\ref{HC5}); поведение, определяемое уравнением (\ref{HC11}), когда
$M^*(T)\propto T^{-2/3}$ и $S (T)\propto M^*(T)T\propto T^{1/3}$; и
при $\alpha\sim 1$ справедливо уравнение (\ref{HC14}),
$M^*(T)\propto1/\sqrt{T}$, в этом режиме энтропия $S(T)\propto
M^*(T)T\propto\sqrt{T}$ и теплоемкость $C (T)=T(\partial
S(T)/\partial T)\propto\sqrt{T}$.
На Рис. \ref{Fig1303} изображена типичная $M^{*}-T$ диаграмма демонстрирующая все четыре диапазона изменения $M^{*}(T)$. 

\begin{figure} [ht]
\begin{center}
\includegraphics [width=0.47\textwidth] {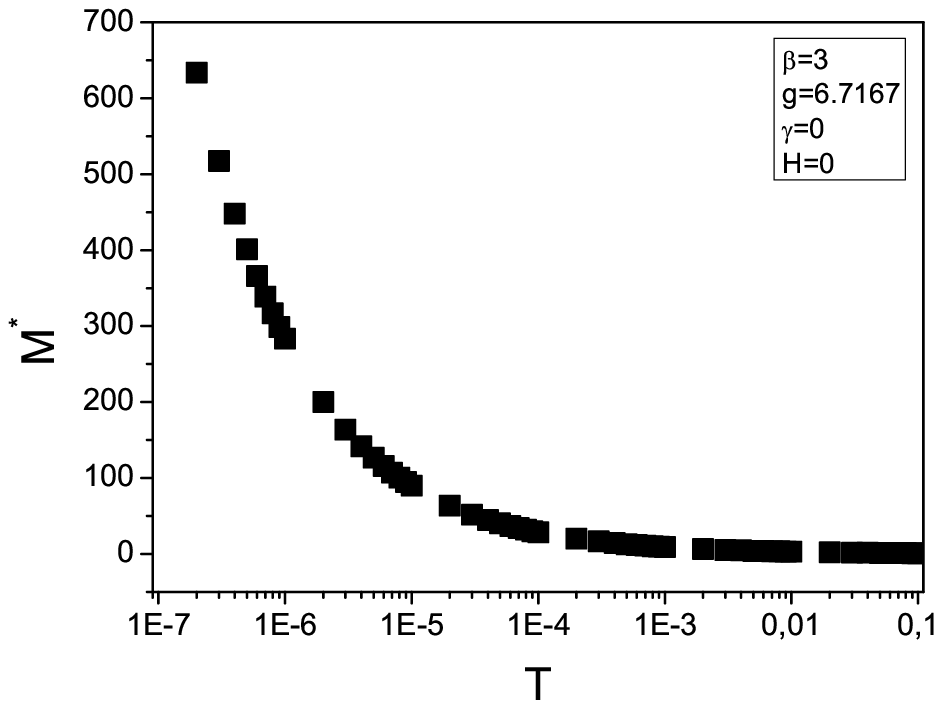}
\end{center}
\caption {Зависимость эффективной массы $M^{*}$ от температуры $T$ для критических значений параметров потенциала взаимодействия. } \label{Fig2228}
\end{figure}

В случае, когда система находится в критической точке ФККФП, то есть $x=x_{FC}$, эффективная масса расходится при уменьшении температуры по закону:
\begin{equation}
M^*(T)\propto\frac{1}{\sqrt{T}}.\label{HC141}
\end{equation}
На Рис. \ref{Fig2228} приведен график функции $M^{*}(T)$, полученный в компьютерном эхперименте для ферми-системы, находящейся в критической точке ФККФП. 
По мере приближения температуры к нолю масса фермионов неограниченно возрастает.
\pagebreak

\begin{figure} [ht]
\begin{center}
\includegraphics [width=0.47\textwidth] {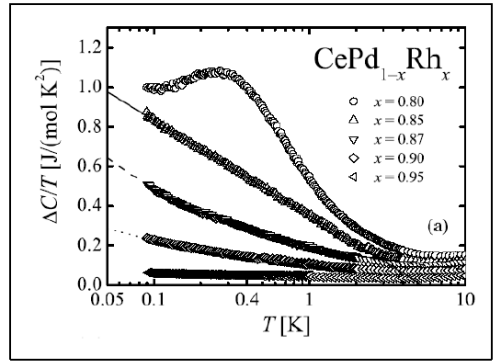}
\end {center}
\caption {Зависимость $\Delta C/T$  от $T$ для $ CePd_{1-x}Rh_{x}$ при различных $x$.} \label {Fig1304}
\end{figure}

\begin{figure} [ht]
\begin{center}
\includegraphics [width=0.47\textwidth] {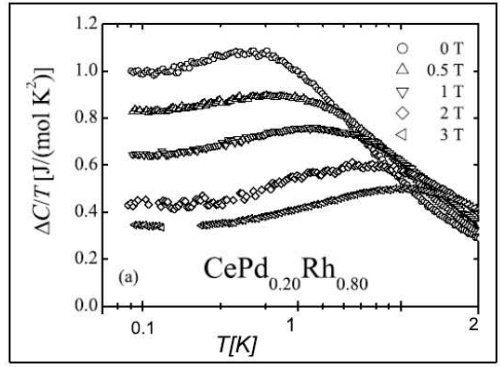}
\end {center}
\caption {Зависимость $\Delta C/T$  от $T$ для $ CePd_{1-x}Rh_{x}$ при  $x=0.80$ и различных значений магнитного поля $B$ от 0 до $3T$.} \label {Fig1305}
\end{figure}

\begin{figure} [ht]
\begin{center}
\includegraphics [width=0.47\textwidth] {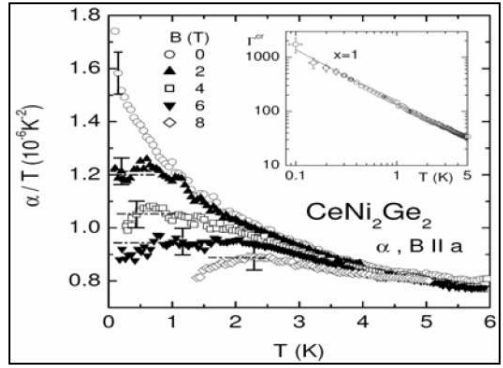}
\end {center}
\caption {Зависимость $\alpha/T$  от $T$ для $ CeNi_{2}Ge_{2}$ при различных значениях индукции магнитного поля.} \label {Fig1306}
\end{figure}

\begin{figure} [ht]
\begin{center}
\includegraphics [width=0.47\textwidth] {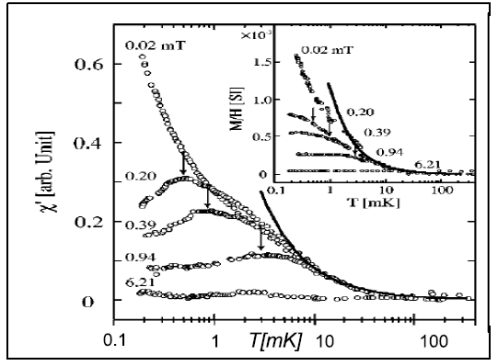}
\end {center}
\caption {Зависимость $\chi(T)$  от $T$ для $ CeRu_{2}Si_{2}$ при различных значениях индукции магнитного поля.} \label {Fig1307}
\end{figure}

\begin{figure} [ht]
\begin{center}
\includegraphics [width=0.47\textwidth] {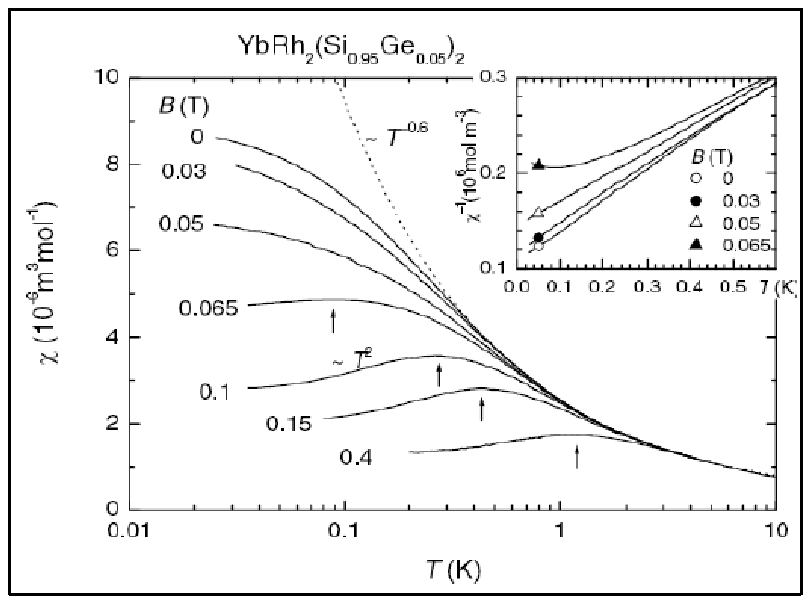}
\end {center}
\caption {Зависимость $\chi(T)$  от $T$ для $ IbRh_{2}(Si_{0.95}Ge_{0.05})_{2}$ при различных значениях индукции магнитного поля.} \label {Fig1308}
\end{figure}

Для иллюстрации представленного выше рассмотрения зависимости $M^{*}$ от температуры $T$ воспользуемся термодинамическим соотношением, связывающим приведенную массу с магнитной восприимчивостью $\chi(T)$, энтропией $S(T)$, удельной электронной теплоемкостью $C(T)$, коэффициентом объемного расширения $\alpha(T)$, удельным сопротивлением $\rho(T)$:
\begin{equation}
M^*(T)\propto \chi(T)\propto \frac{S(T)}{T}\propto \frac{C(T)}{T}\propto \frac{\alpha(T)}{T}\propto \sqrt{\frac{\rho(T)}{T}}.\label{HC15}
\end{equation}
Экспериментально исследованные величины из ряда \ref{HC15}, относящиеся к различных химическим соединениям, демонстрирует все этапы эволюции зависимости массы от температуры для высококоррелированной ферми-жидкости.
На рисунках (\ref{Fig1304} --- \ref{Fig1305}) \cite{pikul} приведены зависимости $\Delta C/T$ для $ CePd_{1-x}Rh_{x}$ при различных $x$ в окрестности точки фазового перехода ферромагнетик-парамагнетик $x_{c}\simeq 0.87$. При $x= 0.80$ электронная система рассматриваемого соединения демонстрирует все этапы эволюции зависимости массы от температуры для высококоррелированной ферми-жидкости. 
На Рис. \ref{Fig1306} \cite{kuch1} приведены зависимости $\alpha/T$ для $ CeNi_{2}Ge_{2}$ при различных значениях индукции, наложенного на систему магнитного поля.  
На Рис. \ref{Fig1307} \cite{takah} приведены зависимости $\chi(T)$ для $ CeRu_{2}Si_{2}$ при различных значениях индукции, наложенного на систему магнитного поля.  
На Рис. \ref{Fig1308} \cite{gegenvart} приведены зависимости $\chi(T)$ для $ IbRh_{2}(Si_{0.95}Ge_{0.05})_{2}$ при различных значениях индукции, наложенного на систему магнитного поля.  

\begin{figure} [ht]
\begin{center}
\includegraphics [width=0.47\textwidth] {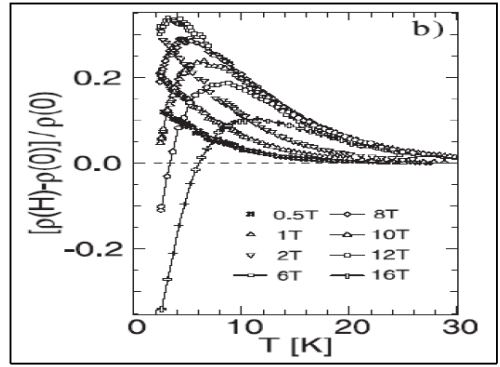}
\end {center}
\caption {зависимости магнетосопротивления $[\rho(H)-\rho(0)]/\rho(0)(T)$ для $ CeCoIn_{5}$ от температуры при различных значениях индукции, наложенного на систему магнитного поля. } \label {Fig1309}
\end{figure}

На Рис. \ref{Fig1309} \cite{paglione} приведены зависимости магнетосопротивления $[\rho(H)-\rho(0)]/\rho(0)(T)$ для $ CeCoIn_{5}$ от температуры при различных значениях индукции, наложенного на систему магнитного поля.
  
\begin{figure} [ht]
\begin{center}
\includegraphics [width=0.47\textwidth] {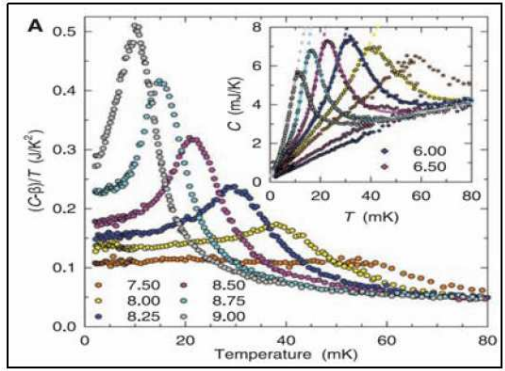}
\end {center}
\caption {Зависимость $\chi(T)$  от $T$ для $ IbRh_{2}(Si_{0.95}Ge_{0.05})_{2}$ при различных значениях индукции магнитного поля.} \label {Fig1310}
\end{figure}

На Рис. \ref{Fig1310} \cite{neumann} приведены зависимости $C/T$ для квазидвухмерного изотопа$ 3^ He$ при различных значениях индукции,  
наложенного на систему магнитного поля. 
\pagebreak[4]



\section* { 4.5. Линейное расширение и закон Грюнайзена. }\label{LRZG}
\addcontentsline{toc}{section}{ 4.5. Линейное расширение и закон Грюнайзена. }

Необычная зависимость энтропии от температуры сильно
коррелированной электронной жидкости, заданная соотношением
(\ref{SL1}), определяет ее особое поведение. Существование не
зависящего от температуры члена $S_0$ может быть проиллюстрировано
вычислением коэффициента теплового расширения $\alpha(T)$
\cite{alp,zver}, который дается выражением \cite{lanl1}
\begin{equation}
\alpha(T)=\frac13\left(\frac{\partial(\log V)}{\partial T}\right)
_P=-\frac{1}{3V}\left(\frac{\partial(S/x)}{\partial
P}\right)_T,\label{SL2}
\end{equation}
здесь $P$ --- давление, $V$ --- объем. Отметим, что сжимаемость
$K=d\mu/d(Vx)$ не имеет сингулярности при ФККФП и в системах с
фермионным конденсатом приблизительно постоянна \cite{noz}.
Подставляя уравнение (\ref{SL1}) в (\ref{SL2}), находим, что
\begin{equation} \alpha_{FC}(T)\simeq
a_0\sim\frac{M_{FC}^{*}T}{p_F^2K},\label{SL3}
\end{equation} где $a_0\sim \partial S_0/\partial P$ --- величина,
не зависящая от температуры. При вычислении выражения (\ref{SL3})
был удержан только главный вклад, связанный с $S_0$. Учтем, что
теплоемкость имеет вид
\begin{equation} C(T)=T\frac{\partial S(T)}{\partial
T}\simeq\frac{a}{2} \sqrt{\frac{T}{T_f}}.\label{SL4}
\end{equation} В результате отношение Грюнайзена $\Gamma(T)$
расходится как \begin{equation} \Gamma(T)=\frac{\alpha(T)}{C(T)}
\simeq2\frac{a_0}{a}\sqrt{\frac{T_f}{T}},\label{SL5}
\end{equation} и мы заключаем, что закон Грюнайзена не выполняется
в случае сильно коррелированных ферми-систем.

\begin{figure} [ht]
\begin{center}
\includegraphics [width=0.47\textwidth] {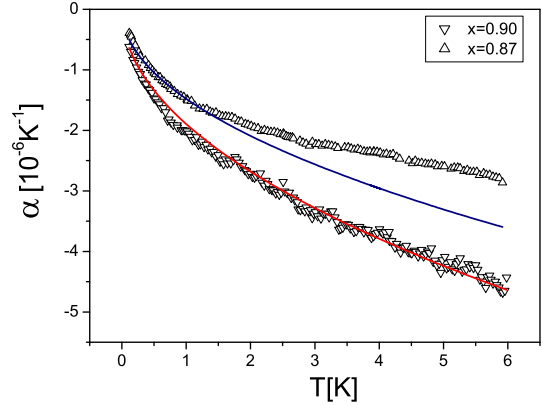}
\end {center}
\caption {Коэффициент теплового расширения $\alpha(T)$ как функция
температуры в интервале $0.1\leq T\leq 6$ K. Экспериментальные
значения при уровнях легирования $x=0.90, 0.87$ взяты из
\cite{sereni}. Сплошные линии представляют аппроксимации
экспериментальных значений $\alpha(T)=c_1\sqrt{T}$, $c_1$ ---
подгоночный параметр.} \label {Fig11}
\end{figure}

Теперь рассмотрим, как поведение эффективной массы, описываемое
уравнением (\ref{HC14}), соотносится с
экспериментальными наблюдениями. Коэффициент теплового расширения
$\alpha(T)$, измеренный для парамагнетика CeNi$_2$Ge$_2$,
демонстрирует поведение вида $\sqrt{T}$ при изменении температуры
от 6 K до 50 мК \cite{geg1}. Такое же поведение $\alpha(T)\propto
\sqrt{T}$ было обнаружено в измерениях на ферромагнетике $\rm
CePd_{1-x}Rh_x$ \cite{sereni}, см. Рис. \ref{Fig11}. Из Рис.
\ref{Fig11} видно, что в критической точке $x=0.90$, в которой
критическая температура ферромагнитного фазового перехода
обращается в ноль, коэффициент теплового расширения хорошо
аппроксимируется корневой зависимостью $\alpha(T)\propto \sqrt{T}$
при изменении температуры почти на два порядка, однако даже
небольшое смещение системы из критической точки резко ухудшает
качество используемой аппроксимации. Отметим, что критическое
поведение двух совершенно различных металлов с ТФ (один ---
парамагнетик, а второй
--- ферромагнетик) удается хорошо описать функцией $\alpha(T)=c_1\sqrt{T}$
с одним подгоночным параметром $c_1$. Это обстоятельство наглядно
показывает, что флуктуации не определяют поведение $\alpha(T)$.
Измерения теплоемкости, проведенные на $\rm CePd_{1-x}Rh_x$ при
$x=0.90$, показали, что $C(T)\propto\sqrt{T}$ \cite{sereni}. Таким
образом, электронные системы обоих металлов можно отнести к высоко
коррелированной электронной жидкости.

Измерения на  ${\rm YbRh_2(Si_{0.95}Ge_{0.05})_2}$ показывают, что
$\alpha/T\propto1/T$, что позволяет отнести электронную систему к
сильно коррелированной жидкости. Отношение Грюнайзена
$\Gamma(T)\simeq T^{-q}$,  $q\simeq0.33$ \cite{geg1}. Наша оценка,
следующая из уравнения (\ref{SL5}), дает $q=0.5$ и находится в
удовлетворительном согласии с этим экспериментальным значением. Оба
типа поведения $\alpha(T)$ не согласуются с выводами теории
ферми-жидкости Ландау, которая при $T\to0$ дает $\alpha(T)/T= M^*=
const$. В нашем случае эффективная масса зависит от $T$, и
зависимость вида $1/\sqrt{T}$ следует из уравнения (\ref{HC14}) и
находится в хорошем согласии с результатом для первой системы
\cite{alp}, а зависимость эффективной массы вида $1/T$, следующая
из уравнения
\begin{equation} M_{FC}^{*} \simeq p_{F}\frac{p_{f}-p_{i}}{4T},\label{SL6}
\end{equation}
и предсказанная в \cite{zver},
соответствует второму результату.

Из (\ref{SL6}) следует, что  $M^*(T\to0)\to\infty$, и сильно
коррелированная электронная жидкость ведет себя так, как будто она
помещена в квантовую критическую точку. Фактически она помещена на
квантовую критическую линию $x<x_{FC}$, и критическое поведение
наблюдается при $T\to0$ для всех $x\leq x_{FC}$. При $T\to0$ сильно коррелированная электронная
жидкость испытывает квантовый фазовый переход первого рода,
поскольку энтропия становится разрывной функцией температуры: при
конечных температурах энтропия дается уравнением (\ref{SL1}), в то
время как $S(T=0)=0$. Поэтому энтропия испытывает скачок $\delta
S=S_0$ при $T\to0$. Отсюда следует, что из-за фазового перехода
первого рода критические флуктуации подавлены вдоль квантовой
критической линии и соответствующие расходимости, например,
расходимость ${\rm\Gamma}(T)$, определяются квазичастицами, а не
критическими флуктуациями, как можно было ожидать в случае обычного
квантового фазового перехода \cite{voj}. Отметим, что согласно
известному неравенству \cite{lanl2}, $\delta Q\leq T\delta S$,
теплота перехода $\delta Q$ от неупорядоченной к упорядоченной фазе
равна нулю, потому что $\delta Q\leq S_0T$ стремится к нулю при
$T\to 0$.

\section* { 4.6. $T-B$ фазовая диаграмма. }\label{TBFD}
\addcontentsline{toc}{section}{ 4.6. $T-B$ фазовая диаграмма. }

\begin{figure} [ht]
\begin{center}
\includegraphics [width=0.47\textwidth] {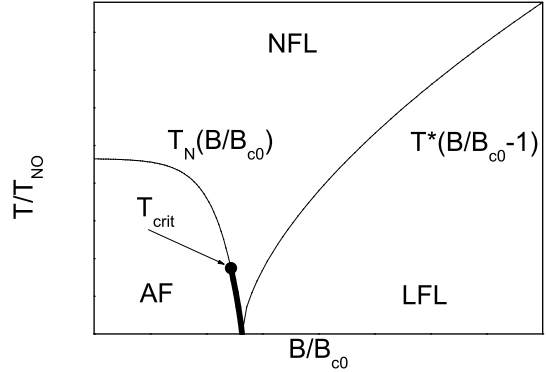}
\end{center}
\caption {Фазовая диаграмма $T-B$ сильно коррелированной
электронной жидкости. Кривая $T_N (B/B_{c0})$ представляет
зависимость от поля $B$ температуры Нееля. Черная точка при
$T=T_{crit}$, отмеченная стрелкой, --- критическая температура, при
которой антиферромагнитный фазовый переход второго рода становится
переходом первого. При $T<T_{crit}$ толстая сплошная линия
представляет зависимость от $B$ температуры Нееля, когда AF фазовый
переход становится переходом первого рода.  Поведение сильно
коррелированной жидкости в области, отмеченной  буквами NFL,
характеризуется энтропией $S_{NFL}$, заданной уравнением
(\ref{SL1}). Линия, разделяющая сильно коррелированную жидкость
(область NFL) и слабо поляризованную электронную жидкость, имеющую
поведение ферми-жидкости Ландау ( область LFL), описывается
функцией $T^*(B/B_{c0}-1)\propto\sqrt {B/B_{c0}-1}$.} \label {Fig16}
\end{figure}

Для изучения фазовой диаграммы $T-B$ сильно коррелированной
электронной жидкости рассмотрим случай, когда не-ферми жидкостное
поведение возникает при подавлении антиферромагнитной фазы внешним
магнитным полем $B$, например, как это происходит в металлах с ТФ
$\rm YbRh_2Si_2$ и YbRh$_2$(Si$_{0.95}$ Ge$_{0.05}$)$_2$
\cite{geg,geg1}. Антиферромагнитная фаза представляет собой
электронную ферми-жидкость Ландау с энтропией, обращающейся в ноль
при $T\to0$. Для магнитных полей, превышающих критическое значение
$B_{c0}$, при котором температура Нееля $T_N(B\to B_{c0})\to0$,
антиферромагнитная фаза преобразуется в слабо поляризованную
парамагнитную сильно коррелированную электронную жидкость. Как было
показано в разделе \ref{HFL}, при $T=0$ наложение магнитного поля
$B$ расщепляет ферми-конденсатное состояние, занимающее область
$(p_f-p_i)$, на уровни Ландау и подавляет сверхпроводящий параметр
порядка $\kappa({\bf p})$. Новое состояние дается многосвязной
сферой Ферми, где гладкая функция распределения квазичастиц
$n_0({\bf p})$ в интервале импульсов $(p_f-p_i)$ заменена
распределением $\nu({\bf p})$, см. Рис. \ref{Fig4}. Поэтому
поведение ферми-жидкости Ландау восстанавливается и характеризуется
квазичастицами с эффективной массой $M^*(B)$, определяемой
уравнением \cite{shag4}
\begin{equation} M^*(B)\propto
\frac{1}{\sqrt{B-B_{c0}}},\label{HF5}
\end{equation}
здесь $B_{c0}$ --- критическое магнитное поле, которое помещает
металл с ТФ в настраиваемую магнитным полем квантовую критическую
точку. При повышении температуры настолько, что
$T>T^*(B-B_{c0})\propto\sqrt{B-B_{c0}}$, энтропия электронной
жидкости дается уравнением (\ref{SL1}). Описанное выше поведение
сильно коррелированной жидкости показано на фазовой диаграмме
$T-B$, приведенной на Рис. \ref{Fig16}.

В соответствии с экспериментом предположим, что при относительно
высоких температурах $T/T_{NO}\sim 1$, антиферромагнитный фазовый
переход является переходом второго рода \cite{geg}, где $T_{NO}$
температура Нееля в отсутствии магнитного поля. В этом случае
энтропия и другие термодинамические функции являются непрерывными
функциями при температуре перехода $T_N(B)$. Это означает, что
энтропия антиферромагнитной фазы $S_{AF}$ совпадает с энтропией
$S_{NFL}$ сильно коррелированной жидкости, определяемой уравнением
(\ref{SL1}),
\begin{equation}
S_{AF}(T\to T_N(B))=S_{NFL}(T\to T_N(B))\label{SL61}.
\end{equation}
Поскольку антиферромагнитная фаза имеет поведение ферми-жидкости
Ландау, т.е. $S_{АФ}(T\to0) \to0$, уравнение (\ref{SL6}) не может
быть удовлетворено при достаточно низких температурах $T\leq
T_{crit}$ из-за независимого от температуры слагаемого $S_0$.
Следовательно, антиферромагнитный фазовый переход второго рода
становится переходом первого рода при $T=T_{crit}$, как это
показано на Рис. \ref{Fig16}.

При $T=0$, критическое поле $B_{c0}$, при котором
антиферромагнитная фаза становится ферми-жидкостью Ландау,
определяется условием, что энергия основного состояния
антиферромагнитной фазы совпадает с энергией основного состояния $E
[n_0({\bf p})]$ электронной ферми-жидкости Ландау. Это значит, что
основное состояние антиферромагнитной фазы становится вырожденным
при $B=B_{c0}$. Поэтому температура Нееля $T_N(B\to B_{c0})\to0$,
поведение эффективной массы $M^*(B\geq B_{c0})$ определено
уравнением (\ref{HF5}), и $M^*(B)$ расходится при $B\to B_{c0}$.
Фазовый переход, разделяющий антиферромагнитную фазу, существующую
при $B\leq B_{c0}$, и ферми-жидкость Ландау при $B\geq B_{c0}$,
есть квантовый фазовый переход первого рода. Контролирующим
параметром этого фазового перехода является магнитное поле $B$.
Отметим, что соответствующие квантовые и тепловые критические
флуктуации исчезают при $T<T_{crit}$, поскольку при этих
температурах имеем дело с антиферромагнитным фазовым  переходом
первого рода. Можно также заключить, что критическое поведение,
наблюдаемое при $T\to0$ и $B\to B_{c0}$, определяется
квазичастицами, а не критическими флуктуациями, сопровождающими
фазовые переходы второго рода. Когда $r\to0$, электронная жидкость
приближается к ФККФП со стороны упорядоченной фазы. Очевидно,
$T_{crit}\to0$ в точке $r=0$, и температура Нееля равна нулю в
точке, где антиферромагнитный фазовый переход второго рода
становится переходом первого рода. Поэтому можно ожидать, что
вклады, определяемые критическими флуктуациями, приводят только к
логарифмическим поправкам к величинам, вычисляемым в теории фазовых
переходов Ландау \cite{lanl2}, а степенные законы критического
поведения снова определяются соответствующими квазичастицами. Таким
образом, парадигма Ландау, основанная на понятии квазичастиц и
параметра порядка, применима при рассмотрении $T-B$ фазовой
диаграммы сильно коррелированной электронной жидкости.

\section* { 4.7. Универсальное поведение ферми-системы в окрестности ФККФП. }\label{UPFS1}
\addcontentsline{toc}{section}{ 4.7. Универсальное поведение ферми-системы в окрестности ФККФП. }

Рассмотрим состояние системы при $r\to0$. Его свойства определяются
магнитным полем $B$ и температурой $T$. При переходной температуре
$T\simeq T_1(B)$ эффективная масса зависит как от $T$, так и от
$B$, в то время как при $T\ll T_1 (B)$ система является
ферми-жидкостью Ландау с эффективной массой, определяемой
уравнением (\ref{HC5}), а при $T\gtrsim T_1(B)$, масса задана
уравнением (\ref{HC11}). Для описания поведения эффективной массы при $T\lesssim T_1$ уравнение
(\ref{HC3}) допускает приближенное решение, которое можно использовать как простую интерполяционную формулу:
\begin{equation}
M^*(B,T)=\frac{c_1(B-B_{c0})^{2}+c_2T^{2}}
{c_3(B-B_{c0})^{8/3}+c_4T^{8/3}},\label{HC21}\end{equation}
Коэффициенты $c_1$, $c_2$, $c_3$ и $c_{4}$ --- некоторые константы.
Разделив обе стороны уравнения (\ref{HC21}) на $M^*(B,T=0)$, и
обозначив $y=(T/(B-B_{c0}))$, получаем
\begin{equation}
\frac{M^*(B,T)}{M^*(B)}=\frac{1+(c_2/c_1)y^2}{1+(c_{4}/c_3)y^{8/3}}.
\label{HC22}\end{equation} Из уравнения (\ref{HC22}) следует, что
поведение эффективной массы может быть представлено универсальной
функцией только одной переменной $y$. 
Для удобства дальнейшего использования перепишем (\ref{HC22}) как зависимость $M^*(B,T)$ от температуры:
\begin{equation}
\frac{M^*(B,T_N)}{M^*_M}={M^*_N(T_N)}\approx
c_0\frac{1+c_1T_N^2}{1+c_2T_N^{8/3}}. \label{UMN}
\end{equation}
Здесь $M^*_N(T_N)$ нормированная эффективная масса, $M^*_M$
максимальное значение $M^*(B,T,x)$, которое эффективная масса
достигает при $T=T_{\rm max}$, и нормированная температура $T_N$
определяется равенством $T_N=T/T_{\rm max}$, величина
$c_0=(1+c_2)/(1+c_1)$, $c_1$ и $c_2$ константы, которые
параметризуют дипольную амплитуду Ландау и определяются из подгонки
экспериментальных данных. Уравнение (\ref{UMN}) находится в согласии
с численными решениями уравнения Ландау (\ref{HC1}) и определяет
универсальное поведение нормированной эффективной массы $M^*_N(T_N)$
сильнокоррелированной ферми-жидкости при переходе из ФЖЛ в НФЖ
состояние \cite{ckhz,sap,prl,epl2}. 
\begin{figure} [ht]
\begin{center}
\vspace*{-0.5cm}
\includegraphics [width=0.47\textwidth]{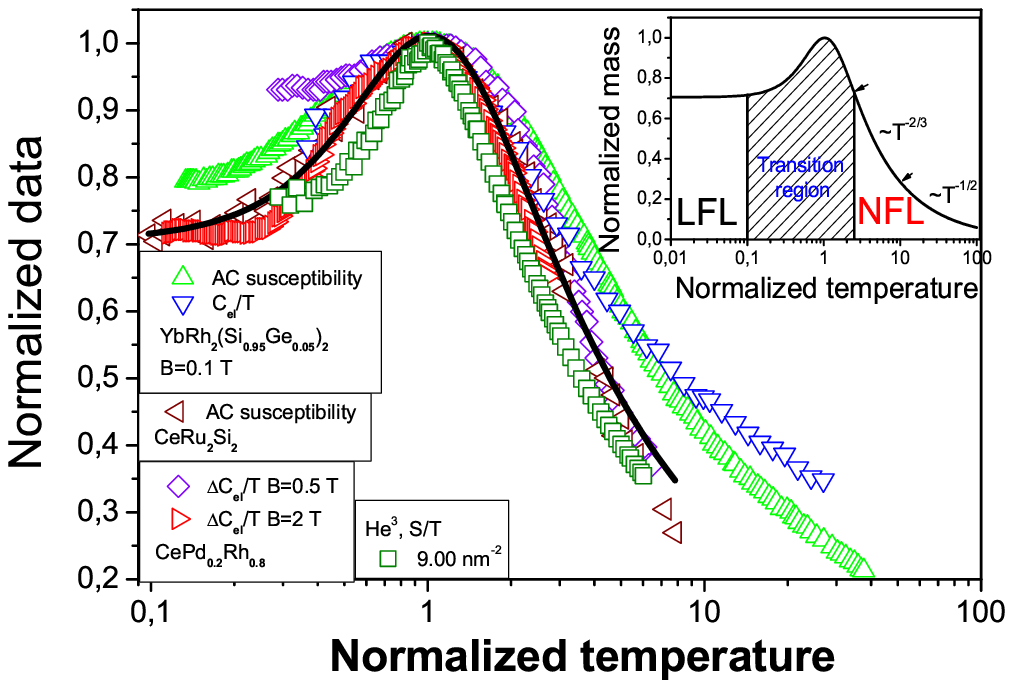}
\end{center}
\vspace*{-0.7cm} \caption{Универсальное поведение нормированной
эффективной массы как функции нормированной температуры. Данные
извлечены из измерений в магнитных полях $B$: теплоемкости $C$ и
$AC$ магнитной восприимчивости на
$\rm{YbRh_2(Si_{0.95}Ge_{0.05})_2}$ \cite{cust,geg3},
восприимчивости на $\rm{CeRu_2Si_2}$ \cite{takah}, теплоемкости на
$\rm{CePd_{1-x}Rh_x}$ при $x=0.80$ \cite{pikul}, и энтропии $S$ 2D
$\rm ^3He$ при плотности $x=9.00$ nm$^2$ \cite{neumann}. Сплошная линия
--- описание при помощи (\ref{UMN}). На вставке показано
универсальное поведение нормированной эффективной массы. Штриховкой
показана переходная область от ФЖЛ (LFL) к НФЖ (NFL)
поведению.}\label{FD}
\end{figure}
Это поведение определяется
указанными выше безразмерными величинами, поскольку система вблизи
ФККФП не имеет внешних характерных шкал для измерения $T$ и $M^*$.
Из (\ref{UMN}) видно, что $M^*_N$ имеет три режима, показанных на
вставке Рис. \ref{FD}. При $T_N\ll 1$ система ведет себя как ФЖЛ.
При $T_N\sim 1$ начинается переходный режим, показанный штриховкой:
$M^*_N(T_N)$ растет, достигает максимума $M^*_N=1$ при $T_N=1$ и
затем начинает убывать. Следы ФЖЛ исчезают при $T_N\geq 1$, когда
$M^*_N$ сначала убывает как $T_N^{-2/3}$, а затем как $T_N^{-1/2}$.

Магнитное поле входит в уравнение (\ref{HC21}) как отношение $\mu_B
B/T$, поэтому температура $T^*(B)$,
\begin{equation}T^*(B)=a_1+a_2B\simeq T_{\rm max}\sim \mu_B
(B-B_{c0})\label{TB},\end{equation} является характерной
температурой перехода из ФЖЛ в НФЖ. Здесь $a_1$ и $a_2$ константы,
$\mu_B$ магнетон Бора. Отметим, что температура перехода $T^*(B)$ не
определяет какой-нибудь фазовый переход, поэтому она связана с
довольно широкой областью температур, показанных штриховкой на
вставке Рис. \ref{FD}, и сильно зависит от метода определения.
Обычно $T^*(B)$ определяется из измерений сопротивления $\rho(T)$ и
соответствует точке, где сопротивление начинает отклоняться от $T^2$
зависимости, характерной для ФЖЛ.

Экспериментальные значения нормированной эффективной массы $M^*_N$
приведены на Рис. \ref{FD}, сплошной линией показана подгонка при
помощи \eqref{UMN} \cite{prl,epl2}. Как видно из рисунка, уравнение
\eqref{UMN} хорошо описывает поведение экспериментальной $M^*_N$,
которое заслуживает комментария. Если бы квазичастицы разрушались в
переходной области или в НФЖ состоянии, как это предполагается в
разнообразных сценариях НФЖ поведения \cite{ste,col2,loh,si},
эффективная масса $M^*_N$ не имела бы смысла, и вместо ее
универсального поведения, описываемого функцией \eqref{UMN},
зависящей от одной переменной $T_N$, мы бы имели функцию от четырех
переменных. Действительно, $M^*_N$ должна зависеть от $T$, $B$,
уровня легирования $x$ металла и от плотности системы, например, от
плотности $x$ 2D $\rm ^3He$. Однако поведение экспериментальной
эффективной массы, выявленное при помощи описанных выше масштабных
преобразований, позволяет нам заключить, что квазичастицы Ландау и
эффективная масса являются реальными физическими объектами как в
ФЖЛ, так и в НФЖ состояниях. Необычной является только сильная
зависимость эффективной массы от нормированной температуры, т.е. от
температуры, магнитного поля, уровня легирования, плотности системы
и других внешних параметров.

\section* { 4.8. Общие свойства индуцированной магнитным полем ферми-жидкости Ландау в ВТСП и металлах с ТФ. }\label{UPFS}
\addcontentsline{toc}{section}{ 4.8. Общие свойства индуцированной магнитным полем ферми-жидкости Ландау в ВТСП и металлах с ТФ. }

В настоящем разделе показано, что индуцированный магнитным полем переход из
не-ферми-жидкостного состояния в ферми-жидкостное состояние в
высокотемпературном сверхпроводнике $\rm Tl_2Ba_2CuO_{6+x}$
аналогичен переходу, наблюдаемому в металлах с тяжелыми фермионами.
\begin{figure} [ht]
\begin{center}
\vspace*{-0.5cm}
\includegraphics [width=0.47\textwidth]{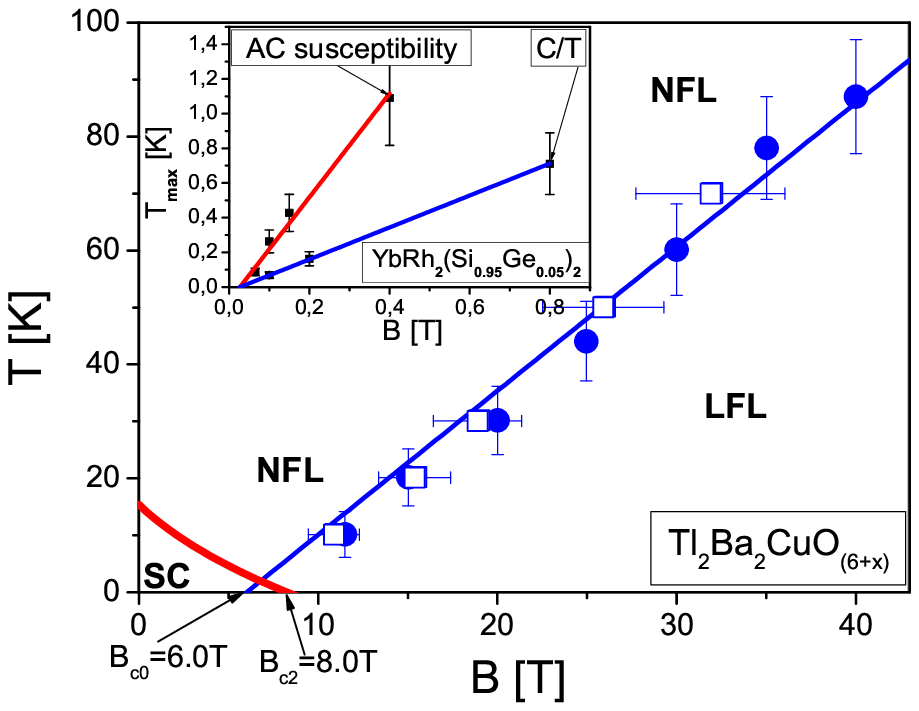}
\end{center}
\vspace*{-0.7cm} \caption{$B-T$ фазовая диаграмма ВТСП $\rm
Tl_2Ba_2CuO_{6+x}$. Измерения температуры перехода $T^*(B)$ из ФЖЛ в
НФЖ  \cite{PNAS} показаны квадратами и кругами, прямая линия задана
формулой \eqref{TB} и разделяет ФЖЛ (LFL) и НФЖ (NFL). Внизу в левом
углу рисунка стрелки показывают критическое поле $B_{c0}=6$ T и
критическое поле $B_{c2}=8$ T, разрушающее сверхпроводимость.
Область, занятая сверхпроводимостью \cite{PNAS}, отделена сплошной
линией и обозначена буквами {\bf SC}. На вставке показана $T_{\rm
max}\simeq T^*(B)$, измерения $C/T$ и $AC$ восприимчивости
проводились на $\rm{YbRh_2(Si_{0.95}Ge_{0.05})_2}$ \cite{cust,geg3}.
Прямые линии, построенные по формуле \eqref{TB}, пересекаются в
точке $B_{c0}\simeq 0.03$ T.}\label{TMM}
\end{figure}
На Рис. \ref{TMM} показана температура перехода $T^*(B)$ из ФЖЛ в
НФЖ, полученная в измерениях на ВТСП $\rm Tl_2Ba_2CuO_{6+x}$
\cite{PNAS}. На вставке показаны значения температуры $T_{\rm
max}\simeq T^*(B)$, при которой имеют место максимумы $C/T$ и $AC$
восприимчивость, в зависимости от величины магнитного поля $B$;
измерения проводились на металле с ТФ
$\rm{YbRh_2(Si_{0.95}Ge_{0.05})_2}$, близкого по своим свойствам к
$\rm YbRh_2Si_2$ \cite{cust,geg3}. Обе прямые линии, показанные на
вставке, при нулевой температуре проходят через точку $B_{c0}\simeq
0.03$ T, соответствующую квантовой критической точке, наведенной
магнитным полем \cite{cust,geg3}. Из рисунка видно, что в
соответствии с \eqref{TB} все данные хорошо описываются линейной
функцией магнитного поля.

\begin{figure} [ht]
\begin{center}
\vspace*{-0.5cm}
\includegraphics [width=0.44\textwidth]{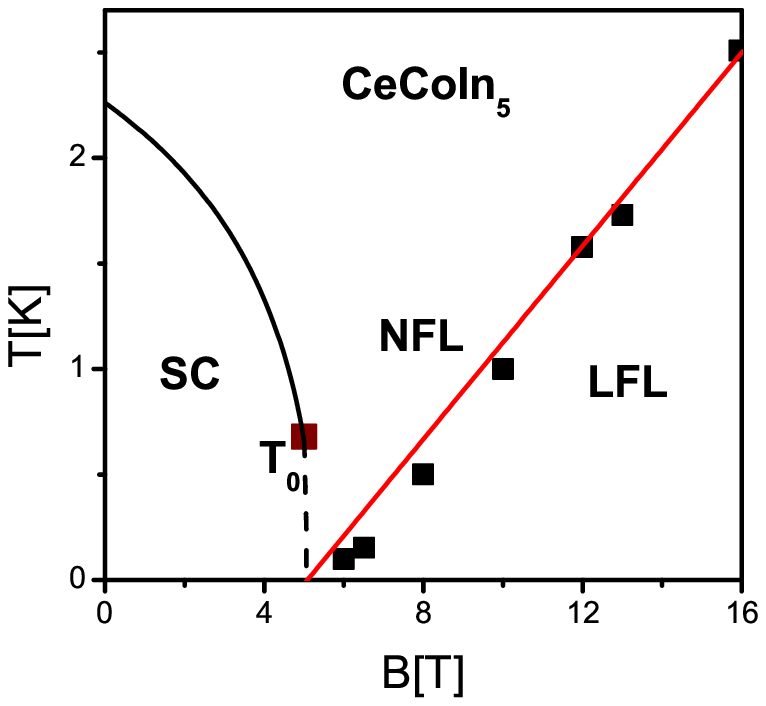}
\end{center}
\vspace*{-0.7cm} \caption{$B-T$ фазовая диаграмма металла с ТФ $\rm
CeCoIn_5$. Граница раздела между сверхпроводящей и нормальной фазами
показана сплошной линией до квадрата, где фазовый переход становится
переходом первого рода. При $T<T_0$ фазовый переход является
переходом первого рода  \cite{bian}, граница между сверхпроводящей и
нормальной фазами показана штриховой линией. Сплошная прямая линия,
заданная формулой \eqref{TB}, с экспериментальными точками
\cite{paglione}, показанными квадратами, обозначает раздел между ЛФЖ
(LFL) и НФЖ (NFL) состоянием.}\label{CeCo}
\end{figure}

Из Рис. \ref{TMM} следует, что критическое поле $B_{c2}=8$ T,
разрушающее сверхпроводимость, близко к критическому полю $B_{c0}=6$
T, помещающему металл в квантовую критическую точку. Это совпадение
не случайно, поскольку, как было показано выше, состояние системы с
ФК характеризуется сверхпроводящим параметром порядка $\kappa({\bf
p})$ и потому имеет большое сродство к сверхпроводимости
\cite{xoshag,obz1,sap}. Однако в магнитных полях $B>B_{c0}$, как
видно из Рис. \ref{Fig4}, ФК разрушается, система переходит в ФЖЛ,
сверхпроводящая щель становится экспоненциально малой при малой
константе связи $\lambda$ \cite{bcs}, $\Delta\propto
\exp{-1/\lambda}$, что ведет при дальнейшем увеличении поля к
разрушению сверхпроводимости. Из нашего рассмотрения следует, что
поля $B_{c0}$ и $B_{c2}$ удовлетворяют соотношению $B_{c2}\geq
B_{c0}$, т.е. магнитная квантовая критическая точка лежит внутри или
на краю $B-T$ области, занятой сверхпроводящей фазой. Отметим, что в
случае $\rm CeCoIn_5$, как видно из Рис. \ref{CeCo}, $B_{c0}\simeq
B_{c2}\simeq 5$ T \cite{pag}, это равенство $B_{c0}\simeq B_{c2}$
справедливо и для $\rm CeCoIn_{5-x}Sn_x$  при $x\leq 0.12$
\cite{bau}. Однако под воздействием внешнего давления равенство
перестает выполняться, и $B_{c2}>B_{c0}$ в случае $\rm CeCoIn_5$
\cite{ron}, что также происходит и для других металлов с ТФ,
например, для $\rm CePd_2Si_2$ \cite{mat}. Заметим, что при слабой
величине спаривательного взаимодействия антиферромагнитная фаза
вытесняет сверхпроводимость и $B_{c2}=0$, как это имеет место в
металлах $\rm YbRh_2Si_2$ и $\rm{YbRh_2(Si_{0.95}Ge_{0.05})_2}$,
которые не переходят в сверхпроводящее состояние при минимальных
достижимых температурах, а поле $B_{c0}$ имеет  конечное значение
\cite{geg,cust,geg3}.

Совпадение критических значений $B_{c0}\simeq B_{c2}$ ведет к
превращению сверхпроводящего фазового перехода второго рода в
фазовый переход первого рода в магнитных полях близких к $B_{c2}$ и
при $T<T_0$, как показано на Рис. \ref{CeCo}. При относительно
высоких температурах, когда фазовый переход является переходом
второго рода, энтропия и другие термодинамические функции являются
непрерывными функциями при температуре перехода $T_c(B)$. Из этого
условия непрерывности получаем
\begin{equation}
S_{SC}(T\to T_c(B))=S_{NFL}(T\to T_c(B))\label{SL},
\end{equation}
где $S_{SC}$ энтропия сверхпроводящей фазы и $S_{NFL}$ энтропия
системы, находящейся в состоянии НФЖ. Поскольку $S_{SC}(T\to0)\to0$,
а $S_{NFL}(T\to0)$ стремится к конечному значению $S_0$ \cite{sap},
уравнение \ref{SL} не может быть удовлетворено при достаточно
низких температурах $T\leq T_{0}$. Следовательно, рассматриваемый
фазовый переход второго рода становится переходом первого рода при
$T=T_{0}$, как это показано на Рис. \ref{CeCo}.

Сравнение $B-T$ фазовых диаграмм $\rm Tl_2Ba_2CuO_{6+x}$ и $\rm
CeCoIn_5$, показанных на Рис. \ref{TMM} и \ref{CeCo} соответственно,
указывает на их близость, что позволяет заключить, что, как и в
случае $\rm CeCoIn_5$ \cite{sap,epl1}, при низких температурах и в
сильных магнитных полях сверхпроводящий фазовый переход в ВТСП может
стать первого рода, а дифференциальная туннельная проводимость,
являющаяся асимметричной функцией напряжения, когда система
находится в НФЖ режиме, становится симметричной функцией при
восстановлении состояния ФЖЛ \cite{sap,shagpopov}. Однако надо иметь в
виду, что в случае $\rm Tl_2Ba_2CuO_{6+x}$ критические поля не
совпадают, $B_{c2}>B_{c0}$, поэтому  в полях $B>B_{c2}$ сплав
демонстрирует поведение ФЖЛ, член $S_0$ обнуляется, и уравнение
\ref{SL} может удовлетворяться при достаточно низких температурах.
Это обстоятельство может воспрепятствовать изменению типа перехода.
\begin{figure} [ht]
\begin{center}
\vspace*{-0.5cm}
\includegraphics [width=0.47\textwidth]{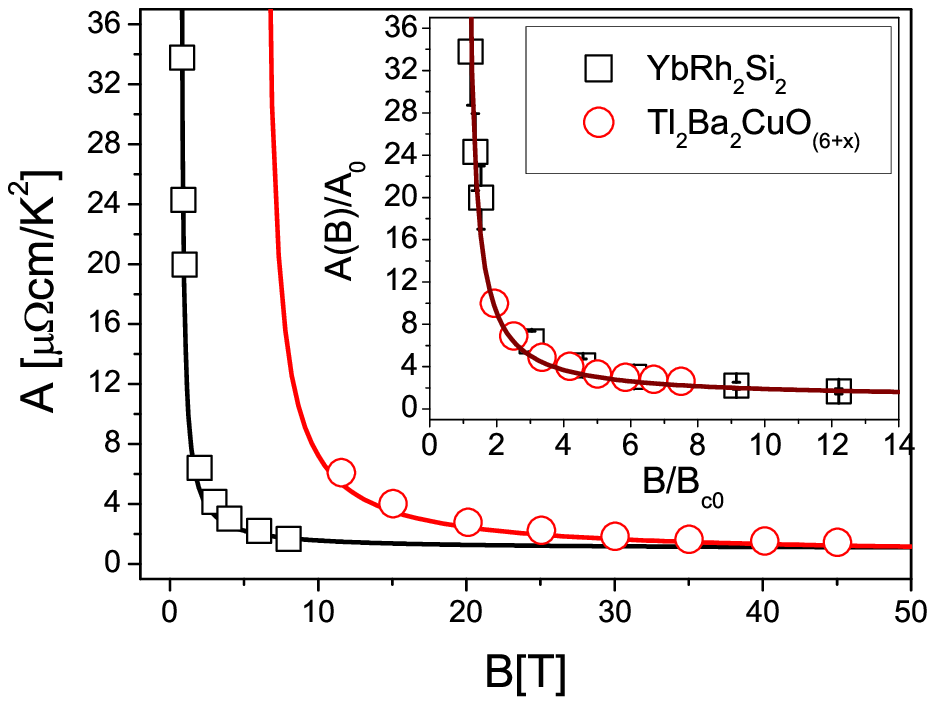}
\end{center}
\vspace*{-0.7cm} \caption{Коэффициент $A(B)$ показан квадратами для
$\rm YbRh_2Si_2$ \cite{geg}, кругами для $\rm Tl_2Ba_2CuO_{6+x}$
\cite{PNAS}, сплошные линии --- подгонка с помощью \eqref{HF6}.
Вставка: нормированный коэффициент  $A(B)/A_0$ как функция
нормированного магнитного поля $B/B_{c0}$. Сплошная линия задана
формулой \eqref{HF7}.}\label{fn2}
\end{figure}

При наложении на систему магнитного поля восстанавливается
квадратичная зависимость сопротивления от температуры:
$\rho(T)\propto AT^2$. Коэффициент $A(B)$ определяет зависящую от
температуры часть сопротивления, $\rho (T)=\rho_0+\Delta\rho$, где
$\rho_0$ --- остаточное сопротивление и $\Delta\rho=A(B)T^2$. Так
как этот коэффициент непосредственно определяется эффективной массой
\cite{ksch}, $A(B)\propto (M^*(B))^2$,
\begin{equation} A(B)\simeq
A_0+\frac{D}{B-B_{c0}},\label{HF6}\end{equation} где $A_0$ и $D$
константы.
На Рис. \ref{fn2} приведена подгонка при помощи формулы \ref{HF6}
коэффициента $A(B)$, полученного в измерениях на $\rm YbRh_2Si_2$
(квадраты) и $\rm Tl_2Ba_2CuO_{6+x}$ (круги). Как видно из рисунка,
оба набора экспериментальных точек хорошо описываются при помощи
формулы (\ref{HF6}), которая задает универсальное поведение $A(B)$ в
режиме ФЖЛ. Чтобы наглядно продемонстрировать это поведение,
перейдем к безразмерным величинам, представив (\ref{HF6}) в виде:
\begin{equation} \frac{A(B)}{A_0}\simeq
1+\frac{D_N}{B/B_{c0}-1},\label{HF7}\end{equation} где
$D_N=D/(A_0B_{c0})$ --- безразмерная константа. Из формулы
(\ref{HF7}) следует, что при помощи масштабного преобразования
значения коэффициента $A(B)$ для ВТСП $\rm Tl_2Ba_2CuO_{6+x}$ и
металла с ТФ $\rm YbRh_2Si_2$ могут быть описаны при помощи функции,
зависящей от одной переменной и одного параметра, если
соответствующие магнитные поля измеряются в единицах $B_{c0}$. Для
$\rm YbRh_2Si_2$ $B_{c0}\simeq 0.66$ T \cite{geg}, а для $\rm
Tl_2Ba_2CuO_{6+x}$, как видно из Рис. \ref{TMM}, $B_{c0}\simeq 6$ T.
На вставке Рис. \ref{fn2} приведены нормированные на $A_0$ данные
для $\rm Tl_2Ba_2CuO_{6+x}$ (показанные кругами) и для $\rm
YbRh_2Si_2$ (квадраты). Сплошной линией показан результат подгонки
при помощи (\ref{HF7}), который позволяет нам заключить, что ВТСП и
металлы с ТФ демонстрируют однотипное поведение вблизи квантовой
магнитной критической точки $B_{c0}$.

Таким образом, мы показали, что два разных сплава $\rm YbRh_2Si_2$ и
$\rm Tl_2Ba_2CuO_{6+x}$, характеризующихся разными микроскопическими
свойствами и принадлежащих к разным классам сильнокоррелированных
ферми-систем --- ВТСП и металлам с ТФ, имеют одну и ту же магнитную
квантовую критическую точку, связанную с ферми-конденсатным
квантовым фазовым переходом. Поэтому этот переход может
рассматриваться как универсальная причина сильнокоррелированного
поведения, наблюдаемого в самых различных металлах и жидкостях,
таких как высокотемпературные сверхпроводники, металлы с тяжелыми
фермионами и двумерные ферми-системы.

\section* { 4.9. Асимметричная проводимость
сильно коррелированных металлов. }\label{TUN}
\addcontentsline{toc}{section}{ 4.9. Асимметричная проводимость
сильно коррелированных металлов. }

В экспериментах на металлах с ТФ  исследуются, главным образом, их
термодинамические свойства. Было бы желательно изучать такие
свойства коррелированных электронных жидкостей, которые
определяются функцией распределения квазичастиц $n({\bf p},T)$, а
не только плотностью состояний или поведением эффективной массы
$M^*$ \cite{tun,shagpopov}. Сканирующая туннельная микроскопия и
контактная спектроскопия, тесно связанная с Андреевским отражением
\cite{andr,guy}, будучи чувствительными к плотности состояний и
функции $n({\bf p},T)$, определяющей вероятность заполнения
квазичастичных состояний, являются идеальной техникой для того,
чтобы изучать аномальное поведение сильно коррелированных
ферми-систем, определяемое функцией $n_0({\bf p},T)$ и энтропией
$S_0$.

\subsection* { 4.9.1. Нормальное состояние. }\label{NS}
\addcontentsline{toc}{subsection}{ 4.9.1. Нормальное состояние. }

Туннельный ток $I$ через точечный контакт между двумя обычными
металлами пропорционален приложенному напряжению $V$ и квадрату
модуля квантовомеханической амплитуды перехода $t$, умноженной на
разность $N_1(0)N_2(0)(n_1(p, T)-n_2(p,T))$ \cite{zag}. Здесь
$N_{1,2}(0)$ --- плотности состояний соответствующих металлов. С
другой стороны, волновая функция, вычисленная в приближении ВКБ и
определяющая амплитуду $t$, пропорциональна
$(N_1(0)N_2(0))^{-1/2}$. В результате плотность состояний выпадает
из ответа, и туннельный ток становится независимым от
$N_1(0)N_2(0)$. Учитывая, что при $T\to0$ распределение
$n(p,T\to0)\to\theta(p_F-p)$, где $\theta(p_F-p)$ --- ступенчатая
функция, можно проверить, что в рамках теории ферми-жидкости Ландау
дифференциальная туннельная проводимость $\sigma_d (V)=dI/dV$
является симметричной или четной функцией напряжения $V$.
Фактически, симметрия $\sigma_d(V)$ соблюдается при условии
существования симметрии дырка-квазичастица, которая есть в теории
ферми-жидкости Ландау. Поэтому наличие симметрии $\sigma_d(V)$
весьма очевидно и естественно в случае контактов металла с
металлом, когда они являются обычными и находятся в нормальном или
сверхпроводящем состояниях.

Рассмотрим туннельный ток при низкой температуре, который для
обычных металлов дается выражением \cite{guy,zag}
\begin{equation}\label{tun1}
I(V)=2|t|^2\int\left[n(\varepsilon-V)-
n(\varepsilon)\right]d\varepsilon.
\end{equation}
Мы используем атомную систему единиц: $e=m=\hbar =1$ и нормируем
амплитуду перехода на единицу, $|t|^2=1$. Так как температуры малы,
аппроксимируем функцию распределения обычного металла
$n(\varepsilon)$ ступенчатой функцией $\theta(\mu-\varepsilon)$ и
получаем из уравнения (\ref{tun1}) $I(V)=a_1V$, так что
дифференциальная проводимость $\sigma_d(V)=dI/dV=a_1=const$ есть
симметричная функция приложенного напряжения $V$.

Чтобы рассмотреть количественно поведение асимметричной части
проводимости $\sigma_d(V)$, продифференцируем обе стороны уравнения
(\ref{tun2}) по $V$ и получим уравнение для проводимости
$\sigma_d(V)$
\begin{equation}
\sigma_d=\frac{1}{T}\int n(\varepsilon(z)-V,T)
(1-n(\varepsilon(z)-V,T)) \frac{\partial \varepsilon}{\partial
z}dz,\label{tun2}
\end{equation} В подынтегральном выражении (\ref{tun2}) мы взяли в
качестве переменной безразмерный импульс $z=p/p_F$ вместо
$\varepsilon$, поскольку в случае сильно коррелированной
электронной жидкости $n(\varepsilon)$ уже не является функцией
переменной $\varepsilon$, а зависит от импульса, как это видно из
Рис. \ref{Fig1}. Действительно, переменная $\varepsilon$ в
интервале $(p_f-p_i)$ равна $\mu$, а функция распределения
квазичастиц изменяется в этой области. Из уравнения (\ref{tun2})
после несложных преобразований  получаем, что асимметричная часть
$\Delta \sigma_d(V)=(\sigma_d(V)-\sigma_d(-V))/2$ дифференциальной
проводимости равна
\begin{eqnarray}
&&\Delta\sigma_d(V)=\frac{1}{2}\int\frac{\alpha(1-\alpha^2)}
{[n(z,T)+\alpha[1-n(z,T)]^2} \nonumber\\
&\times&\frac{\partial n(z,T)}{\partial z}\frac{1-2n(z,T)}{[\alpha
n(z,T)+[1-n(z,T)]]^2}dz,\label{tun3}
\end{eqnarray}
здесь $\alpha=\exp(-V/T)$.

Асимметричную  туннельную проводимость можно наблюдать в измерениях
на металлах, электронная система которых содержит ФК. К таким
металлам можно отнести высокотемпературные сверхпроводники и
металлы с ТФ, например, YbRh$_2$(Si$_{0.95}$Ge$_{0.05}$)$_2$, $\rm
CeCoIn_5$ или YbRh$_2$Si$_2$. Измерения следует проводить, когда
металл с ТФ находится в сверхпроводящем  или нормальном состояниях.
Если металл находится в нормальном состоянии, то измерения
$\Delta\sigma_d(V)$ могут выполняться в магнитном поле $B>B_{c0}$
при температурах $T^*(B)<T\leq T_f$, или без магнитного поля при
температурах выше соответствующей критической температуры, когда
электронная система находится в парамагнитном состоянии и ее
поведение определяется энтропией $S_0$. Отметим, что при достаточно
низкой температуре $T<T^*(B)$ наложение магнитного поля $B>B_{c0}$
ведет к восстановлению поведения ферми-жидкости Ландау с $M^*(B)$,
определяемой уравнением (4.5), и асимметричное поведение
дифференциальной проводимости исчезает \cite{tun,shagpopov}.

\begin{figure} [ht]
\begin{center}
\includegraphics [width=0.47\textwidth] {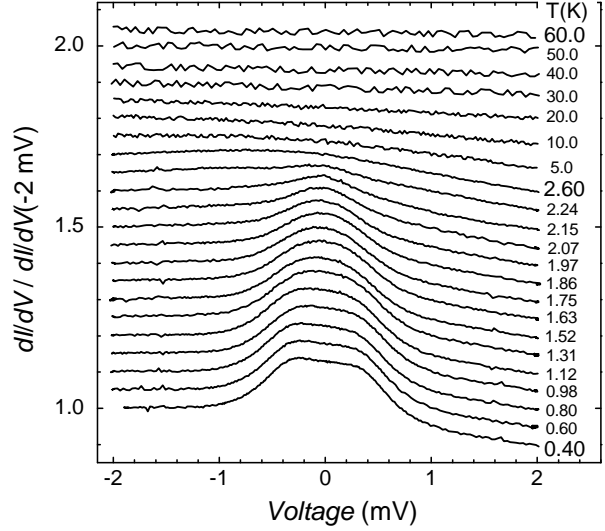}
\end {center}
\caption{Дифференциальная проводимость $\sigma_d(V)$, измеренная в
случае точечных контактов Au/CeCoIn$_5$. Графики функции
$\sigma_d(V)$ сдвинуты вдоль вертикальной оси на 0.05. Проводимость
нормирована на значение проводимости при -2 mV. Асимметрия
становится заметной при температуре $T<45$ К и возрастает с
уменьшением температуры \cite{park}.} \label{Fig1_t}
\end{figure}
Недавние измерения дифференциальной проводимости, проведенные на
CeCoIn$_5$ в технике спектроскопии с точечными контактами
\cite{park}, наглядно выявили асимметрию дифференциальной
проводимости в сверхпроводящем ($T_c=2.3$ K) и нормальном
состояниях. Результаты измерений приведены на Рис. \ref{Fig1_t}.
Как видно из Рис. \ref{Fig1_t}, $\sigma_d(V)$ почти постоянна,
когда металл с ТФ находится в сверхпроводящем состоянии, она не
испытывает заметного изменения вблизи $T_c$, затем, с ростом
температуры, монотонно убывает \cite{park}.
\begin{figure} [ht]
\begin{center}
\includegraphics [width=0.47\textwidth] {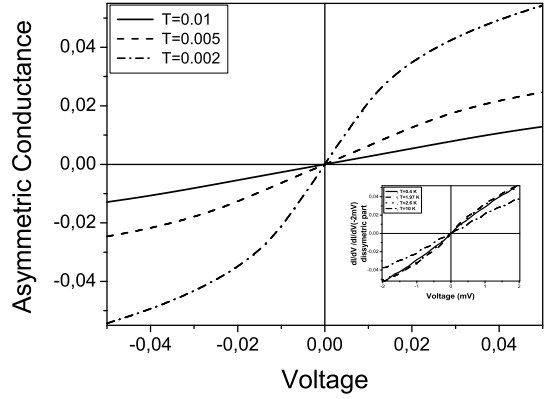}
\end {center}
\caption{Асимметричная проводимость $\Delta \sigma_d(V)$ как
функция нормированного на $\mu$,  $V/\mu$, приведена для трех
значений нормированной на $\mu$ температуры, $T/\mu$. Значения
нормированной температуры приведены в верхнем левом углу рисунка.
Асимметричная проводимость, полученная из данных приведенных на
Рис. 9.2, показана на вставке внизу.} \label{Fig2_t}
\end{figure}
Результаты вычислений асимметричной части $\Delta\sigma_d(V)$
проводимости $\sigma_d(V)$ по (\ref{tun3}) приведены на Рис.
\ref{Fig2_t}. Для вычисления функции распределения $n(z,T)$
использовался функционал (\ref{TRMKS1}), параметры $\beta=3$  и $g=8$.
В этом случае $(p_f-p_i)/p_F\simeq 0.1$. Как видно из Рис.
\ref{Fig2_t}, асимметричная часть проводимости $\Delta \sigma_d(V)$
линейно зависит от $V$ при малых значениях напряжения и убывает с
ростом температуры что соответствует поведению экспериментальных
кривых, показанных на вставке Рис. \ref{Fig2_t}.

Получим оценочную формулу для анализа асимметричной части
дифференциальной проводимости. Как следует из (\ref{tun3}), при
малых величинах напряжения $V$ асимметричная часть ведет себя как
$\Delta\sigma_d(V)\propto V$. Здесь уместно напомнить, что
асимметричная часть туннельной проводимости является нечетной
функцией $V$, поэтому $\Delta\sigma_d(V)$ должна менять знак при
изменении знака напряжения: $\Delta\sigma_d(V)\propto V$.
Естественной единицей для измерения напряжения является $2T$,
поскольку эта величина определяет характерную энергию для ФК, как
это видно из уравнения $E_{0}=4T$. Фактически, асимметричная
часть должна быть пропорциональна области, занятой $(p_f-p_i)/p_F$
ФК:
\begin{equation}
\Delta\sigma_d(V)\simeq c\frac{V}{2T}\frac{p_f-p_i} {p_F}\simeq
c\frac{V}{2T}\frac{S_0}{x_{FC}}.\label{tun4}
\end{equation}
Здесь, $S_0/x_{FC}\sim (p_f-p_i)/p_F$ --- не зависящая от
температуры часть энтропии (см. уравнение (\ref{SL1})), $c$ ---
константа порядка  единицы. Эта константа может быть оценена путем
использования аналитически решаемых моделей. Например, вычисления
$c$ в рамках простой модели с функционалом Ландау $E[n(p)]$ вида
\cite{ksk}
\begin{equation}
E[n(p)]=\int\frac{p^2}{2M}\frac{d{\bf p}}{(2\pi)^3} +V_1\int
n(p)n(p)\frac{d{\bf p}}{(2\pi)^3},\label{tun5}
\end{equation}
дают $c\simeq 1/2$. Из уравнения (\ref{tun4}) следует, что когда
$V\simeq 2T$ и ФК занимает заметную часть объема Ферми,
$(p_f-p_i)/p_F\simeq1$, асимметричная часть становится сопоставимой
с дифференциальной туннельной проводимостью $\Delta\sigma_d (V)\sim
V_d(V)$.

\subsection* { 4.9.2. Сверхпроводящее состояние. }\label{SC}
\addcontentsline{toc}{subsection}{ 4.9.2. Сверхпроводящее состояние. }

Туннельная проводимость может оставаться асимметричной, когда
высокотемпературный сверхпроводник или рассматриваемый металл с ТФ
переходят из нормального в сверхпроводящее состояние. Причина этого
состоит в том, что функция $n_0({\bf p})$ снова определяет
дифференциальную проводимость. Как мы видели в разделе \ref{SC},
$n_0({\bf p})$ заметно не искажается спаривательным
взаимодействием, которое является относительно слабым по сравнению
с взаимодействием Ландау, формирующим функцию распределения
$n_0({\bf p})$. Поэтому асимметричная часть проводимости почти не
изменяется при  $T\leq T_c$, что находится в согласии с
экспериментом, см. Рис. \ref{Fig1_t}. При вычислении проводимости,
измеряемой при помощи туннельного микроскопа, необходимо учесть,
что плотность состояний,
\begin{equation}  N_s(E)=
N(\varepsilon-\mu)\frac{E}{\sqrt{E^2-\Delta^2}}, \label{tun6}
\end{equation}
определяет проводимость, которая при $E\leq|\Delta|$ равна нулю.
Здесь $E$ --- энергия квазичастицы, заданная соотношением
\begin{equation}E({\bf p})=\sqrt{\xi^2({\bf
p})+\Delta^2({\bf p})},\label{SC3.1}\end{equation} 
где $\xi({\bf p})=\varepsilon({\bf p})-\mu$, 
$\Delta({\bf p})$ --- сверхпроводящая щель и $\varepsilon-\mu=\sqrt{E^2-\Delta^2}$. Из
уравнения (\ref{tun6}) следует, что туннельная проводимость может
быть асимметричной, если плотность состояний $N(\varepsilon)$
асимметрична по отношению к ферми-уровню \cite{pand}, как это имеет
место в случае сильно коррелированных ферми-систем с ФК. Наши
вычисления плотности состояний, основанные на модельном функционале
(\ref{NMC1}) и с теми же параметрами, какие использовались при
вычислении проводимости $\Delta \sigma_d(V)$, приведенной на Рис.
\ref{Fig2_t}, подтверждают этот вывод.
\begin{figure} [ht]
\begin{center}
\includegraphics [width=0.47\textwidth] {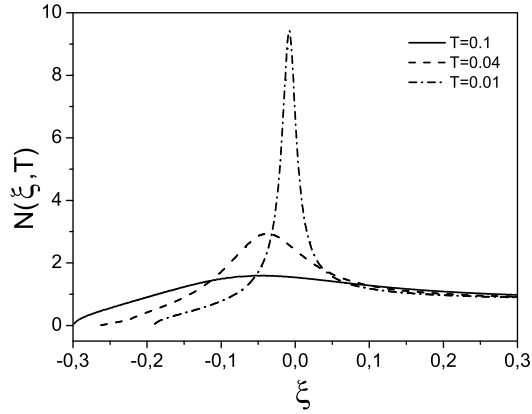}
\end{center}
\caption {Плотность состояний $N(\xi,T)$ как функция переменной
$\xi=(\varepsilon-\mu)/\mu$, вычисленная при трех значениях
нормированной на $\mu$ температуры $T$. Значения нормированной
температуры показаны в правом верхнем углу рисунка.} \label {Fig7}
\end{figure}
На Рис. \ref{Fig7} представлены вычисления плотности состояний
$N(\xi,T)$. Из Рис. \ref{Fig7} видно, что $N(\xi,T)$ существенно
асимметрична относительно уровня Ферми. Если рассматриваемая
система находится в сверхпроводящем состоянии, то значения
нормированной температуры, приведенные в правом верхнем углу Рис.
\ref{Fig7}, можно связать с $\Delta_1$. Принимая во внимание, что
$\Delta_1\simeq 2T_c$, получаем $2T/\mu\simeq \Delta_1/\mu$. Так
как $N(\xi,T)$ асимметрична, первая производная $\partial
N(\xi,T)/\partial\xi$ у ферми-уровня не равна нулю, и при малых
значениях  $\xi$, $N(\xi,T)$ может быть записана как
$N(\xi,T)\simeq a_0+a_1\xi$. Коэффициент $a_0$ не дает вклада в
асимметричную часть. $\Delta \sigma_d(V)$, очевидно, определяется
коэффициентом $a_1$, который $a_1\propto M^*(\xi=0)$. В свою
очередь, $ M^*(\xi=0)$ определяется соотношением 
\begin{equation} M^*_{FC}\simeq
p_F\frac{p_f-p_i}{2\Delta_1}.\label{SC7}\end{equation}
В результате из уравнения (\ref{tun6}) получаем:
\begin{equation}\Delta \sigma_d(V)\sim c_1\frac{V}{|\Delta|}\frac{S_0}{x_{FC}},
\label{tun7}\end{equation} поскольку $(p_f-p_i)/p_F\simeq
S_0/x_{FC}$, энергию $E$ заменяем на $V$, и
$\xi=\sqrt{V^2-\Delta^2}$. В (\ref{tun7}) энтропия $S_0$ относится
к нормальному состоянию металла с ТФ. Отметим, что (\ref{tun7}) по
сути совпадает с (\ref{tun4}), если учтем, что характерная энергия $E_0$
для сверхпроводящего состояния определяется равенством :
\begin{equation} E_0=\varepsilon({\bf p}_f)- \varepsilon({\bf
p}_i)\simeq p_F \frac{(p_f-p_i)}{M^*_{FC}}\simeq
2\Delta_1.\label{SC8}\end{equation}
и не зависит от температуры. При изучении универсального поведения
асимметричной проводимости формула (\ref{tun7}) удобнее, чем
расчеты, основанные на формуле (\ref{tun6}). Из формул (\ref{tun4})
и (\ref{tun7}) следует, что измерения транспортных свойств
(асимметричной части проводимости) позволяют определять
термодинамические свойства нормальной фазы, связанные с энтропией
$S_0$. Из уравнения (\ref{tun7}) видно, что асимметричная часть
дифференциальной туннельной проводимости становится сравнимой с
дифференциальной туннельной проводимостью при $V\sim 2|\Delta|$,
если ФК занимает заметную часть объема Ферми,
$(p_f-p_i)/p_F\simeq1$. В случае $d$-волновой симметрии щели правая
часть (\ref{tun7}) должна быть усреднена по распределению щели
$\Delta(\phi)$, где $\phi$ --- угол. Эта процедура не представляет
труда и сводится к переопределению величины щели или константы
$c_1$. В результате (\ref{tun7}) будет применимо и при
$V<\Delta_1$, где $\Delta_1$ --- максимальное значение $d$-волновой
щели \cite{tun}. При Андреевском отражении, когда ток отличен от
нуля при любых малых значениях $V$, формула (\ref{tun7}) применима
при $V<\Delta_1$ и в случае $s$-волновой щели.
\begin{figure} [ht]
\begin{center}
\includegraphics [width=0.47\textwidth] {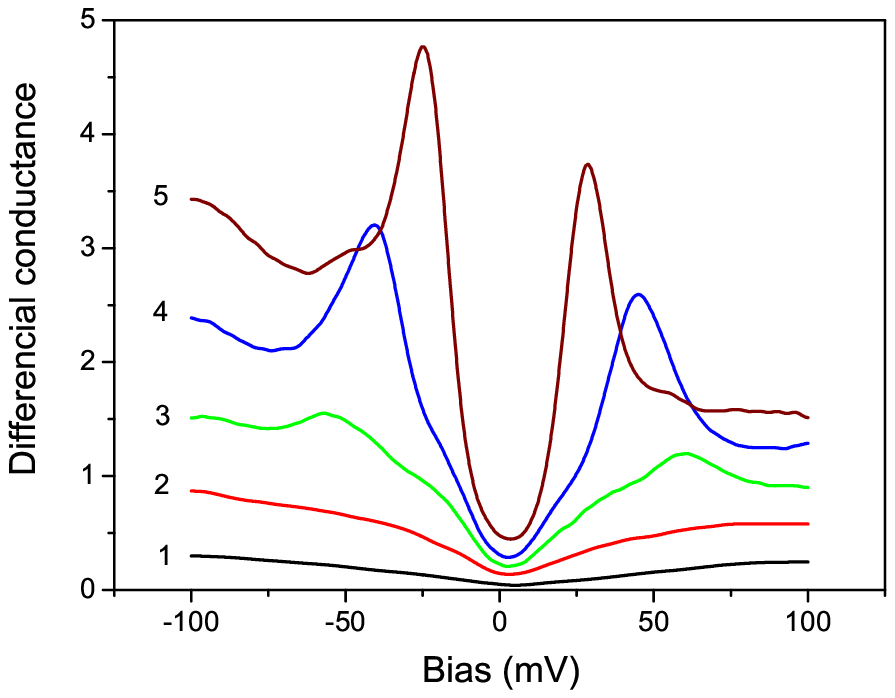}
\end {center}
\caption{Пространственное изменение спектров туннельной
дифференциальной проводимости, измеренной на $\rm
Bi_2Sr_2CaCu_2O_{8+x}$. Кривые 1 и 2 относятся к областям, где
интегральная локальная плотность состояний является очень малой.
Низкая дифференциальная проводимость и отсутствие сверхпроводящей
щели являются показателями изолятора. Кривая 3 соответствует
большой щели величиной 65 meV со слабо выраженными пиками.
Интегральная величина локальной плотности состояний для кривой 3
является малой, но больше, чем для кривых 1 и 2. Кривая 4
--- для щели размером 40 meV, такая величина является близкой к
средней величине распределения щелей. Кривая 5 соответствует
максимальной интегральной локальной плотности состояний и
наименьшей щели 25 meV и имеет два острых когерентных пика
\cite{pan}.}\label{Fig3_t}
\end{figure}

Неоднородность в плотности распределения электронов в $\rm
Bi_2Sr_2CaCu_2O_{8+x}$ была обнаружена в низкотемпературных
измерениях с использованием туннельной микроскопии и спектроскопии
\cite{pan}. Эта неоднородность проявляется как пространственное
изменение в локальной плотности состояний в низко энергетической
части спектра и в величине сверхпроводящей щели. Неоднородность,
наблюдаемая в интегральной локальной плотности состояний, вызвана
не примесями, а свойственна природе системы. Наблюдения позволили
связать величину интегральной локальной плотности состояний с
концентрацией $x$ локальных добавок кислорода. Пространственные
изменения спектра туннельной дифференциальной проводимости показаны
на Рис. \ref{Fig3_t}, из которого видно, что туннельная
дифференциальная проводимость сильно асимметрична в сверхпроводящем
состоянии соединения $\rm Bi_2Sr_2CaCu_2O_{8+x}$. Туннельная
дифференциальная проводимость, показанная на Рис. \ref{Fig3_t},
может рассматриваться как измеренная при разных значениях
$\Delta_1(x)$, но при одной и той же температуре, что позволяет
изучить зависимость $\Delta \sigma_d(V)$ от величины $\Delta_1(x)$.
\begin{figure} [ht]
\begin{center}
\includegraphics [width=0.47\textwidth] {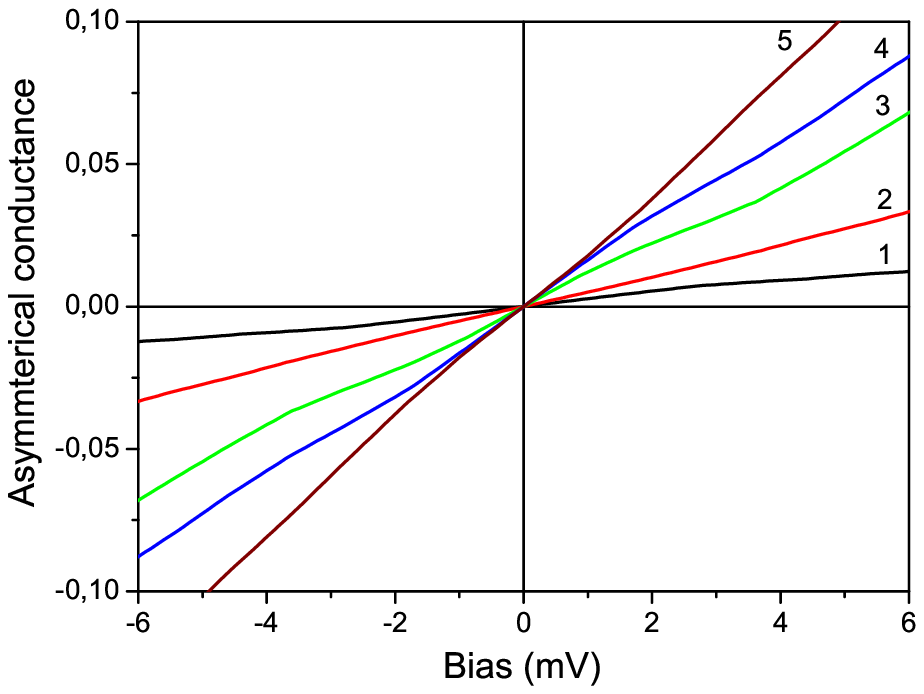}
\end {center}
\caption{Асимметричная часть $\Delta\sigma_d(V)$ туннельной
дифференциальной проводимости, измеренная на ВТСП $\rm
Bi_2Sr_2CaCu_2O_{8+x}$ и извлеченная из данных, приведенных на Рис.
\ref{Fig3_t}, показана как функция напряжения $V$ (mV). Номер
графика, представляющего асимметричную часть, соответствует номеру
графика на Рис. \ref{Fig3_t}, из которого были получены данные для
построения этого графика.}\label{Fig4_t}
\end{figure}
На Рис. \ref{Fig4_t}  приведены графики асимметричной проводимости,
полученные из данных, показанных на Рис. \ref{Fig3_t}. Видно, что
при малых значениях $V$ и в соответствии с (\ref{tun7})
$\Delta\sigma_d(V)$ есть линейная функция напряжения, а наклон
соответствующих прямых линий $\Delta\sigma_d(V)$ обратно
пропорционален величине щели $\Delta_1$. Величины $\Delta_1$
указаны в подписи к Рис. \ref{Fig3_t}.
\begin{figure} [ht]
\begin{center}
\includegraphics [width=0.47\textwidth] {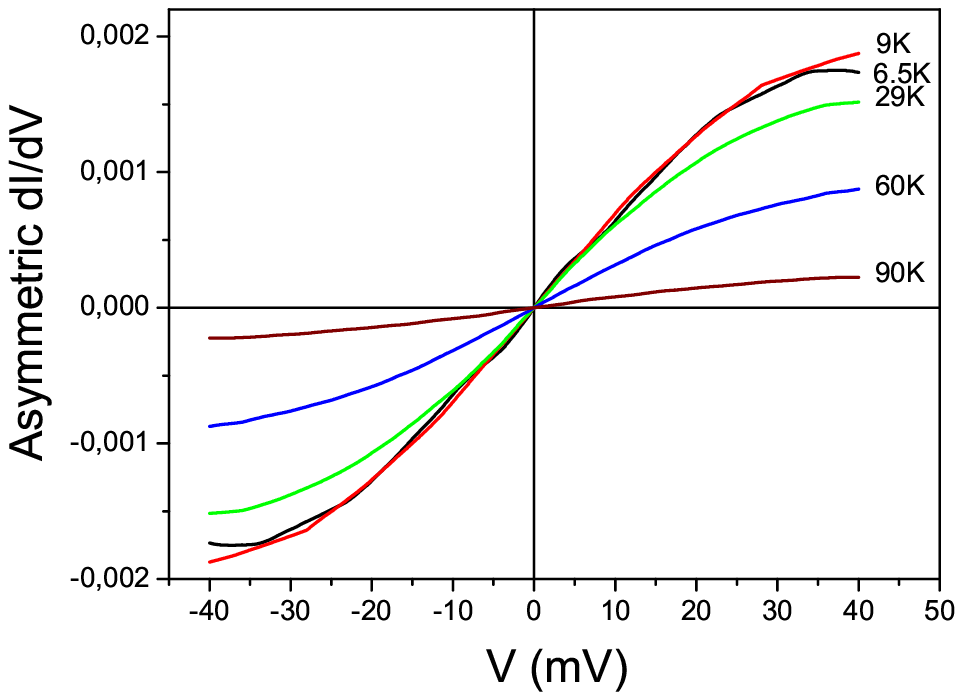}
\end {center}
\caption{ Температурная зависимость асимметричной части
$\Delta\sigma_d(V)$ спектров проводимости, полученных в измерениях
на $\rm YBa_2Cu_3O_{7-x}/La_{0.7}Ca_{0.3}MnO_3$ в рамках контактной
спектроскопии, критическая температура $T_c\simeq 30$
\cite{samanta}. Величины температур, при которых проводились
измерения, приведены в верхнем правом углу рисунка.} \label{Fig5_t}
\end{figure}
На Рис. \ref{Fig5_t} показано изменение асимметричной части
проводимости $\Delta\sigma_d(V)$ с ростом температуры. Измерения
были проведены на $\rm YBa_2Cu_3O_{7-x}/La_{0.7}Ca_{0.3}MnO_3$ с
$T_c\simeq 30$ \cite{samanta}. Видно, что при $T<T_c$ и в области
линейной зависимости от $V$, $\Delta\sigma_d(V)$ слабо зависит от
температуры, такое поведение соответствует (\ref{tun7}). При
$T>T_c$, наклон прямых участков графиков $\Delta\sigma_d(V)$
убывает с ростом температуры, это поведение воспроизводится
формулой (\ref{tun4}). Можно заключить, что описание универсального
поведения асимметричной части проводимости $\Delta\sigma_d(V)$ при
помощи формул (\ref{tun4}) и (\ref{tun7}) находится в хорошем
согласии с экспериментами, представленными на Рис. (\ref{Fig2_t},
\ref{Fig4_t} и \ref{Fig5_t}).
\pagebreak

\newpage
\section* { 4.10. Выводы и заключение. }\label{VZ}
\addcontentsline{toc}{section}{ 4.10. Выводы и заключение. }

В докладе рассмотрена теория ферми-конденсатного квантового
фазового перехода и его влияние на свойства различных ферми-систем. Представлено
значительное число свидетельств, основанных как на теоретических исследованиях, так и на компьютерных экспериментах, в поддержку его существования.
Продемонстрировано, что многочисленные экспериментальные факты,
собранные в исследованиях разнообразных материалов, таких как
высокотемпературные сверхпроводники, металлы с тяжелыми фермионами
и коррелированные двумерные ферми-структуры, могут быть объяснены в
рамках теории, основанной на ферми-конденсатном квантовом фазовом
переходе.


\tableofcontents
\addcontentsline{toc}{chapter}{\bf Оглавление}

\end{document}